\documentclass[times]{TRR}
\makeatletter
\renewcommand{\endpage}{\pageref{FirstPage}} 
\makeatother

\usepackage{threeparttable}
\usepackage{placeins}
\usepackage{svg}
\usepackage{moreverb,url}
\usepackage[pagewise, columnwise]{lineno}
\usepackage{booktabs}

\newcommand\BibTeX{{\rmfamily B\kern-.05em \textsc{i\kern-.025em b}\kern-.08em
T\kern-.1667em\lower.7ex\hbox{E}\kern-.125emX}}

\usepackage{xcolor}
\newcommand{\rev}[1]{\textcolor{black}{#1}}

\begin{document}

\runninghead{Ahmed and Hawkins}

\title{Residential Price Modeling using Spatially Validated Machine Learning Methods: A Comparison Across Geographic Contexts}

\author{Usman Ahmed\affilnum{1} and Jason Hawkins\affilnum{2}}

\affiliation{\affilnum{1}University of Tennessee, Knoxville, TN, USA\\
\affilnum{2}University of Calgary, Calgary, Canada}

\corrauth{Usman Ahmed, uahmed@utk.edu}

\begin{abstract}
Land use (i.e., buildings) and transportation infrastructure are tightly coupled systems, such that residential property prices play a critical role in transportation planning. In a similar manner, transportation infrastructure influences property prices such that accurate forecasts of both systems are related research problems. \rev{The objective of this study is to examine these interactions and evaluate the effectiveness of machine learning methods in modeling residential real estate prices across different urban contexts. Specifically,} we first examine the impact of land and transportation infrastructure on residential real estate prices using XGBoost and Random Forest machine learning methods, and compare results between two cities representing diverse geographic and socioeconomic contexts. Second, we investigate the application of machine learning methods on spatial data and provide a comparison of non-spatial and spatial cross-validation on the performance of machine learning methods. We use SHAP values to study the impact of land use and transportation infrastructure on real estate prices. The models are applied, and results are compared between the Rawalpindi and Islamabad Metropolitan Area in Pakistan and the City of Toronto in Canada. We find that, despite differences in demographics and economic development, the two cities exhibit similarities in the effect of transportation infrastructure and local amenities on dwelling prices. Proximity to the major central city cores (i.e., downtowns) increases sale price. The effect of transportation infrastructure is differentiated, with high quality transit (e.g., subway and bus rapid transit) increasing and conventional bus stop proximity decreasing sale price, respectively. We confirm the previous finding in other fields that non-spatial cross-validation over-estimates the prediction accuracy of machine learning algorithms on spatially referenced datasets. We find that XGBoost model has slightly higher performance than the Random Forest model. We recommend careful use of machine learning methods in the case of spatial data specifically in modeling of land prices.


\end{abstract}

\maketitle

\section{Introduction}
The interaction of land use and transportation systems has been studied by many researchers over the past few decades, resulting in the development of computational integrated land use and transportation simulation models \citep{hawkins2019integrated}. Residential property price is an important component of the integrated modeling paradigm. Price plays an important role in residential (re)location choice of households. Residential location has a profound impact on travel behavior, specifically in terms of number and length of trips. Many studies have confirmed the impacts of the transportation system on residential property prices \citep{hawkins2018spatio,waddell2018integrated}. We aim to address two research gaps in the existing literature with this study. First, the interaction of property price with the transportation system is mostly studied in the context of developed cities, while these intricacies are less studied for and compared with developing cities. We contrast results for a developed city (Toronto, Canada) with those found for developing cities (Rawalpindi and Islamabad, Pakistan). Second, most machine learning (ML) applications to property price modeling do not adequately treat the spatial nature of property price data.

Machine learning methods are used across quantitative research domains for their data-driven basis and flexibility. This flexibility provides strong predictive accuracy. However, there remain many open questions for machine learning applications in regional planning. One set of questions stems from the spatial nature of most planning problems. A core component of many machine learning algorithms and validation methods is k-fold cross-validation. Cross-validation is used to test the transferability of a model trained on one dataset to predict the same outcome variable for an independent sample drawn from the same population. Standard k-fold cross-validation partitions randomly drawn observations into k equally-sized samples. The model is trained on k-1 samples and tested on the hold-out sample until all samples are used in both training and testing. In the case of spatial data, the independence assumption underlying k-folds cross-validation is violated due to spatial autocorrelation among the observations. Several studies \citep{ploton2020spatial, meyerImportanceSpatialPredictor2019,robertsCrossvalidationStrategiesData2017} in ecology and other fields find that ignoring spatial autocorrelation leads to over-prediction bias. Spatial k-folds addresses this bias by sampling data based on spatial clustering. Machine learning methods also have the advantage of representing non-linear relationships, which are harder to capture using classical regression techniques.

The objectives of this study are: 1) application of machine learning to model residential property prices, with a focus on investigating the impact of land use and transportation infrastructure on prices; 2) compare the impact of non-spatial and spatial cross-validation methods on model performance to guide accurate use of the model and interpretation of model results; 3) compare two machine learning methods XGBoost and Random Forest (RF); 4) explore non-linear relationships between property price and its predictors using SHAP values; 5) compare the factors impacting property price between a developed city and a developing city. In this study, we train XGBoost and RF models on residential price data for two metropolitan regions representing diverse conditions - Rawalpindi and Islamabad Metropolitan Area (RIMA) in Pakistan and the City of Toronto in Canada. A 5-fold cross-validation method is used, wherein spatial folds are created using k-means clustering. SHAP plots are used to identify and explain the most important price determinants, along with their non-linear relationships and make comparisons between metropolitan areas. Machine learning methods can overcome some of the challenges in property price modeling when investigating the impact of transportation infrastructure. Some studies have found non-linear relationships between property prices and transportation infrastructure. For instance, Mohammad et al. \citep{mohammad2013meta} found that the impact of distance to rail stations on property prices shows a non-linear behavior. Therefore, machine learning models are promising for investigating the impact of transportation infrastructure on property prices.

\section{Study Area}
\begin{figure*}[]
   \centering
    \includegraphics[width=1\linewidth]{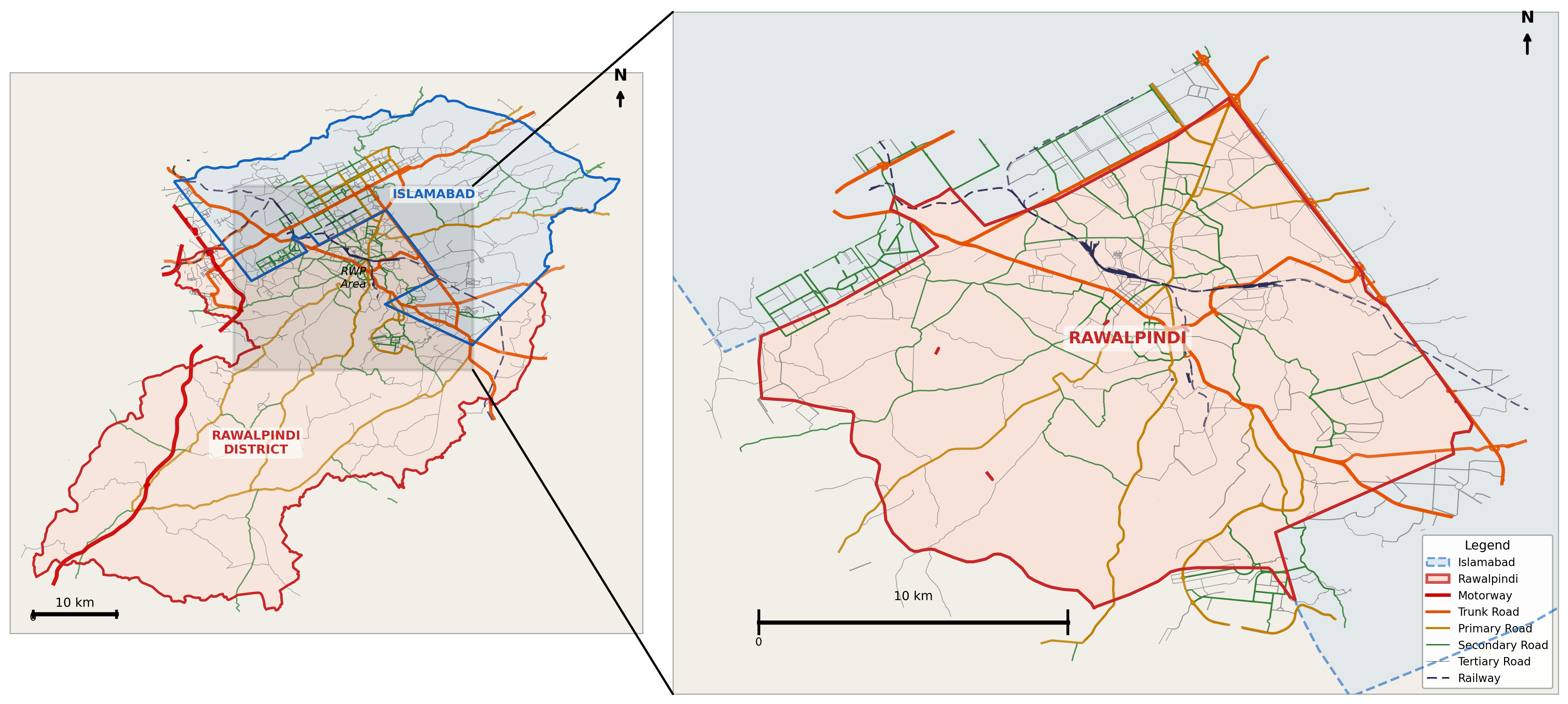}
    \caption{Rawalpindi Islamabad Metropolitan Area (RIMA)}
    \label{fig:rima_map}
\end{figure*}

\begin{figure*}[]
   \centering
    \includegraphics[width=1\linewidth]{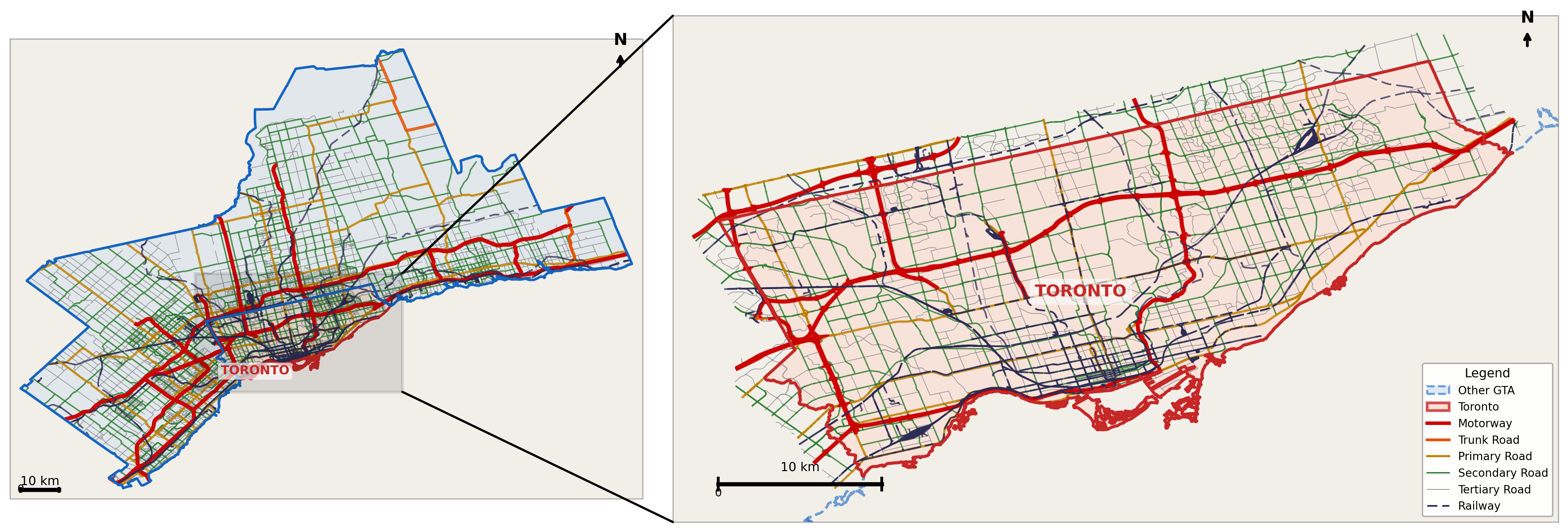}
    \caption{Greater Toronto Area (GTA)}
    \label{fig:gta_map}
\end{figure*}

The two metropolitan areas chosen for this study differ significantly in their real estate and transportation features. The Greater Toronto Area (GTA) is consistently among the ten least affordable real estate markets in the world \citep{Demographia_2023}, while RIMA does not appear in the rankings due to its lower prices. In this study, we focus on the City of Toronto proper due to real estate data availability, excluding its surrounding suburbs from analysis. However, we provide some statistics for the larger GTA as a metropolitan area comparison to RIMA. \rev{Figure \ref{fig:rima_map} shows the Rawalpindi District, Rawalpindi City, and the City of Islamabad, including major transportation infrastructure. In this study, our analysis focuses on the RIMA region, which comprises the cities of Rawalpindi and Islamabad. Figure \ref{fig:gta_map} shows the City of Toronto and its surrounding area (i.e., the GTA), including major transportation infrastructure.} In 2016, the average dwelling in Toronto sold for 730,000 CAD (351 USD per sqft) \citep{the_canadian_press_toronto_2019}, while a dwelling in RIMA cost about 273 USD per sqft (adjusting for purchasing power parity) \citep{GlobalPropertyGuide_2019,WorldBank_2024}. Table \ref{tab:sum_stats} provides additional comparison statistics. The GTA has roughly twice the population of RIMA but at half the population density. With respect to transportation infrastructure, Toronto is a developed city with an extensive multi-modal regional transit system that includes subway, streetcar, bus, and commuter rail connections to the GTA and beyond. RIMA introduced a bus rapid transit network in 2015, which has since expanded to 84 km of service \citep{Rawalpindi-IslamabadMetrobus_2024}. RIMA has plans for an integrated transportation system, including adding more feeder routes in phase 1 of the project \citep{PunjabMassTransitAuthority_2024} and anticipated enhanced transit service in phase 2 which make this study timely to study the impacts of such services on property prices.

\begin{table}[]
\centering
\begin{threeparttable}
\caption{Comparison of GTA and RIMA key statistics
\cite{CityofToronto_2017,Knoema_2023, Wikipedia_2023, worldbankWorldBankOpen2024}}
\label{tab:sum_stats}
\begin{tabular}{@{}lll@{}}
\toprule
Indicators                                                         & GTA       & RIMA      \\ \midrule
Land Area (km\textsuperscript{2})                                                    & 5,903     & 1,165     \\ 
Population                                                         & 6,471,850 & 3,108,063 \\ 
Population density (persons per km2)                               & 1,096     & 2,668     \\ 
Unemployment Rate (Population 15+)                                 & 6.3\%     & 3.8\%     \\ 
National GDP (\$ per capita)\tnote{1} & \$47,458  & \$4,747   \\
Average Household Income (\$)\tnote{1}                   & \$71,802  & \$651  \\ \bottomrule
\end{tabular}
\begin{tablenotes}
\item[1] All dollar amounts are converted to 2016 international dollars
\end{tablenotes}
\end{threeparttable}
\end{table}

The following section summarizes the two relevant literature. The first topic is machine learning applications to real estate price prediction. The second topic is cross-validation methods applied to spatial datasets. The data used in the empirical study are then provided for each study area. We describe the real estate price model and contrast non-spatial and spatial results. Real estate price model results are provided for each study area. Important variables are compared using SHapley Additive exPlanations (SHAP) plots. We then conclude with a discussion of the implications of model setup on results and avenues for future research.

\section{Literature Review}
Hedonics, as introduction by Rosen \cite{Rosen_1974}, provides a theoretical basis for decomposing prices as arising from the combination of their attributes. The hedonic price literature for land and property is also informed by bid-rent theory \citep{Alonso_1964}, which posits that households trade-off between property and transportation costs. We structure the literature review around three themes. First, machine learning (ML) applications in real estate price hedonics. Second, the relationship between transportation and locational variables with property prices. Third, the use of cross-validation methods in spatial models across diverse fields.

\subsection{Machine Learning in Real Estate Price Modeling}
Machine learning methods have seen increased use in land and property price prediction to address the limitations of classical regression models, including low prediction accuracy and multicollinearity \citep{binoy2022factors}. Property price exhibits a strong spatial association, such that any model assessment must include a spatial component. For example, Ho et al. \citep{ho2021predicting} compared three machine learning methods: support vector machines (SVM), random forest (RF), and gradient boosting machine (GBM). They found that SVM provides the highest prediction accuracy among these methods. However, as summarized below, prediction accuracy is strongly influenced by the inclusion of spatial dependency in the goodness-of-fit statistic. Depending on the modeling approach and data, it is conceivable that SVM was not the highest performing method in the Ho et al. \citep{ho2021predicting} study if one were to consider spatial dependency. Soltani et al. \cite{Soltani_Zali_Aghajani_Hashemzadeh_Rahimi_Heydari_2024} estimate artificial neural network (ANN) models for Tehran using non-spatial cross-validation for comparison with a linear regression model. Both Genc and Colak \cite{Genc_Colak_2026} and Zhang et al. \cite{Zhang_Li_Lownes_Zhang_2021} highlight the importance of integrating spatially explicit methods (e.g., geographically weighted regression (GWR)) into hedonic price models to better understand localized positive and negative impacts of transit investments, urban form, and neighborhood characteristics on housing markets.

Louati et al. \citep{louatiPriceForecastingReal2022} highlighted the use of property price as an economic indicator. They estimated various machine learning models for price forecasting in Riyadh, Saudi Arabia. Their model used a district indicator, land and floor area, latitude and longitude, and type indicator (e.g., land, house, flat). Random forest provided the strongest predictive accuracy based on median squared error (MSE) and mean absolute error (MAE). Limsombunc et al. \citep{limsombuncHousePricePrediction2004} followed a similar approach in Christchurch, New Zealand using artificial neural networks (ANN) to predict residential property price based on property features and community amenities. Their price dataset was relatively small, only 200 observations, so statistical power was low. They based their goodness-of-fit analysis on the $R^2$ and root mean squared error (RMSE) statistics. Based on out-of-sample prediction against a linear regression model, they concluded that ANN provides a higher prediction accuracy. Another study used support vector machines (SVM) and grey model (GM) methods to predict average housing prices by district in China \citep{jirongHousingPriceForecasting2011}. They found that SVM significantly outperforms GM based on absolute relative error (ARE). A study in Chongqing, China found that SVM outperforms neural nets based on prediction accuracy \citep{wangRealEstatePrice2014}. A study in Turkey compared adaptive neuro-fuzzy (ANFS) with grid partition (ANFS-GP) and sub-clustering (ANFS-SC) \citep{gerekHouseSellingPrice2014}. Despite the variation in methods, all studies have commonalities in their treatment of space being based solely on neighborhood attributes and district fixed effects. There is no consideration of spatial dependence in either models or goodness-of-fit statistics.

Mathotaarachchi et al. \cite{Mathotaarachchi_Hasan_Mahmood_2024} found overfitting in property price predictions with XGBoost and RF when evaluated using standard cross-validation. \cite{Genc_Colak_Ozbilgin_2025} applied various machine learning models to property prices in Trabzon, T\"{u}rkiye, from which they measured local spatial prediction using inverse distance weighted (IDW) spatial interpolation. They found RF performed the best in terms of predicting adjacent prices.

Overall, we found that existing property price ML models fail to adequately consider spatial dependence in their design. Many studies found overfitting using non-spatial cross-validation but did not conduct suitable adjustments, such as incorporating spatial lag dependence or spatial cross-validation.

\subsection{Impact of Transportation Infrastructure on Real Estate}

There are numerous studies that examine the relationship between transportation variables and real estate prices \citep{hawkins2018spatio,Choi_Park_Dewald_2021,Soltani_Heydari_Aghaei_Pettit_2022,Efthymiou_Antoniou_2013,mohammad2013meta,Debrezion_Pels_Rietveld_2011,Bowes_Ihlanfeldt_2001,Genc_Colak_2026,Zhang_Li_Lownes_Zhang_2021}. Among the studies of transportation access and real estate prices using machine learning methods, most focus on prediction accuracy rather than variable interpretation. Soltani et al. \cite{Soltani_Heydari_Aghaei_Pettit_2022} used several standard algorithms (decision trees, random forest, and gradient-boosted trees) to predict residential prices in Adelaide, Australia. They  incorporated a spatial-temporal lag. However, the lag variable is invalid as they apply it to the dependent variable (price) and the lack of a likelihood function precludes the isolation of the dependent variable terms to a single side of the equation. While they include a wide range of accessibility variables, interpretation is limited to parameter signs. \cite{TrindadeNeves_Aparicio_deCastroNeto_2024} used XGBoost to study residential prices in Lisbon, Portugal. They used SHAP plots to provide a more robust interpretation of results but used non-spatial cross-validation, the limits of which we outline below. Kang et al. \cite{Kang_Zhang_Peng_Gao_Rao_Duarte_Ratti_2021} used gradient-boosting machines with non-spatial cross-validation and used Uber-derived transportation variables to predict housing prices in the Boston area. Testing several machine learning algorithms (long short-term memory, gated recurrent units, and recurrent neural networks), Afandizadeh et al. \cite{Afandizadeh_Sedighi_Kalantari_Mirzahossein_2024} found that accessibility leads to lower residential prices, counter to the standard literature finding. They attribute this result to noise pollution and decreased privacy, though it is equally likely due to omitted variable bias or their accessibility specification. Chen et al. \cite{Chen_Jiao_Farahi_2023} considered public transit effects on single-family housing prices in Austin, Texas. They found that vulnerable (i.e., low-income and minority) communities place more value on access to public transit based on a variety of machine learning models and SHAP plots. Bowes and Ihlanfeldt \cite{Bowes_Ihlanfeldt_2001} highlighted the competing effects of rail transportation stations; access and retail employment effects are generally positive, while the induction of crime tends to reduce real estate prices.

Focusing on the regions in our case study, Toronto has a long history of hedonic property price studies. A 1976 paper by Dewees \cite{Dewees_1976} contrasted the effect of a new subway line on property prices parallel and perpendicular to its alignment. He found the major effect was on the rent surface perpendicular to the subway and that walking time was a greater factor than subway travel time. Haider and Miller \cite{Haider_Miller_2000} used a standard spatial autoregressive (SAR) model to examine the effects of transportation and location variables on property prices in Toronto. They found that, after controlling for dwelling and neighborhood features, locational and transportation factors were not strong predictors of property price. Distance to the central business district (CBD) and shopping centers were the only variables found to statistically influence property prices. Zhang et al. \cite{Zhang_Zhang_Miller_2021} proposed a systematic framework for decomposing prices into land price and dwelling price. They found that neighborhood socioeconomic environment had a larger influence on prices than either dwelling attributes or transportation access. In the most recent study of Toronto real estate and transportation, Kasraian et al. \cite{Kasraian_Li_Raghav_Shalaby_Miller_2023} focused on comparison of regional and local transit accessibility indicators. They used a gravity-based measure to represent regional accessibility and distance to commuter and subway rail stations to represent local accessibility. They found that regional accessibility plays an important role beyond local accessibility.  

Contrasting the focus in Toronto, Naeem and Rana \cite{Naeem_Rana_2024} found that water availability and proper drainage are critical factors when purchasing a property in Islamabad, followed by proximity to public transport, workplaces, shopping centers, and airports. Rehman and Jamil \cite{Rehman_Jamil_2021} found that rents and commuting costs are generally higher in Islamabad than Rawalpindi. Rawalpindi residents are more likely to use motorbikes or public transit, while wealthier Islamabad residents are more likely to own an automobile. Pakistani cities also face a higher rate of property disputes than Canadian cities. There were an estimated 5,500 unregistered or illegal housing developments in Pakistan as of 2018 \cite{AuditorGeneralPakistan2017}. Land-related disputes account for 50–70\% of all court cases in Pakistan \cite{Rauf_Weber_2021}. Islamabad faces an inability to finance public infrastructure development and maintenance \cite{Javid_2019}, increased commuting costs due to urban sprawl \cite{Adeel_2010,O’Donovan-Iland_2025}, and affordability issues \cite{Malik_Roosli_Tariq_Yusof_2020}.

\subsection{Cross-Validation Methods}
\begin{figure}[]
   \centering
    \includegraphics[width=0.9\linewidth]{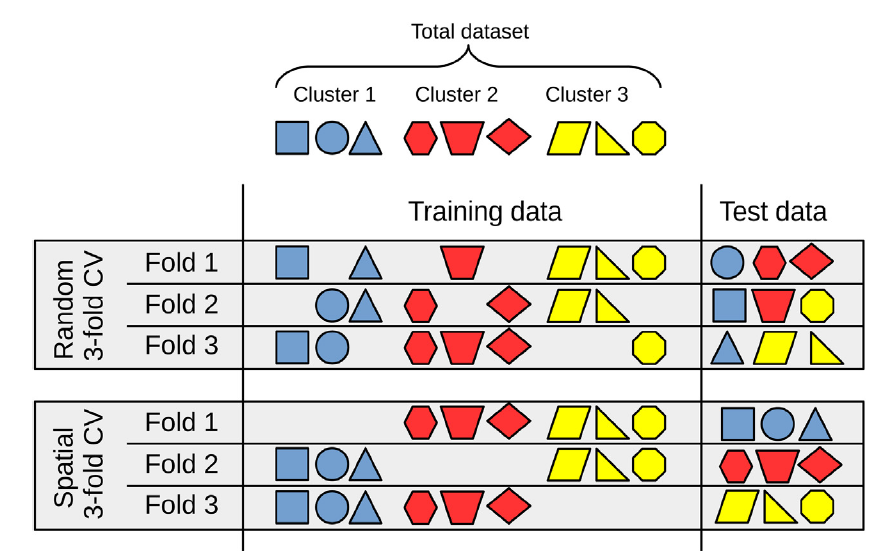}
    \caption{Comparing Non-Spatial and Spatial Cross-Validation \citep{meyerImportanceSpatialPredictor2019}}
    \label{fig:crossval}
\end{figure}

Cross-validation is a resampling technique used in modeling to assess the predictive accuracy of a model and to tune the hyperparameters of the machine learning models. Conventional k-folds cross-validation partitions the data into k equally-sized samples, then trains the model on k-1 samples. The held out sample is then used to test the predictive accuracy of the model, with the process being repeated for each of the k samples. The implicit assumption made by k-folds cross-validation is that observations are independent. That is, there is no dependence structure among observations. This assumption is typically violated in spatially referenced datasets by Tobler's First Law of Geography - everything is related to everything else, but near things are more related than distant things \citep{toblerComputerMovieSimulating1970}. There are several approaches to overcome this limitation of cross-validation. The first approach is to impose a spatial contiguity constraint on the samples (spatial k-folds). \textbf{Figure \ref{fig:crossval}}  illustrates the results for a 3-fold cross-validation example. An important feature of the spatial k-folds approach is that samples may not be drawn from the same data support. A lack of common data support among samples will lead to extrapolation. There are two perspectives of the implications of spatial k-fold cross-validation. If the modeling goal is inferential then the potential lack of a common data support violates a central tenet of causal inference \citep{gelmanRegressionOtherStories2020}. Conversely, in purely predictive modeling tasks, training and testing the model on spatially distinct datasets acts as a \emph{stress test}. A second approach to cross-validation, accounting for spatial dependence, is buffered leave-one-out (B-LOO CV). In this approach, a single observations is withheld from each spatial fold \citep{meyerImportanceSpatialPredictor2019}. The advantage of this approach over spatial k-folds is that B-LOO CV maintains a constant data support across the training and testing datasets. However, We used spatial k-means rather than B-LOO to ensure roughly equal sample sizes and balanced spatial coverage across folds; B-LOO evaluates one test location at a time and can leave large areas out of the training set, producing unstable variance estimates \cite{robertsCrossvalidationStrategiesData2017}

An important question regarding cross-validation on spatial data is the magnitude of the error associated with ignoring spatial dependence. Several studies find that standard non-spatial cross-validation provides overoptimistic error estimates due to the problem of spatial dependence (autocorrelation) \citep{bahn2013testing,micheletti2014machine,juelSpatialApplicationRandom2015,gaschSpatiotemporalInterpolationSoil2015,gudmundsson2015towards,robertsCrossvalidationStrategiesData2017,meyer2018improving,henglRandomForestGeneric2018}. Pohjankukka \citep{pohjankukkaEstimatingPredictionPerformance2017} estimated the effect size at 40 percent overprediction based on three empirical case studies. Ploton et al. \citep{ploton2020spatial} trained a random forest model on environmental variables. They found that non-spatial cross-validation methods overpredict outcomes. Furthermore, they highlighted the close link between geographical and environmental domains, such that removing training observations in the spatial neighborhood of a test observation may remove the environmental conditions found at the test location from the model's calibration domain. This removal may then influence the inferential validity of the model. Zhang et al. highlight the importance of spatial dependence in property price modeling, stating "housing units are prone to form a spatially aggregated cluster, which represent the
regional price of a neighborhood" \cite[p. 2]{Zhang_Zhang_Miller_2021}.

We contribute to the literature by contrasting non-spatial and spatial cross-validation for machine learning-based prediction. Our specific application is property prices in two diverse metropolitan areas. Contrasting results between these areas helps to identify the spatial and contextual extent of cross-validation variation. We use an interpretation approach, SHAP (SHapley Additive exPlanations), to consider land use and transportation variables effects on prices.

\section{Data}

\subsection{Rawalpindi Islamabad Metropolitan Area (RIMA)}

\begin{figure*}[]
   \centering
    \includegraphics[width=1\textwidth]
    {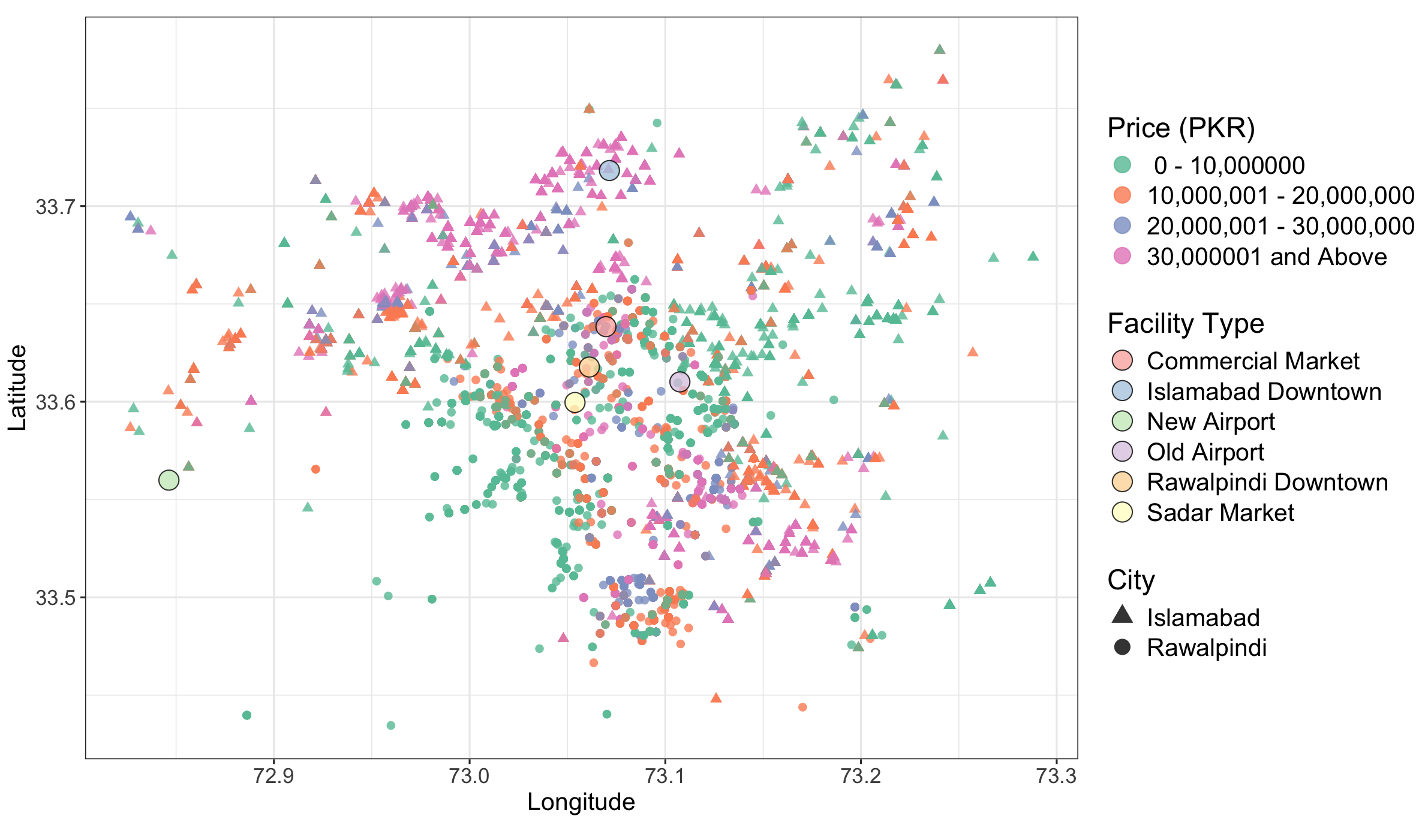}
    \caption{RIMA Price Distribution}
    \label{fig:rima}
\end{figure*}

We use residential sale price data for the twin cities of Islamabad (capital city) and Rawalpindi, Pakistan for the year 2019. The data comprised about 19,417 records, with about 53\% from Islamabad and 47\% from Rawalpindi. The data include location, price, area, and dwelling type information. This dataset is combined with land use and transportation related variables, chiefly derived from Openstreetmap layers. Land use variables include distance to the old and new airports, old Rawalpindi city center, and major markets (Sadar Bazar in Rawalpindi and the Blue Area Market in Islamabad). \textbf{Figure \ref{fig:rima}} illustrates the spatial distribution of RIMA data points and the location of major facilities used in our study. The color gradient represents the variation in price values. The prices are relatively higher in the Islamabad region than in Rawalpindi city. This is because Islamabad is one of the most developed cities in the country with high quality of living and a concentration of major businesses in the downtown area. Prices are relatively higher near major economic hubs. Overall, the spatial distribution of the data appears to be well represented across the study area.

Transportation related variables included distance to bus stops, highways, roads, motorway (freeway) interchanges, and railway stations. Bus stops are divided into two types: 1) metro bus stops where buses have physically separated dedicated lanes and 2) regular bus stops. Distance to highways is included via two dummy variable that indicate whether the dwelling is located  within 1 km or 2 km of the nearest highway, respectively. For primary and secondary roads, dummy variables are used to identify whether the dwelling is located within 500 m or 1 km of these types of roads. \textbf{Table \ref{tab:rima_des}} shows the descriptive statistics of all variables used in our model. The prices in \textbf{Table \ref{tab:rima_des}} are converted to Canadian Dollars for comparison with Toronto. For model training, dwelling floor area is converted to square meters and all distances are measured in meters. All continuous variables, including sale price, are log transformed as a normalization (\textbf{Table \ref{tab:rima_des}}).

\subsection{City of Toronto}
\begin{figure*}[]
   \centering
    \includegraphics[width=0.8\textwidth]
    {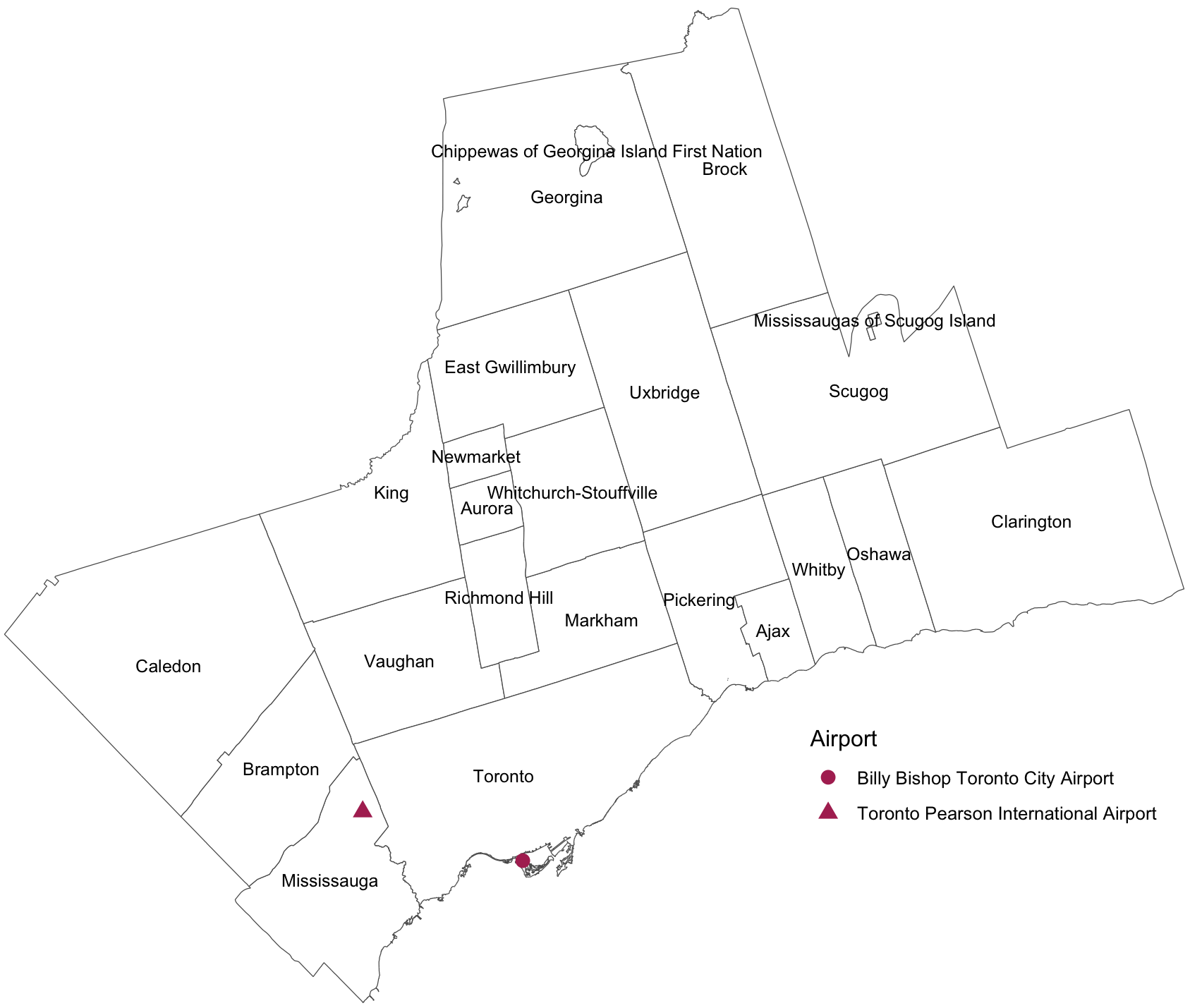}
    \caption{Major Cities and Airports in GTA}
    \label{fig:gta}
\end{figure*}

\begin{figure*}[]
   \centering
    \includegraphics[width=0.7\textwidth]
    {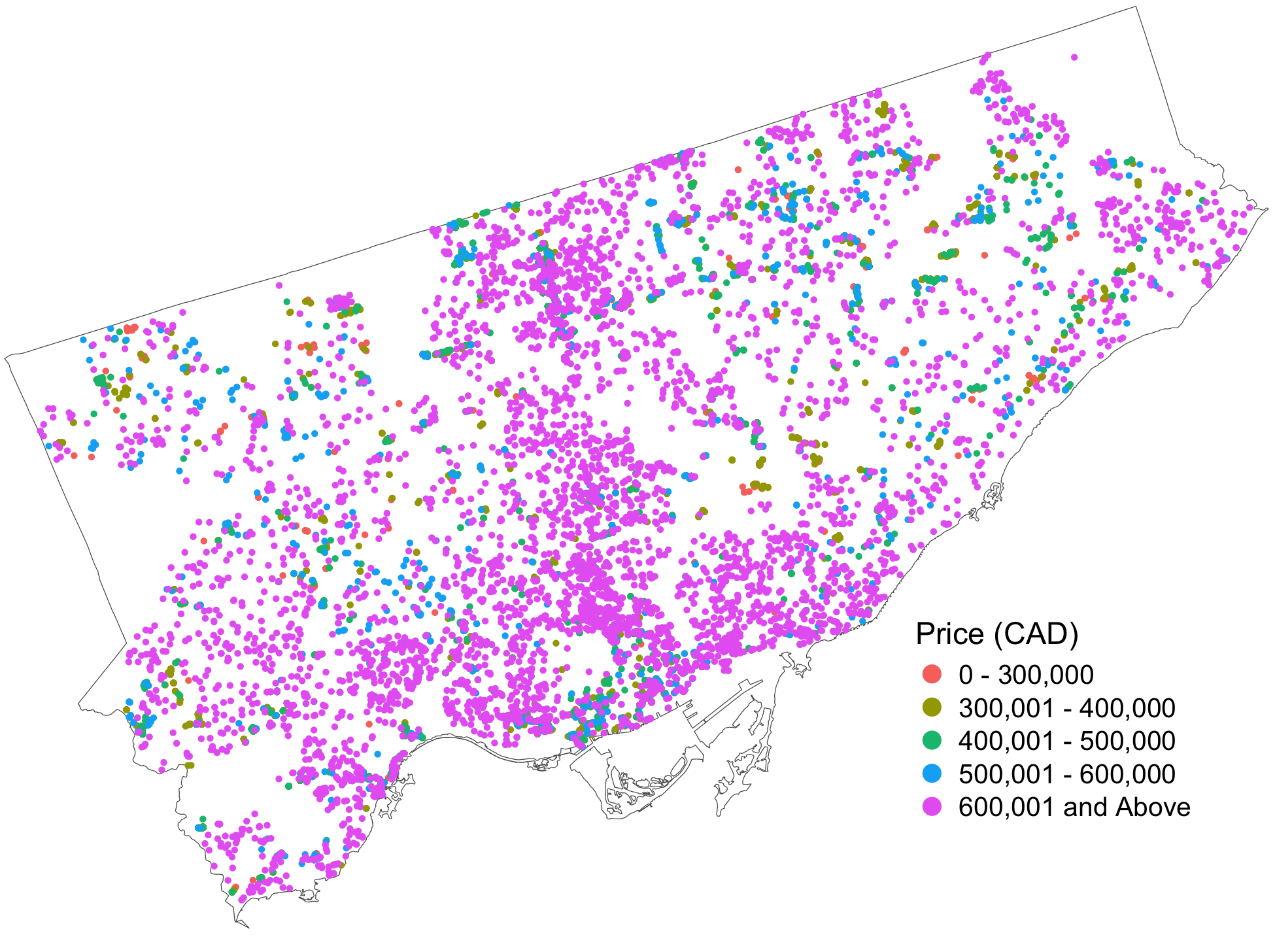}
    \caption{Toronto Price Distribution}
    \label{fig:tor}
\end{figure*}

For Toronto, we use real estate data for the City of Toronto only but consider the transportation system within the larger GTA metropolitan area. \textbf{Figure \ref{fig:gta}} shows the map of GTA outlining the boundaries of major cities (census subdivisions) in the metropolitan region. Major cities in the region include Toronto, Mississauga, Brampton, Vaughan, and Markham. These cities are the population and economic hubs in the GTA region and impact land prices in the City of Toronto. The Toronto data are chosen to approximately match variables available for RIMA. Residential sale price data for 2016 were available from the Toronto Real Estate Board (TREB) for 32,000 transactions within the City of Toronto. The data include location, price, area, and dwelling type information similar to that available for RIMA. \textbf{Figure \ref{fig:tor}} shows the spatial distribution of prices throughout the City of Toronto. The color gradient represents the variation in price values. Similar to RIMA, the spatial distribution of the data appears to be well represented across the City of Toronto.

OpenStreetMap land use and transportation data provided comparable variables to those available for RIMA. Transit station (subway, streetcar, and bus) distances parallel transit proximity variables calculated for RIMA. Distances to airports (Lester B. Toronto Pearson International and Billy Bishop Toronto City \textbf{Figure \ref{fig:gta}}). and major interchanges are calculated in a similar manner. The RIMA model used proximity to markets. To approximate this amenity variable, we calculated the distance to the city center (i.e., downtown) of major cities in the GTA \textbf{Figure \ref{fig:gta}}. \textbf{Table \ref{tab:tor_des}} shows the descriptive statistics of all the variables used in this study. For both study areas, we excluded records outside of the study area, records with outlier values, and excluded all buildings other than houses.

The data for RIMA and Toronto were collected at relatively similar times. The 2016-2019 period is not affected by either the global financial crisis (2008-2009) or the COVID-19 pandemic (2020-2022). Given the research objectives of comparing cross-validation methods and transportation infrastructure effects on real estate prices, the three year difference should have minimal effect on interpretation. If our objective was to examine the effect of an exogenous shock across real estate markets, it would be necessary to collect panel data and introduce year fixed effects.

\begin{table*}[]
\caption{RIMA Descriptive Statistics}
\begin{tabular}{l|c|c|c|c|c}
\toprule
\textbf{Variable}                         & \textbf{Type}        & \textbf{Minimum}     & \textbf{Mean}        & \textbf{Medium}      & \textbf{Maximum}     \\
\midrule
Price (CAD)*                              & Continuous           & 13350                & 258047               & 146850               & 8010000              \\
\multicolumn{6}{l}{\textbf{Land-use Related Variables}} \\
Area (sqm)                                & Continuous           & 62.7                 & 235.4                & 167.2                & 4180.6               \\
Distance to Islamabad Downtown            & Continuous           & 71.5                 & 15091.7              & 15247.1              & 35349.3              \\
Distance to Rawalpindi Downtown           & Continuous           & 150.7                & 9665.8               & 9391.4               & 25576.0              \\
Distance to Sadar Market                  & Continuous           & 389.7                & 9606.0               & 9067.8               & 26405.3              \\
Distance to Rawalpindi Commercial Market  & Continuous           & 7.8                  & 10111.0              & 9916.4               & 27855.7              \\
\multicolumn{6}{l}{\textbf{Transportation Related Variables}} \\
Distance to Old Airport                   & Continuous           & 614.0                & 9682.5               & 9232.2               & 27734.7              \\
Distance to New Airport                   & Continuous           & 877.1                & 22996.0              & 23668.8              & 43582.3              \\
Distance to the Nearest Bus Stop          & Continuous           & 11.4                 & 2070.3               & 1352.2               & 17620.2              \\
Distance to the Nearest Metro Bus Station & Continuous           & 54.0                 & 6906.6               & 6753.1               & 23223.5              \\
Distance to Railway Station               & Continuous           & 153.8                & 5312.1               & 4159.3               & 22803.7              \\
Distance to Nearest Motorway Interchange  & Continuous           & 536.1                & 18907.7              & 19134.8              & 40010.0              \\
Highway Proximity within 1 Km             & Dummy                & 0                    &  -                   &  -                   & 1                    \\
Major Road Proximity within 1 Km          & Dummy                & 0                    &  -                   &  -                   & 1                    \\
\midrule
\multicolumn{3}{l}{*Based on average exchange rate in 2019. 1 PKR = 0.0089 CAD}        & \multicolumn{1}{l}{} & \multicolumn{1}{l}{} & \multicolumn{1}{l}{} \\
\multicolumn{3}{l}{ Note: Distance values are in meters}    & \multicolumn{1}{l}{} & \multicolumn{1}{l}{} & \multicolumn{1}{l}{}   \\
\bottomrule
\end{tabular}
\label{tab:rima_des}
\end{table*}

\begin{table*}[]
\caption{Toronto Descriptive Statistics}
\begin{tabular}{l|c|c|c|c|c}
\toprule
\textbf{Variable}                         & \textbf{Type}        & \textbf{Minimum}     & \textbf{Mean}        & \textbf{Medium}      & \textbf{Maximum}     \\
\midrule
Price (CAD)                               & Continuous           & 240088               & 649385               & 454000               & 8740000              \\
\multicolumn{6}{l}{\textbf{Land-use Related Variables}} \\
Area (sqm)                                & Continuous           & 46.00                & 116.23               & 88.00                & 465.00               \\
Distance to Toronto Downtown              & Continuous           & 42.20                & 10244.78             & 10741.63             & 26257.91             \\
Distance to Vaughan Downtown              & Continuous           & 2487.02              & 16246.51             & 17327.76             & 32491.27             \\
Distance to Brampton Downtown             & Continuous           & 11888.31             & 30020.05             & 29594.94             & 52718.59             \\
Distance to Richmond Hill Downtown        & Continuous           & 9210.95              & 20709.80             & 21657.08             & 33516.81             \\
Distance to Markham Downtown              & Continuous           & 4388.41              & 18327.13             & 18541.48             & 34061.41             \\
Distance to Mississauga Downtown          & Continuous           & 6981.69              & 24553.62             & 24021.56             & 47829.09             \\
Distance to Pickering Downtown            & Continuous           & 5216.61              & 28868.65             & 29287.25             & 45686.78             \\
\multicolumn{6}{l}{\textbf{Transportation Related Variables}} \\
Distance to Pearson Airport               & Continuous           & 2713.26              & 18669.57             & 18421.77             & 41226.70             \\
Distance to Bishop Airport                & Continuous           & 299.03               & 11527.84             & 11791.92             & 28485.46             \\
Distance to Nearest Bus Station           & Continuous           & 3.04                 & 147.41               & 116.87               & 2091.11              \\
Distance to the Nearest Streetcar Station & Continuous           & 6.84                 & 5361.02              & 4484.33              & 19029.38             \\
Distance to the Nearest Subway Station    & Continuous           & 20.58                & 2345.39              & 1661.34              & 13298.13             \\
Distance to the Nearest Road              & Continuous           & 0.82                 & 185.21               & 130.84               & 1568.04              \\
Distance to Motorway                      & Continuous           & 18.12                & 1566.15              & 1233.70              & 8787.99              \\
Distance to the Nearest Interchange       & Continuous           & 70.91                & 1841.75              & 1624.18              & 8895.56              \\
\midrule
\multicolumn{3}{l}{Note: Distance values are in meters}    & \multicolumn{1}{l}{} & \multicolumn{1}{l}{} & \multicolumn{1}{l}{}                              \\
\bottomrule
\end{tabular}
\label{tab:tor_des}
\end{table*}

\section{Methods of Analysis}
We first assess the presence of spatial autocorrelation in prices using the Moran's I test \cite{moran1950notes} to determine whether spatial effects should be considered in the modeling framework. The Moran I statistic is given by \cite{moran1950notes}

\begin{equation}
\rho = \frac{N}{W} \frac{\sum_{i=1}^N\sum_{j=1}^N w_{ij}(x_i-\hat{x})(x_j-\hat{x})}{\sum_{i=1}^N (x_i-\hat{x})^2}
\end{equation}
where $N$ is the number of spatial units indexed by $i$ and $j$, $x$ is the variable tested for spatial dependence, $\hat{x}$ is the mean value of $x$, and $w_ij$ are the elements of a spatial weight matrix.

We then use Extreme Gradient Boosting (XGBoost) and Random Forest (RF) models to measure variations in prediction accuracy based on non-spatial and spatial cross-validation frameworks. Both models are ensemble tree-based methods, such that spatial k-folds can be incorporated into the model training stage rather than prediction accuracy analysis only. 
The RF method belongs to the bagging class of ensemble learning algorithms that builds multiple decision trees and combines their outputs to improve prediction accuracy and reduce overfitting \citep{breiman2001random}. The diversity among trees is created through two randomization mechanisms (i) each tree is trained on a bootstrap sample from the training data, and (ii) a random subset of predictors is selected at each node split. The final prediction is obtained by aggregating the outputs from all trees—through averaging in regression or majority voting in classification tasks. The RF method has three hyperparameters:

\begin{itemize}
    \item The total number of decision trees (\emph{trees}) in the ensemble model. More trees generally decrease the error but increase the computational time.

    \item The size of the random subset of predictors at each node split (\emph{mtry}).

    \item Node size (\emph{min\_n}) which represents the smallest allowable number of observations in a terminal node.
    
\end{itemize}

For RF, iterative training via bagging generates a \emph{strong learner} from multiple decision trees based on bootstrap sampling (which may incorporate spatial dependence). Roberts et al. \citep{robertsCrossvalidationStrategiesData2017} and Valavi et al. \citep{valaviBlockCVPackageGenerating2018} recommend such a spatial blocking approach for RF training to mitigate autocorrelation errors and ensure independence between training and testing sets. 

XGBoost is also a tree-based ensemble method, wherein trees are built in a sequential manner to reduce the error of former trees via a boosting approach \cite{friedman2001greedy}. Whereas RF addresses variance through bagging, XGBoost focuses on bias reduction by boosting weake learners. XGBoost is more prone to overfitting, requring careful parameter tuning. In addition to (\emph{trees}),(\emph{mtry}), and (\emph{min\_n})  the hyperparameters of the XGBoost model are:

\begin{itemize}
    \item The maximum depth of a tree (\emph{tree\_depth}).
    \item Learning rating that scales the contribution of each tree to prevent overfitting (\emph{learn\_rate}).
    \item The minimum loss reduction required to create an additional split on a leaf node (\emph{loss\_reduction}).
    \item Subsample ratio of the training data (\emph{sample\_size}).
\end{itemize}

In this study, we fix the number of trees to 1000, because model performance does not improve beyond 1000 trees and computational time increases substantially. The optimal values of all other hyperparameters are selected using a hyperparameter tuning process. The process involved generating a random grid of hyperparameters and testing the values using a cross-validation process. The optimal values of hyperparameters are selected based on the best performing model. In this study, we use a grid size of 20 and 5-fold cross-validation procedure. We use k-means clustering for spatial cross-validation and random sampling for non-spatial cross-validation. The model performance is evaluated using the Root Mean Squared Error (RMSE), Mean Absolute Error (MAE), Mean Absolute Percentage Error (MAPE), and R\textsuperscript{2}. The optimal values of the hyperparameters are selected based on the RMSE value.

\subsection{Model Development}
\begin{figure}[]
   \centering
    \includegraphics[width=0.5\textwidth]
    {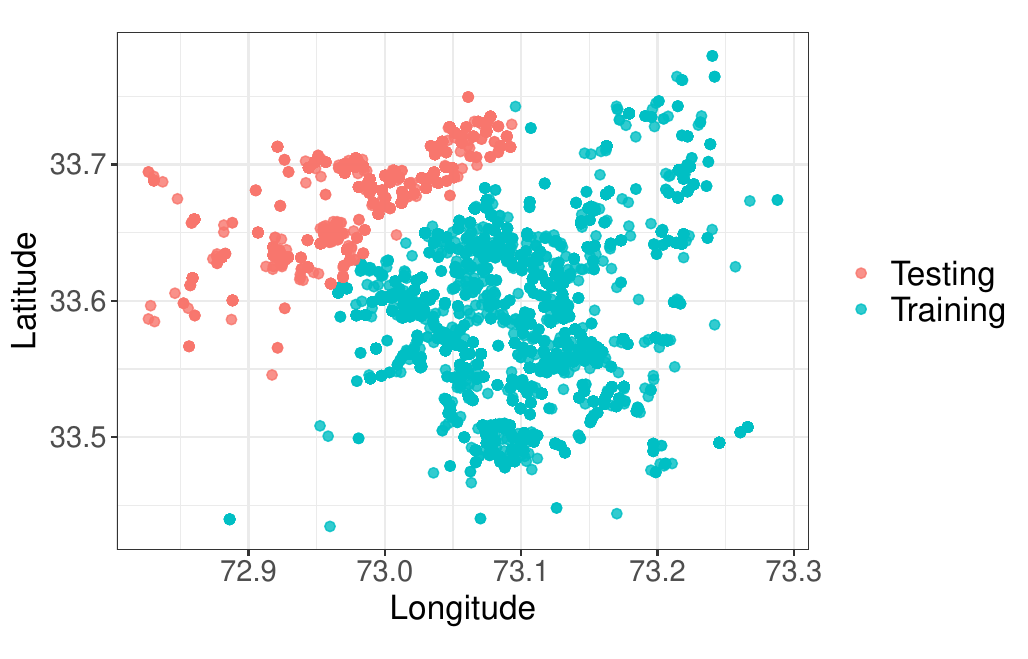}
    \caption{Training and Testing Set 1 Based on Spatial Clustering (RIMA)}
    \label{fig:rima_spatialfold1}
\end{figure}

\begin{figure}[]
   \centering
    \includegraphics[width=0.5\textwidth]
    {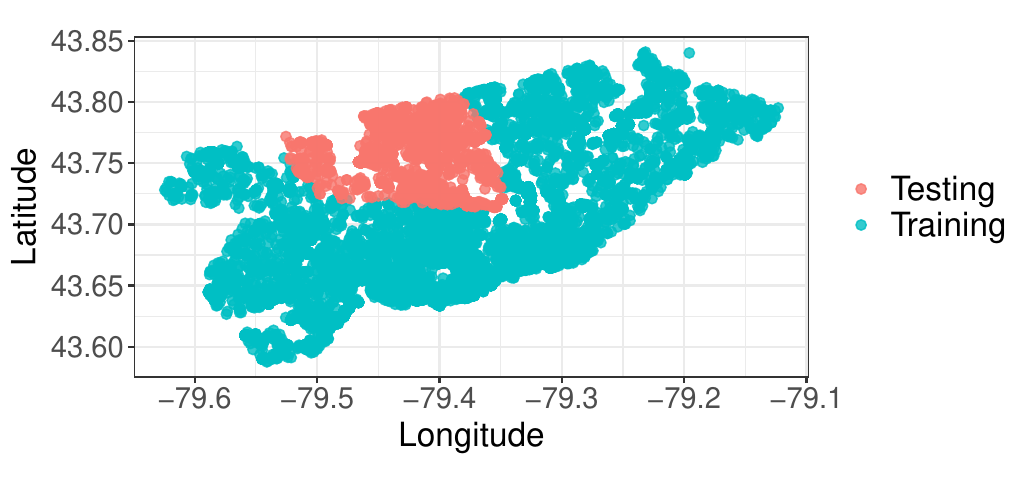}
    \caption{Training and Testing Set 1 Based on Spatial Clustering (Toronto)}
    \label{fig:gta_spatialfold1}
\end{figure}

\begin{figure}[]
   \centering
    \includegraphics[width=0.4\textwidth]
    {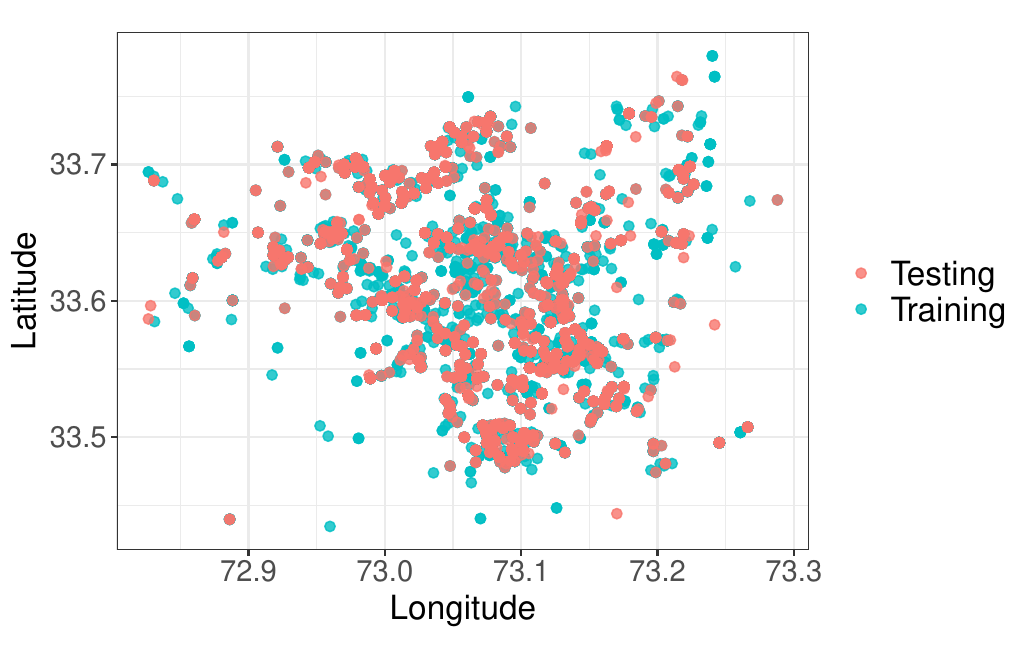}
    \caption{Training and Testing Set 1 Based on Non-Spatial Clustering (RIMA)}
    \label{fig:rima_non-spatialfold1}
\end{figure}

\begin{figure}[]
   \centering
    \includegraphics[width=0.5\textwidth]
    {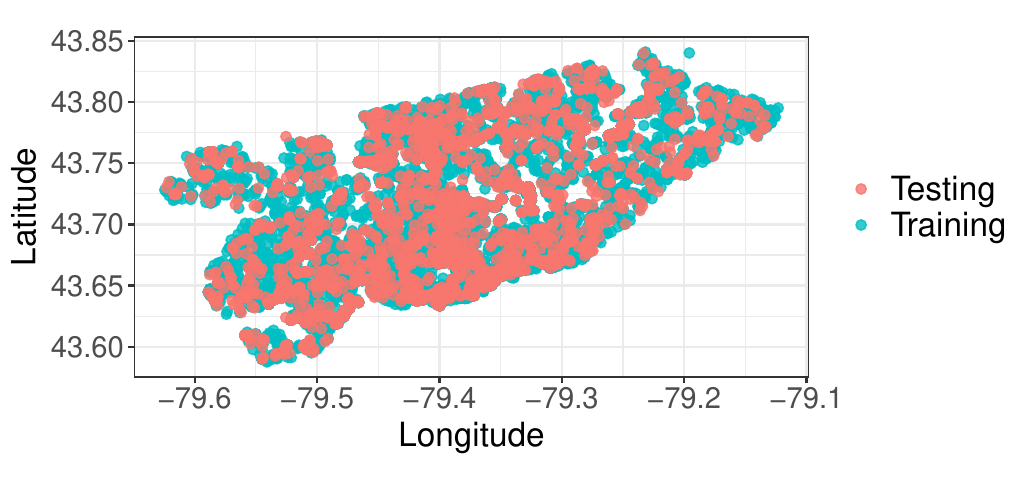}
    \caption{Training and Testing Set 1 Based on Non-Spatial Clustering (Toronto)}
    \label{fig:gta_non-spatialfold1}
\end{figure}

The model development is comprised of two steps. In the first step, for both spatial and non-spatial approaches, the data are divided into five samples. For the spatial approach, the data are divided using a spatial clustering approach. Specifically, we apply k-means clustering based on the spatial coordinates to create spatially contiguous folds for cross-validation. In contrast, the non-spatial approach involves random partitioning of the data. We develop five models for each approach (spatial and non-spatial) using both XGBoost and RF, resulting in a total of 20 models. The reason for developing five models for each approach is to tease out the variability in model performance due to differences in data partitioning. Each sample contains a \emph{training set} and a \emph{testing set}. Figures \ref{fig:rima_spatialfold1} and \ref{fig:gta_spatialfold1} illustrate how the data are divided into two spatially distinct regions for each study area. In contrast, Figures \ref{fig:rima_non-spatialfold1} and \ref{fig:gta_non-spatialfold1} show that training and testing sets are divided randomly for non-spatial cross-validation, with the model being essentially trained and tested on similar data. The rest of the spatial and non-spatial folds are provided in Appendix A.

In the second step, each model is trained on the training set using five-fold cross validation (using k-means clustering with spatial coordinates in the case of spatial models and random sampling in the case of non-spatial models). The optimal values of the hyperparameters are selected during the five-fold cross validation procedure.

One challenge with using machine learning methods for inferential analysis is parameter interpretation. SHAP (SHapley Additive exPlanations) plots \citep{lundberg_unified_2017} are a useful tool for this purpose. SHAP is a game theoretic approach that assigns an importance value to each model feature. The  appealing properties of SHAP are that it is 1) additive such that feature contributions are independent and additive to the cumulative effect on the outcome, 2) locally accuracy such that SHAP provides local interpretation of model prediction for a given input, 3) robust to missing values, and 4) consistent under changes in model specification. SHAP parameters are used in this study to compare feature importance.
\FloatBarrier

\section{Results}

Moran I statistics for Toronto and RIMA property prices were  0.70 (p-value $<$ 0.001) and 0.76 (p-value $<$ 0.001), respectively. These results indicate strong spatial autocorrelation in the dependent variable.

\subsection{Hyperparameter Tuning}

\textbf{Table \ref{tab:rima_xgb_hyp}} presents the optimal hyperparameters of the XGBoost models developed for RIMA under both spatial and non-spatial approaches, while \textbf{Table \ref{tab:tor_xgb_hyp}} provides the corresponding results for Toronto. The variability in hyperparameter values is high in spatial models as compared to non-spatial models. This further shows the impact of data variability among spatial folds as compared to non-spatial folds. Similar results are observed in the case of RF models (\textbf{Tables \ref{tab:rima_rf_hyp}} and \textbf{\ref{tab:tor_rf_hyp}}). These parameters are used to develop the final models on the respective training sets and model performance is assessed on the corresponding testing sets.

\begin{table*}[]
\caption{RIMA XGBoost Model Hyperparameter Values}
\begin{tabular}{c|c|c|c|c|c|c|c}
\toprule
\textbf{Model} & \textbf{Mtry} & \textbf{Min\_n} & \textbf{Tree Depth} & \textbf{Learn Rate} & \textbf{Loss Reduction} & \textbf{Sample Size} \\ \midrule
Spatial 1    & 6             & 32              & 8.000               & 0.016               & 0.001                   & 0.617                \\
Spatial 2    & 14            & 18              & 15.000              & 0.059               & 0.009                   & 0.129                \\
Spatial 3    & 6             & 32              & 8.000               & 0.016               & 0.001                   & 0.617                \\
Spatial 4    & 10            & 7               & 14.000              & 0.007               & 0.002                   & 0.317                \\
Spatial 5    & 10            & 7               & 14.000              & 0.007               & 0.002                   & 0.317                \\
\midrule
Non-Spatial 1   & 14            & 18              & 15.000              & 0.059               & 0.009                   & 0.129                \\
Non-Spatial 2   & 10            & 7               & 14.000              & 0.007               & 0.002                   & 0.317                \\
Non-Spatial 3   & 14            & 18              & 15.000              & 0.059               & 0.009                   & 0.129                \\
Non-Spatial 4   & 14            & 18              & 15.000              & 0.059               & 0.009                   & 0.129                \\
Non-Spatial 5   & 10            & 7               & 14.000              & 0.007               & 0.002                   & 0.317                \\ \bottomrule
\end{tabular}
\label{tab:rima_xgb_hyp}
\end{table*}

\begin{table*}[h]
\caption{Toronto XGBoost Model Hyperparameter Values}
\begin{tabular}{c|c|c|c|c|c|c|c}
\toprule
\textbf{Model} & \textbf{Mtry} & \textbf{Min\_n} & \textbf{Tree Depth} & \textbf{Learn Rate} & \textbf{Loss Reduction} & \textbf{Sample Size} \\
\midrule
Spatial 1   & 14            & 32              & 5                   & 0.0047              & 0.0199                  & 0.7859               \\
Spatial 2   & 2             & 38              & 7                   & 0.0132              & 0.0000                  & 0.9670               \\
Spatial 3   & 14            & 32              & 5                   & 0.0047              & 0.0199                  & 0.7859               \\
Spatial 4   & 14            & 32              & 5                   & 0.0047              & 0.0199                  & 0.7859               \\
Spatial 5   & 14            & 32              & 5                   & 0.0047              & 0.0199                  & 0.7859               \\
\midrule
Non-Spatial 1  & 1             & 27              & 10                  & 0.0914              & 0.0004                  & 0.3546               \\
Non-Spatial 2  & 1             & 27              & 10                  & 0.0914              & 0.0004                  & 0.3546               \\
Non-Spatial 3  & 1             & 27              & 10                  & 0.0914              & 0.0004                  & 0.3546               \\
Non-Spatial 4  & 1             & 27              & 10                  & 0.0914              & 0.0004                  & 0.3546               \\
Non-Spatial 5  & 1             & 27              & 10                  & 0.0914              & 0.0004                  & 0.3546         \\     
\bottomrule
\end{tabular}
\label{tab:tor_xgb_hyp}
\end{table*}

\begin{table}[]
\centering
\caption{RIMA RF Model Hyperparameter Values}
\begin{tabular}{c|c|c}
\toprule
\textbf{Model} & \textbf{mtry} & \textbf{min\_n} \\ \midrule
Spatial 1           & 5             & 36              \\
Spatial 2           & 6             & 20              \\
Spatial 3           & 7             & 12              \\
Spatial 4           & 9             & 22              \\
Spatial 5           & 10            & 6               \\
\midrule
Non-Spatial 1          & 5             & 10              \\
Non-Spatial 2          & 5             & 10              \\
Non-Spatial 3          & 4             & 2               \\
Non-Spatial 4          & 5             & 10              \\
Non-Spatial 5          & 5             & 10              \\ 
\bottomrule
\end{tabular}
\label{tab:rima_rf_hyp}
\end{table}

\begin{table}[]
\centering
\caption{Toronto RF Model Hyperparameter Values}
\begin{tabular}{c|c|c}
\toprule
\textbf{Model}  & \textbf{Mtry} & \textbf{Min\_n} \\ 
\midrule
Spatial 1   & 13            & 22  \\
Spatial 2   & 14            & 6   \\
Spatial 3   & 12            & 40  \\
Spatial 4   & 13            & 22  \\
Spatial 5   & 11            & 12  \\
\midrule
Non-Spatial 1  & 4             & 2   \\
Non-Spatial 2  & 4             & 2   \\
Non-Spatial 3  & 4             & 2   \\
Non-Spatial 4  & 4             & 2   \\
Non-Spatial 5  & 9             & 4   \\ 

\bottomrule
\end{tabular}
\label{tab:tor_rf_hyp}
\end{table}

\subsection{Model Performance}
This section summarizes the model performance results for the ten XGBoost and and ten RF models, five spatial and five non-spatial models, trained on data from each study area. The spatial models used k-means clustering using spatial coordinates to split the data for cross-validation, while the non-spatial model used random sampling for cross-validation.

Table \ref{tab:rima_xgb_performance} and \ref{tab:rima_rf_performance} shows model performance results of XGBoost and RF for the RIMA training data. Overall, the results show higher model performance for non-spatial models than spatial models. These results parallel prior findings in the literature \citep{bahn2013testing,micheletti2014machine,juelSpatialApplicationRandom2015,gaschSpatiotemporalInterpolationSoil2015,gudmundsson2015towards,robertsCrossvalidationStrategiesData2017,meyer2018improving,henglRandomForestGeneric2018} that non-spatial models over-predict model accuracy because they do not consider spatial autocorrelation in the cross-validation process. According to the non-spatial model results, prediction performance is not influenced by how the data are divided into training and testing sets. Performance metrics are consistent across all five non-spatial models. However, this pattern does not hold true for the spatial models. For instance, for the XGBoost model (Table \ref{tab:rima_xgb_performance}), the mean $R^2$ is 0.70 but varies across models within the range 0.63 to 0.74. Similarly, RMSE fluctuates between 0.43 and 0.62 (Table \ref{tab:rima_xgb_performance}). Similar observations are noted in the case of RF models for RIMA (Table \ref{tab:rima_rf_performance}). Therefore, spatial splitting of the data influences the model performance. 

Tables \ref{tab:tor_xgb_performance} and \ref{tab:tor_rf_performance} provide similar results for Toronto for the XGBoost and RF models, respectively. As with the RIMA model performance results, non-spatial models exhibit no variation in their results for either modeling methods. In the same way, spatial performance results vary across models. These results strengthen the argument that non-spatial models consistently overpredict model performance and neglect inter-model variation arising from spatial heterogeneity. This conclusion is further supported by applying the models to two different cities.

\begin{table}[]
\centering
\caption{RIMA XGBoost Model Performance on Training Data}
\begin{tabular}{@{}ccccc@{}}
\toprule
\textbf{Model Type} & \textbf{RMSE} & \textbf{R$^2$} & \textbf{MAE} & \textbf{MAPE} \\
\midrule
Spatial 1           & 0.43          & 0.74         & 0.32         & 1.95          \\
Spatial 2           & 0.60          & 0.63         & 0.54         & 3.20          \\
Spatial 3           & 0.58          & 0.73         & 0.53         & 3.09          \\
Spatial 4           & 0.57          & 0.72         & 0.44         & 2.60          \\
Spatial 5           & 0.62          & 0.67         & 0.52         & 3.06          \\
\midrule
Non-Spatial 1          & 0.25          & 0.92         & 0.17         & 1.03          \\
Non-Spatial 2          & 0.25          & 0.92         & 0.17         & 1.04          \\
Non-Spatial 3          & 0.25          & 0.92         & 0.17         & 1.01          \\
Non-Spatial 4          & 0.25          & 0.92         & 0.17         & 1.03          \\
Non-Spatial 5          & 0.25          & 0.92         & 0.17         & 1.04          \\
\bottomrule
\end{tabular}
\label{tab:rima_xgb_performance}
\end{table}

\begin{table}[]
\centering
\caption{RIMA RF Model Performance on Training Data}
\begin{tabular}{@{}ccccc@{}}
\toprule
\textbf{Model} & \textbf{RMSE} & \textbf{R$^2$} & \textbf{MAE} & \textbf{MAPE} \\
\midrule
Spatial 1           & 0.45          & 0.73         & 0.34         & 2.06          \\
Spatial 2           & 0.66          & 0.62         & 0.56         & 3.30          \\
Spatial 3           & 0.57          & 0.73         & 0.51         & 2.99          \\
Spatial 4           & 0.55          & 0.71         & 0.40         & 2.41          \\
Spatial 5           & 0.56          & 0.69         & 0.50         & 2.98          \\
\midrule
Non-Spatial 1          & 0.25          & 0.93         & 0.17         & 1.01          \\
Non-Spatial 2          & 0.25          & 0.93         & 0.17         & 1.01          \\
Non-Spatial 3          & 0.25          & 0.92         & 0.17         & 1.00          \\
Non-Spatial 4          & 0.25          & 0.92         & 0.17         & 1.02          \\
Non-Spatial 5          & 0.25          & 0.93         & 0.17         & 1.01          \\
\bottomrule
\end{tabular}
\label{tab:rima_rf_performance}
\end{table}

\begin{table}[]
\centering
\caption{Toronto XGBoost Model Performance on Training Data}
\begin{tabular}{@{}ccccc@{}}
\toprule
\textbf{Model} & \textbf{RMSE} & \textbf{R$^2$} & \textbf{MAE} & \textbf{MAPE} \\
\midrule
Spatial 1      & 0.28          & 0.76         & 0.23         & 1.71          \\
Spatial 2      & 0.28          & 0.77         & 0.21         & 1.61          \\
Spatial 3      & 0.27          & 0.80         & 0.21         & 1.60          \\
Spatial 4      & 0.27          & 0.77         & 0.22         & 1.65          \\
Spatial 5      & 0.28          & 0.79         & 0.23         & 1.75          \\
\midrule
Non-Spatial 1     & 0.15          & 0.93         & 0.10         & 0.78          \\
Non-Spatial 2     & 0.15          & 0.93         & 0.10         & 0.79          \\
Non-Spatial 3     & 0.15          & 0.93         & 0.10         & 0.78          \\
Non-Spatial 4     & 0.15          & 0.93         & 0.10         & 0.78          \\
Non-Spatial 5     & 0.15          & 0.93         & 0.10         & 0.78         \\
\bottomrule
\end{tabular}
\label{tab:tor_xgb_performance}
\end{table}

\begin{table}[]
\centering
\caption{Toronto RF Model Performance on Training Data}
\begin{tabular}{@{}ccccc@{}}
\toprule
\textbf{Model} & \textbf{RMSE} & \textbf{R$^2$} & \textbf{MAE} & \textbf{MAPE} \\
\midrule
Spatial 1      & 0.27          & 0.77         & 0.21         & 1.58          \\
Spatial 2      & 0.28          & 0.79         & 0.21         & 1.58          \\
Spatial 3      & 0.28          & 0.77         & 0.20         & 1.55          \\
Spatial 4      & 0.28          & 0.74         & 0.21         & 1.62          \\
Spatial 5      & 0.27          & 0.78         & 0.20         & 1.53          \\
\midrule
Non-Spatial 1     & 0.15          & 0.93         & 0.11         & 0.79          \\
Non-Spatial 2     & 0.15          & 0.93         & 0.11         & 0.79          \\
Non-Spatial 3     & 0.15          & 0.93         & 0.10         & 0.78          \\
Non-Spatial 4     & 0.15          & 0.93         & 0.11         & 0.79          \\
Non-Spatial 5     & 0.15          & 0.93         & 0.10         & 0.78          \\
\bottomrule
\end{tabular}
\label{tab:tor_rf_performance}
\end{table}

\begin{figure}
    \centering
    \includegraphics[width=1\linewidth]{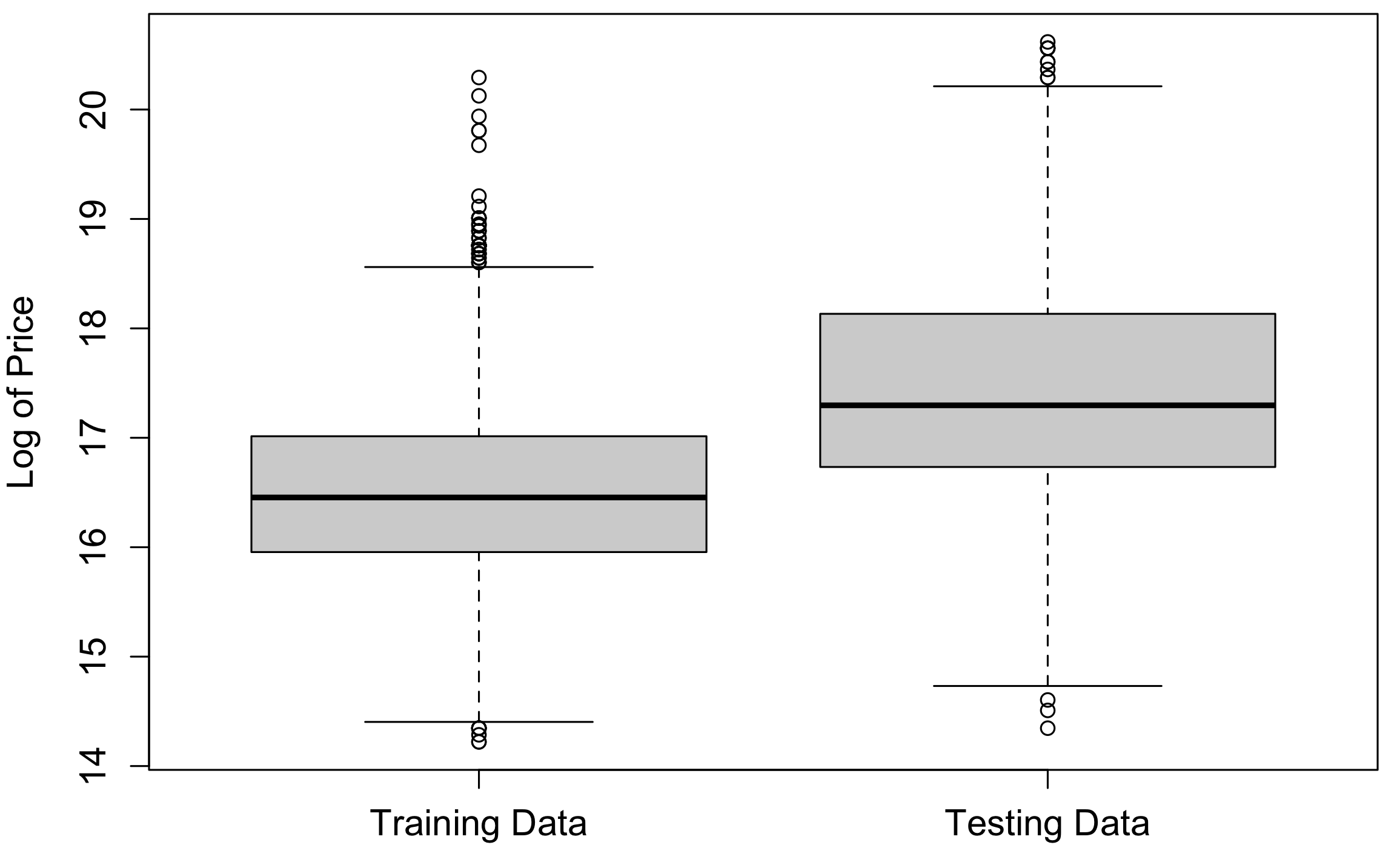}
    \caption{Distribution of Prices in RIMA Spatial Fold 1}
    \label{fig:rima_spatial_boxplot1}
\end{figure}

Each model is trained on the training dataset using its corresponding optimal hyperparameters and subsequently applied to the testing dataset to evaluate predictive performance. Table \ref{tab:rima_xgb_testdata} shows the XGBoost models performance and Table \ref{tab:rima_rf_testdata} shows the RF models performance on the testing data for RIMA. Overall, we observed that the performance of non-spatial models is higher than spatial models. The variation in model performance in spatial models is high as compared to non-spatial models which shows that the five spatial models are different models while the five non-spatial models are almost the same models. The difference in model performance between training and testing data among spatial and non-spatial XGBoost (Table \ref{tab:rima_xgb_performance} and Table \ref{tab:rima_xgb_testdata}) and RF (Table \ref{tab:rima_rf_performance} and Table \ref{tab:rima_rf_testdata})  models also reinforce these findings. Specifically, the RF model performance on the training data and testing data of the first spatial model shows high model overfitting on the training data while the XGBoost model performed relatively better than the RF model. To further examine this performance difference, we plot a box plot comparing the distribution of prices across the two datasets. \textbf{Figure \ref{fig:rima_spatial_boxplot1}} shows the box plot of price (log-transformed) for the first spatial fold. \textbf{Figure \ref{fig:rima_spatial_boxplot1}} shows the difference in prices between the training and testing data indicating that the RF model did not perform well for out-of-sample prediction. The box plots for other folds are presented in Appendix B. The Toronto models show better performance among the training and testing datasets. This difference suggests the Toronto real estate market is more clearly described by the chosen variables, while RIMA exhibits greater residual unexplained variations. 

\begin{figure*}
    \centering
    \includegraphics[width=1\linewidth]{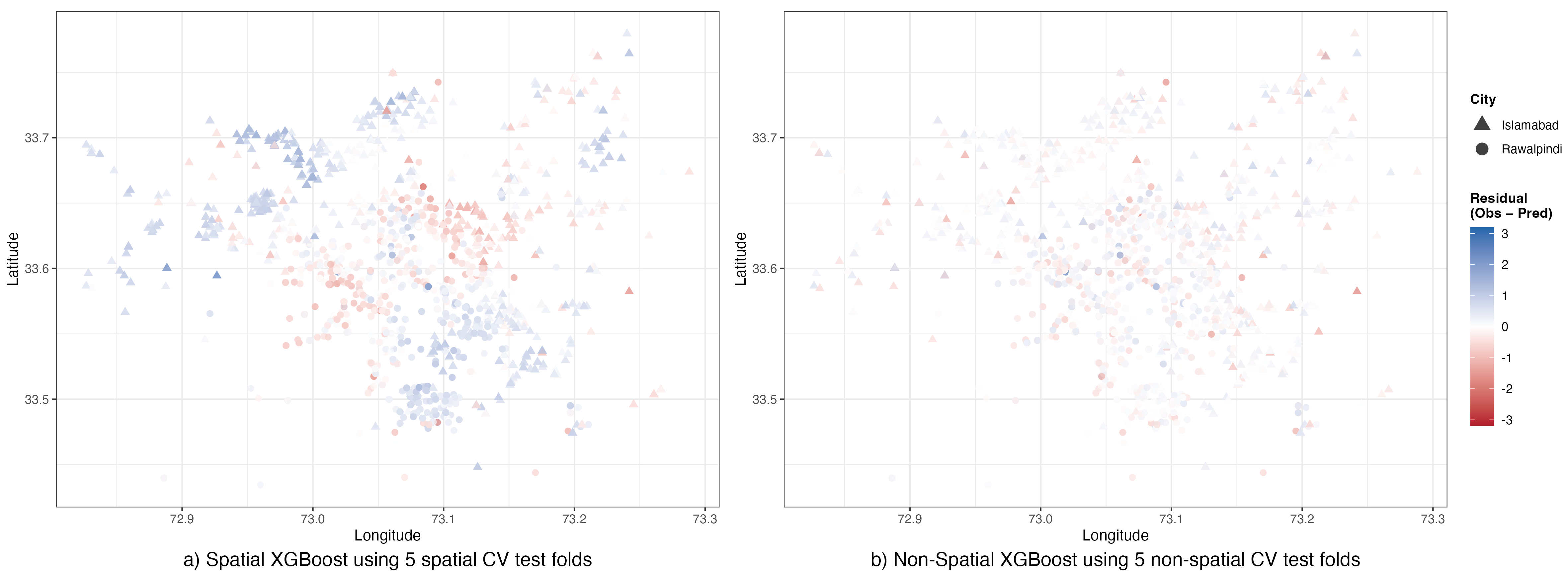}
    \caption{RIMA XGBoost Model Residuals}
    \label{fig:rima_resi}
\end{figure*}

\begin{figure*}
    \centering
    \includegraphics[width=1\linewidth]{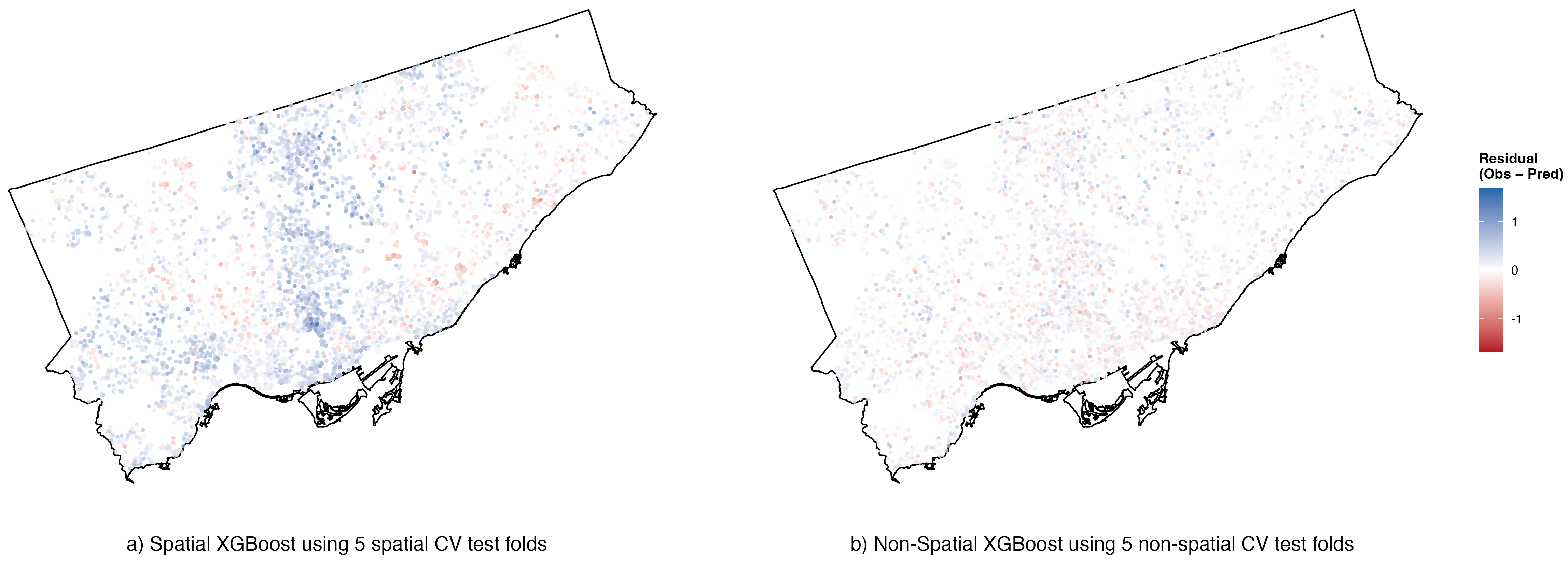}
    \caption{Toronto XGBoost Model Residuals}
    \label{fig:tor_resi}
\end{figure*}

Due to the overfitting of the spatial RF models in the case of RIMA, we chose XGBoost models as the preferred models. We also conducted residual analysis on the XGBoost models using testing data for both RIMA and Toronto. Figures \ref{fig:rima_resi} and \ref{fig:tor_resi} show XGBoost model residual results for both spatial and non-spatial models for RIMA and Toronto, respectively. Results for both study areas confirm that non-spatial XGBoost outperforms spatial XGBoost under random cross-validation prediction accuracy. Residual magnitudes are substantially larger in RIMA than Toronto, reflecting a more heterogeneous and data-sparse market where distance-based features alone cannot fully capture hyper-local price determinants, such as gated-community premiums. Spatially, Toronto's residuals exhibit clear neighborhood-level clustering, especially along the waterfront and outer suburbs, whereas RIMA's spatial clustering is more concentrated in the dense Rawalpindi core.

\begin{table}[]
\centering
\caption{RIMA XGBoost Model Performance on Testing Data}
\begin{tabular}{@{}lllll@{}}
\toprule
\textbf{Model} & \textbf{RMSE} & \textbf{R2} & \textbf{MAE} & \textbf{MAPE} \\
\midrule
Spatial 1      & 0.86          & 0.11        & 0.77         & 4.42          \\
Spatial 2      & 0.49          & 0.68        & 0.4          & 2.42          \\
Spatial 3      & 0.44          & 0.66        & 0.33         & 2.09          \\
Spatial 4      & 0.61          & 0.47        & 0.51         & 3.11          \\
Spatial 5      & 0.54          & 0.4         & 0.47         & 2.79          \\
\midrule
Non-Spatial 1     & 0.25          & 0.92        & 0.17         & 1.01          \\
Non-Spatial 2     & 0.25          & 0.93        & 0.17         & 1.04          \\
Non-Spatial 3     & 0.25          & 0.93        & 0.17         & 1.02          \\
Non-Spatial 4     & 0.24          & 0.93        & 0.16         & 0.97          \\
Non-Spatial 5     & 0.25          & 0.92        & 0.17         & 1.04          \\
\bottomrule
\end{tabular}
\label{tab:rima_xgb_testdata}
\end{table}

\begin{table}[]
\centering
\caption{RIMA RF Model Performance on Testing Data}
\begin{tabular}{@{}lllll@{}}
\toprule
\textbf{Model} & \textbf{RMSE} & \textbf{R2} & \textbf{MAE} & \textbf{MAPE} \\
\midrule
Spatial 1      & 0.93          & -0.03       & 0.84         & 4.77          \\
Spatial 2      & 0.44          & 0.74        & 0.35         & 2.12          \\
Spatial 3      & 0.44          & 0.65        & 0.34         & 2.11          \\
Spatial 4      & 0.56          & 0.56        & 0.49         & 2.97          \\
Spatial 5      & 0.5           & 0.48        & 0.42         & 2.51          \\
\midrule
Non-Spatial 1     & 0.24          & 0.93        & 0.16         & 0.99          \\
Non-Spatial 2     & 0.24          & 0.93        & 0.17         & 1.01          \\
Non-Spatial 3     & 0.25          & 0.93        & 0.17         & 1.01          \\
Non-Spatial 4     & 0.24          & 0.93        & 0.16         & 0.96          \\
Non-Spatial 5     & 0.25          & 0.93        & 0.17         & 1             \\
\bottomrule
\end{tabular}
\label{tab:rima_rf_testdata}
\end{table}

\begin{table}[]
\centering
\caption{Toronto XGBoost Model Performance on Testing Data}
\begin{tabular}{@{}c|c|c|c|c@{}}
\toprule
\textbf{Model} & \textbf{RMSE} & \textbf{R$^2$} & \textbf{MAE} & \textbf{MAPE} \\
\midrule
Spatial 1      & 0.32          & 0.76        & 0.25         & 1.85          \\
Spatial 2      & 0.27          & 0.67        & 0.22         & 1.68          \\
Spatial 3      & 0.27          & 0.73        & 0.2          & 1.52          \\
Spatial 4      & 0.32          & 0.69        & 0.26         & 1.94          \\
Spatial 5      & 0.27          & 0.7         & 0.21         & 1.59          \\
\midrule
Non-Spatial 1     & 0.14          & 0.94        & 0.1          & 0.74          \\
Non-Spatial 2     & 0.14          & 0.94        & 0.1          & 0.75          \\
Non-Spatial 3     & 0.15          & 0.93        & 0.1          & 0.76          \\
Non-Spatial 4     & 0.14          & 0.94        & 0.1          & 0.75          \\
Non-Spatial 5     & 0.15          & 0.93        & 0.11         & 0.79          \\
\bottomrule
\end{tabular}
\label{tab:tor_xgb_testdata}
\end{table}

\begin{table}[]
\centering
\caption{Toronto RF Model Performance on Testing Data}
\begin{tabular}{@{}c|c|c|c|c@{}}
\toprule
\textbf{Model} & \textbf{RMSE} & \textbf{R$^2$} & \textbf{MAE} & \textbf{MAPE} \\
\midrule
Spatial 1      & 0.27          & 0.83        & 0.2          & 1.46          \\
Spatial 2      & 0.23          & 0.76        & 0.17         & 1.34          \\
Spatial 3      & 0.28          & 0.7         & 0.22         & 1.67          \\
Spatial 4      & 0.25          & 0.81        & 0.19         & 1.42          \\
Spatial 5      & 0.37          & 0.42        & 0.29         & 2.22          \\
\midrule
Non-Spatial 1     & 0.14          & 0.94        & 0.1          & 0.73          \\
Non-Spatial 2     & 0.15          & 0.94        & 0.1          & 0.75          \\
Non-Spatial 3     & 0.15          & 0.93        & 0.1          & 0.77          \\
Non-Spatial 4     & 0.15          & 0.94        & 0.1          & 0.75          \\
Non-Spatial 5     & 0.16          & 0.93        & 0.1          & 0.78          \\
\bottomrule
\end{tabular}
\label{tab:tor_rf_testdata}
\end{table}

\section*{Discussion}
\subsection*{Impact of Land Use and Transportation on Property Prices in RIMA}
\begin{figure*}[h]
   \centering
    \includegraphics[width=0.8\linewidth]{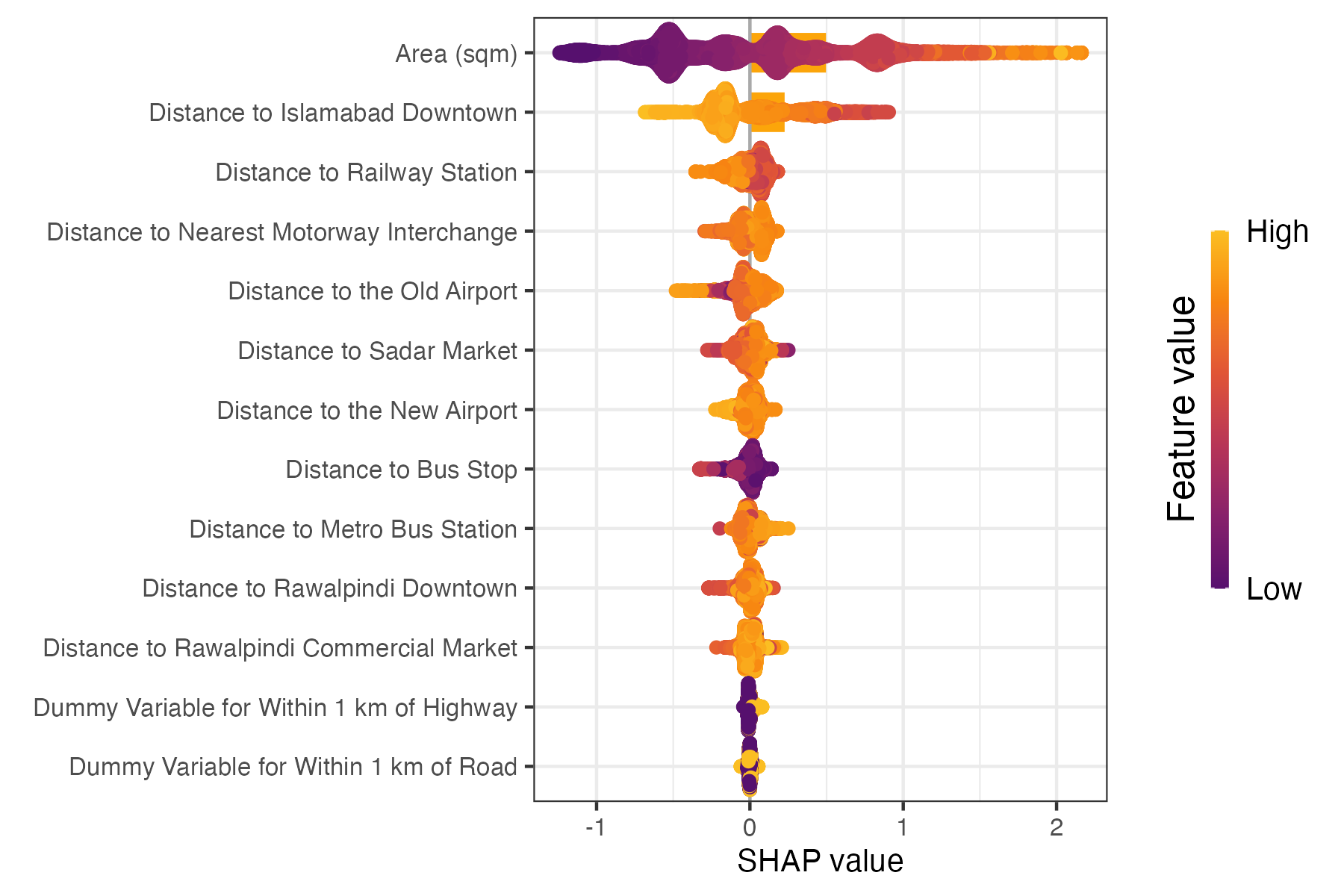}
    \caption{SHAP Values for Non-Spatial Model 1 (RIMA)}
    \label{fig:rima_shap_aspatial1}
\end{figure*}

In this section, we present SHAP plots only for the XGBoost models. SHAP beeswarm plots provide a rich summary of feature importance. Each dot represents a single data point, with its horizontal position showing the impact (SHAP value) of a feature on the model's prediction, and the color representing the value of that feature. Features are ranked on the y-axis by their overall importance (mean absolute SHAP value). \textbf{Figure \ref{fig:rima_shap_aspatial1}} shows SHAP values for the first non-spatial model results for RIMA (SHAP plots for the remaining four folds are included in Appendix C). The variables are arranged based on their importance from high to low. For instance, area is the most important variable and proximity to a highway is the least important variable in predicting the property sale price. The color shows the value of a variable (or feature), which in the case of dummy variables only has two values. \textbf{Figure  \ref{fig:rima_shap_aspatial1}} also shows non-linear relationships, longer tails in one direction than the other, between dependent and independent variables.

\begin{figure*}[h]
   \centering
    \includegraphics[width=0.8\linewidth]{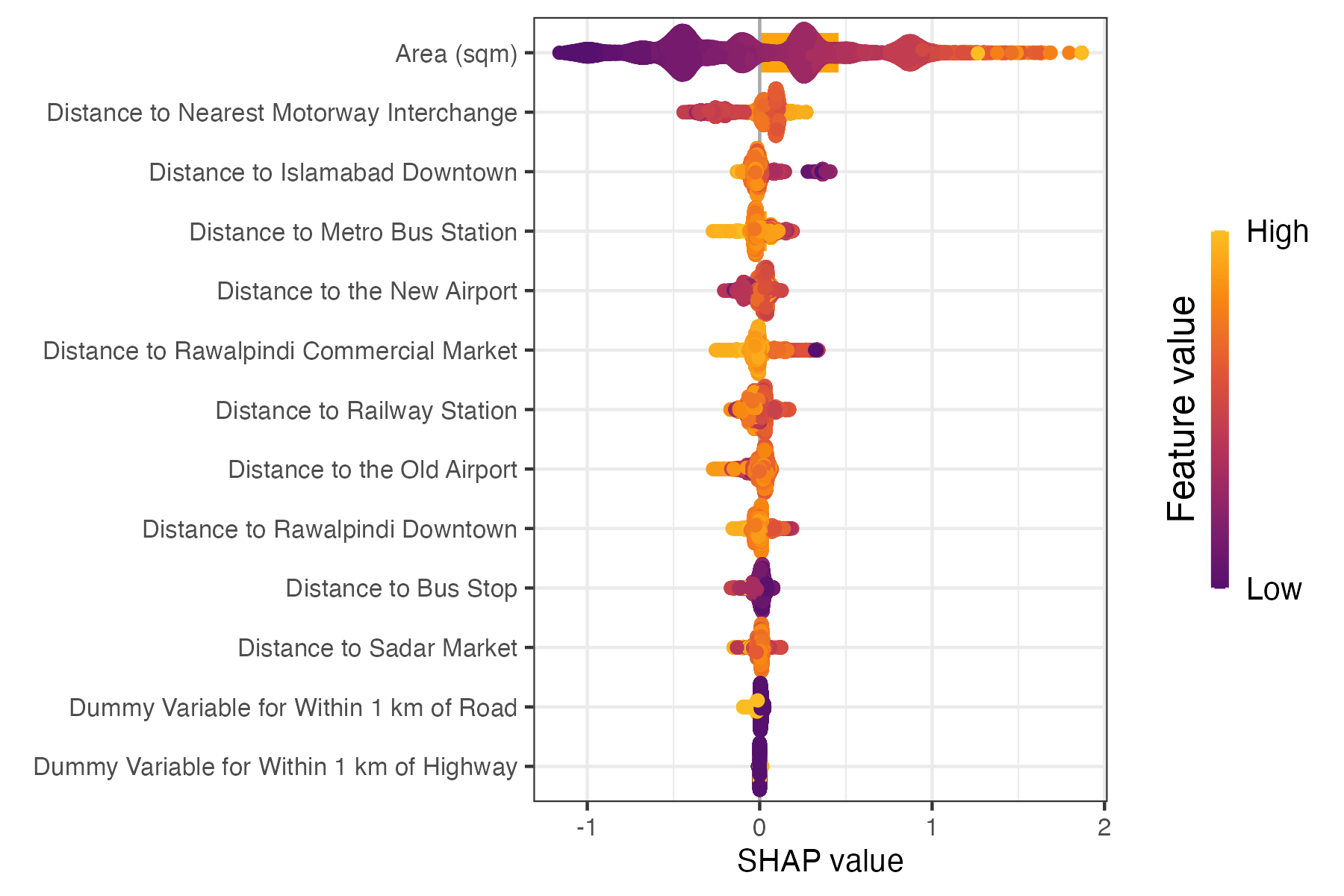}
    \caption{SHAP Values for Spatial Model 1 (RIMA)}
    \label{fig:rima_shap_spatial1}
\end{figure*}

Floor area is the most important variable. As expected, a larger building area correlates with a higher sale price. Accessibility to transportation infrastructure also impacts prices. For instance, houses located closer to motorway ramps have relatively lower sale prices. This result is because motorways in Pakistan are typically for inter-city travel and are tolled. Motorway interchanges are mostly located at a city's outskirts. However, we note that the  non-spatial model exhibits a weaker effect than observed in the spatial model (see \textbf{Figure  \ref{fig:rima_shap_spatial1}}). That is, using non-spatial cross-validation tends to miss the effect size (i.e., importance ranking) and the negative effect on property price captured by the spatial model. 

Based on the spatial model, proximity to metro stations and regular bus stops has a negative impact on sale price, but the impact of metro stations is larger than regular bus stops. This difference could be because metro bus has fixed stops and connects all major commercial centers in Rawalpindi and Islamabad. The overall quality of the service (e.g., comfort and reliability) is far superior than the other bus service. As expected, sale prices are also higher near the major commercial centers, as well as both downtown Rawalpindi and Islamabad. These areas are major employment centers and have higher accessibility to services. Proximity to airports impacts sale price, but the impact is unclear. The old airport is located closer to the Sadar Market, while the new airport is very far from both the cities. Distance to highways has a weak effect on prices.

\subsection*{Impact of Land Use and Transportation on Property Prices in Toronto}
\begin{figure*}[]
   \centering
    \includegraphics[width=0.8\linewidth]{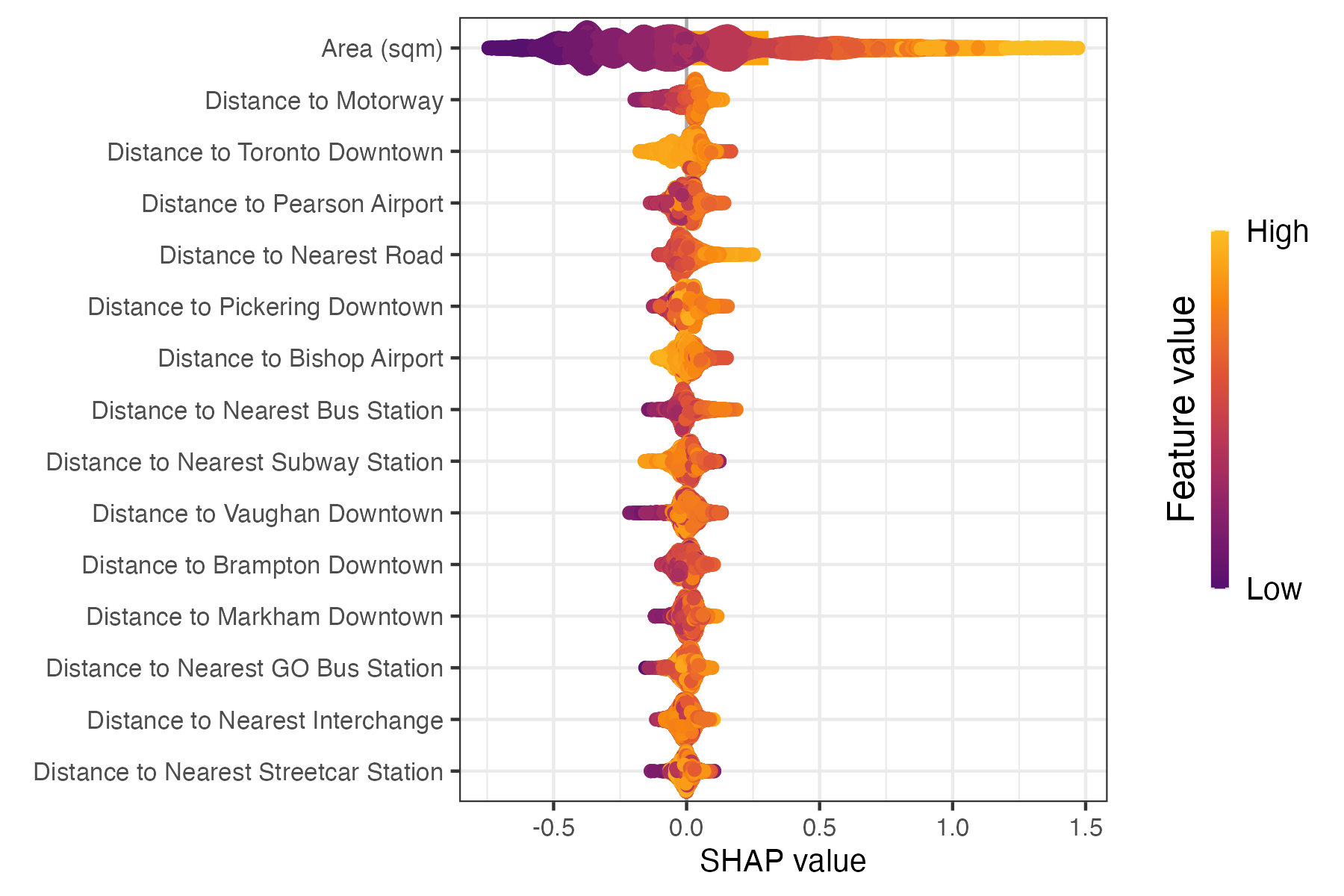}
    \caption{SHAP Values for Non-Spatial Model 1 (Toronto)}
    \label{fig:gta_shap_aspatial1}
\end{figure*}

\textbf{Figure \ref{fig:gta_shap_aspatial1}} shows SHAP values for the first non-spatial model for Toronto. The variables are arranged based on their importance from high to low. Floor area is the most important variable and proximity to the nearest streetcar station is the least important variable in predicting the property sale price. \textbf{Figure  \ref{fig:gta_shap_aspatial1}} also shows non-linear relationships, longer tails in one direction than the other, between dependent and independent variables. Distance to the nearest motorway has a positive influence on property sale price, while distance to the Toronto downtown has a negative effect, meaning that higher priced properties tend to be located closer to the Toronto downtown. Accessibility to transportation infrastructure also impacts prices. Proximity to a subway station tends to be associated with higher prices, while proximity to bus stops tends to decrease price. This result is likely capturing the location of bus stops along major roadways that are viewed as undesirable, whereas subway stations tend to be located in desirable areas with high amenity concentrations (e.g., theaters and restaurants). Proximity to the major airport, Pearson International Airport, is also perceived as undesirable.

\begin{figure*}[]
   \centering
    \includegraphics[width=0.8\linewidth]{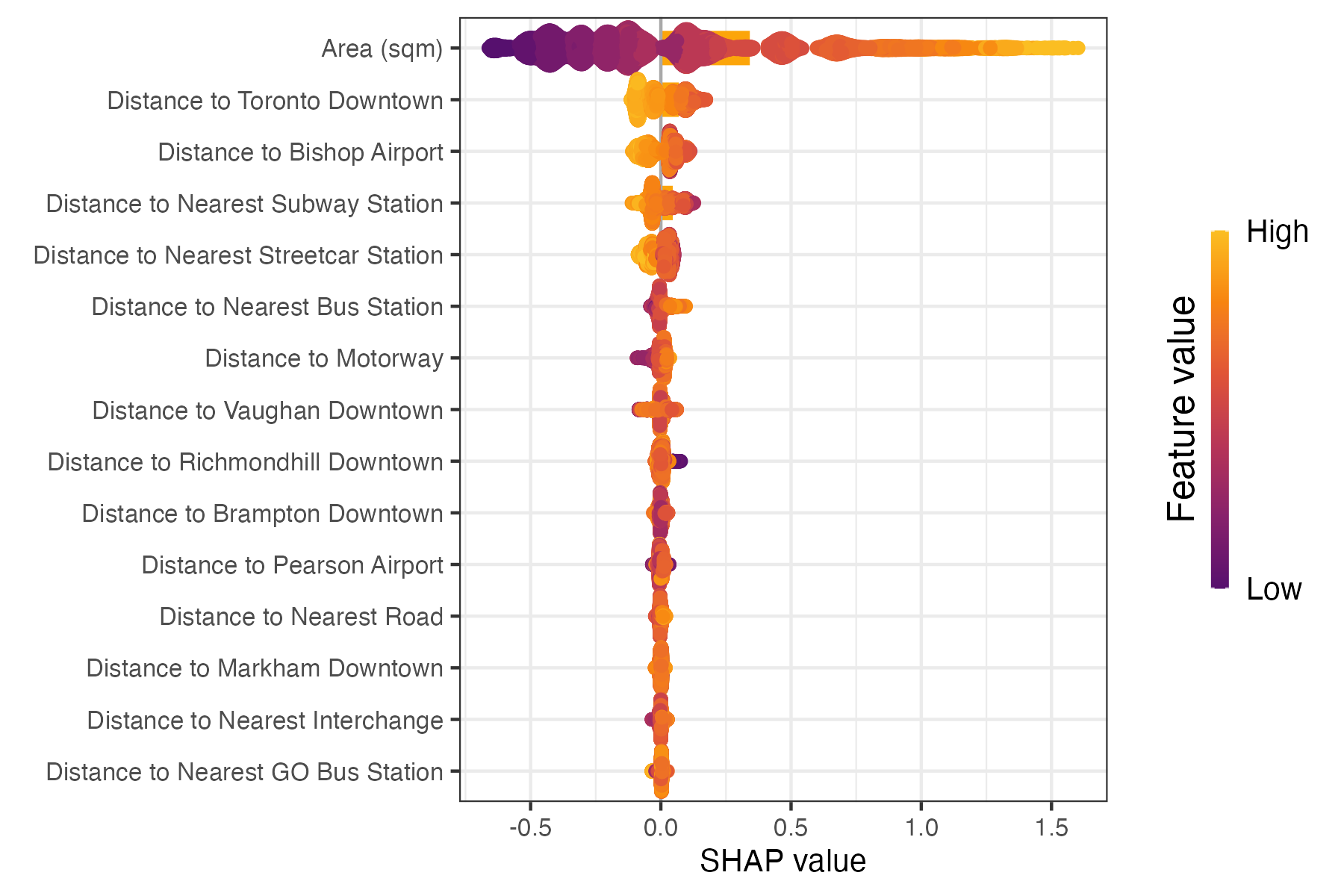}
    \caption{SHAP Values for Spatial Model 1 (Toronto)}
    \label{fig:gta_shap_spatial1}
\end{figure*}

The strength of many features is weaker in the spatial model, as compared to the non-spatial model, as shown in \textbf{Figure \ref{fig:gta_shap_spatial1}} wherein the SHAP values are less variable. The effect of distance to motorways is weaker in this model, while the effect of proximity to the Toronto downtown remains strong.

\subsection*{Comparison of RIMA and Toronto Results}

The Toronto market exhibits several similarities to that of RIMA, despite their geographic and socioeconomic differences. Both models found that floorspace area is the most important factor in determining sale price. Higher order transit (BRT for RIMA and subway for Toronto) have strong positive effects on property price. Proximity to the major central city is desirable in both areas. Results for airport proximity differ between the two areas depending on whether one considers non-spatial or spatial model results. Distance from the new airport is desired in RIMA, to avoid the noise and associated traffic. This effect is also strong for Pearson Airport in the GTA in the non-spatial model but much weaker when using spatial cross-validation. Proximity to Billy Bishop Airport arises as a desirable feature in the spatial Toronto model; however, it must be noted that this airport is relatively small and located on an island connected to downtown Toronto by a tunnel. Distance to non-Toronto city centers have fairly weak effects, similar to non-Islamabad markets in RIMA. With respect to roadway variables, RIMA has a stronger feature importance for motorways, whereas Toronto shows a stronger influence from subway and streetcar proximity. Higher order transit (rail and metro bus) station proximity have stronger effects on sale price than local bus station proximity in both regions. Some of the differences in the impact of transportation infrastructure, specifically the public transit system, are due to cultural and planning differences between the two cities. For example, the spatial coverage of transit infrastructure in the GTA is far higher than RIMA. The transit infrastructure in RIMA is not well connected, especially in Islamabad. However, modes such as rickshaws and motorcycles are common in RIMA but not the GTA.

Our results align with prior research in both regions, finding that proximity to the main CBD tends to increase property price, aligned with the monocentric city model of Alonso and subsequent research in urban economics \cite{Alonso_1964,Brueckner_2011}. Based on SHAP analysis, we recommend prioritizing high-capacity, dedicated right-of-way transit investments (i.e., BRT/Subway expansion) to maximize positive property value impacts and fund transit through land value capture \cite{Kasraian_Li_Raghav_Shalaby_Miller_2023}. Methodological findings also have implications for policy. Non-spatial models overestimate accuracy; thus, policy analysts should use spatial validation techniques in real estate appraisal and land use planning models to ensure reliable and generalizable predictions for investment decisions.

\section*{Conclusions and Future Work}
Land use and transportation systems interact with each other in cities. Residential property prices are an important component of urban systems that are influenced by the integrated nature of land use and transportation. In this study, we examined the impact of land use and transportation systems on property prices using machine learning methods and compare the results between two diverse areas (Toronto, Canada and RIMA, Pakistan). We also investigated the differences between non-spatial and spatial cross-validation methods arising from the presence of spatial auto-correlation. We confirmed prior findings in ecology and other fields \cite{robertsCrossvalidationStrategiesData2017} that non-spatial cross-validation leads to erroneous prediction accuracy results applying machine learning models in two diverse regions and comparing multiple model diagnostic statistics. These findings are important to highlight for transportation researchers as they increasingly apply machine learning methods developed for non-spatial uses in spatial contexts.

We found similar impacts of land use and transportation infrastructure for both Toronto and RIMA, despite their socioeconomic differences. We also found non-linear effects of these variables in predicting property prices for both regions using SHAP based variable importance. These non-linear impacts are crucial to capture, especially for accurate policy analysis. For instance, such results are useful to test the impact of infrastructure investments by type of transit system such as bus, subway, or BRT. While we provide results for both developed and developing economy cities, robust generalization of our results will require applications across more cities. We also emphasis that our study focused on high-level associations rather than causal identification of a specific treatment effect. Causal machine learning is a growing field \cite{Wager_Athey_2018}, but one we did not explore in this work.

Turning to the cross-validation question, we used a spatial blocking approach to cross-validation, which has some trade-offs associated with its application. This approach addresses the independence assumption of cross-validation, that hold-out testing samples are independent from the training sample, but may lead to the testing sample outcome values not existing within the same support as the training data. The interpretation of this assumption depends on whether one is interested in inference or extrapolation in predictor space. An alternative approach would be to block the data in the predictor space based on a dissimilarity criteria. That is, spatially cluster observations to maximize the dissimilarity within each fold (i.e., similarity across folds) across predictor variables. To our knowledge, there are few examples in the literature of cross-validations using blocks defined in the predictor space in this way.

A second challenge with marrying machine learning and spatial econometrics is the lack of a likelihood function in most machine learning approaches. This difference precludes the consistent introduction of spatial lag or error terms. Flexible non-parametric statistical models, such as Bayesian Additive Regression Trees (BART), should be explored more \cite{Krueger_Bansal_Buddhavarapu_2020}. They offer the advantage of maintaining a likelihood function for integration with standard spatial econometric specifications, while maintaining the flexibility of algorithmic machine learning models.

Other future work might include the use of panel data to study the dynamics of price changes. Such dynamics are important to understand how transit investments impact socio-spatial equity, a challenge identified in both Toronto and RIMA. Our transit variables also did not consider reliability or service quality variation across the two study regions. Such information is available for Toronto but may be more challenging to obtain for RIMA or similarly developed areas.

\begin{dci}
The authors declared no potential conflicts of interest with respect to the research, authorship, and/or publication of this article.
\end{dci}

\begin{ac}
The authors confirm contribution to
the paper as follows: study conception
and design: U. Ahmed, J. Hawkins; 
data collection: U. Ahmed, J. Hawkins; 
analysis and interpretation of 
results: U. Ahmed, J. Hawkins; draft manuscript preparation:
U. Ahmed, J. Hawkins. 
All authors reviewed the results and
approved the final version 
of the manuscript.
\end{ac}

\begin{funding}
This research received no specific grant from any funding agency in the public, commercial, or not-for-profit sectors.
\end{funding}

\bibliographystyle{TRR}

\bibliography{references}

\appendix

\renewcommand\thefigure{\thesection A\arabic{figure}}
\setcounter{figure}{0} 

\section{Appendix A. Non-Spatial and Spatial Clusters}

\begin{figure}[htbp]
    \centering
    \includegraphics[width=1\linewidth]{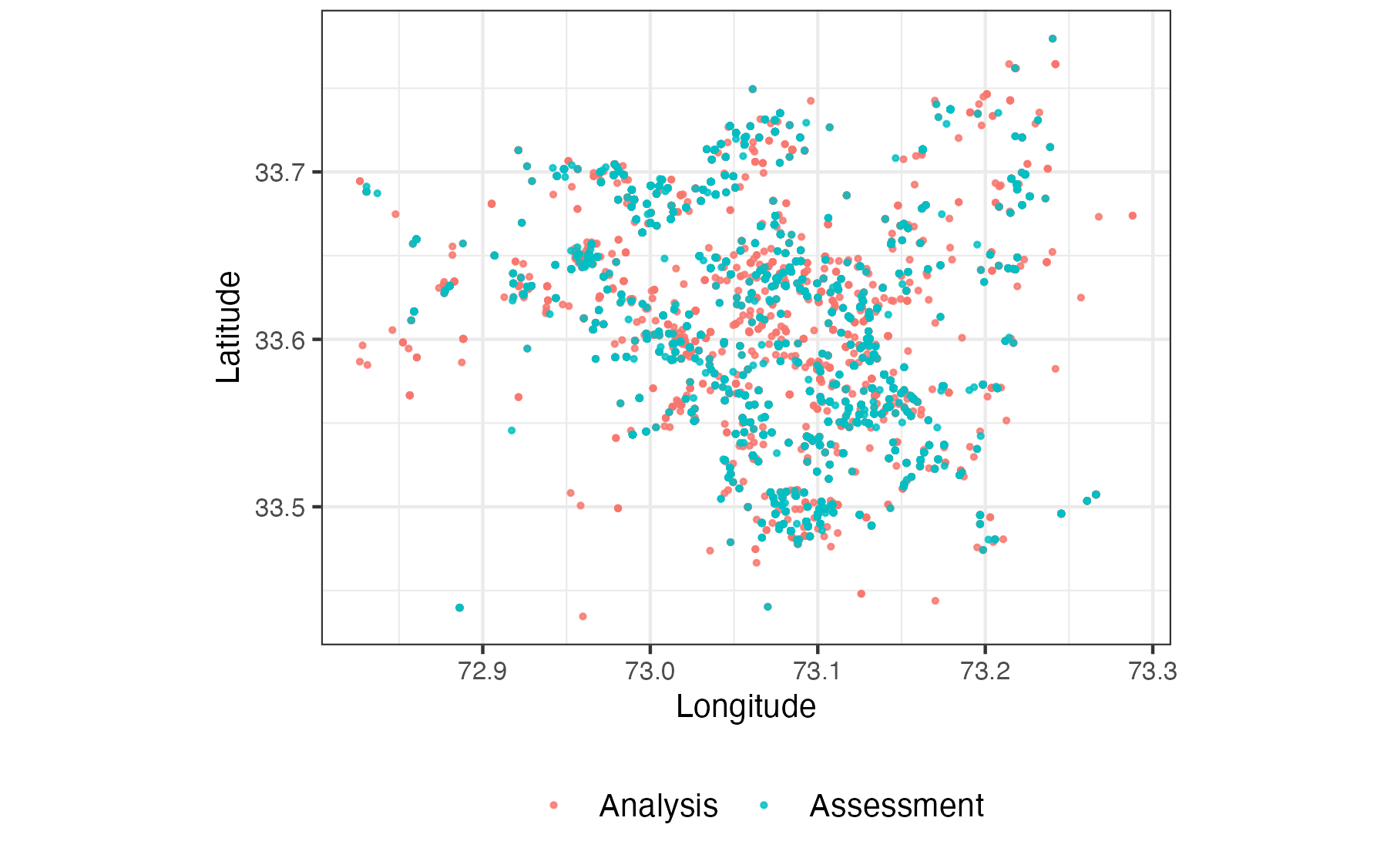}
    \caption{RIMA Non-Spatial Fold 2}
    \label{fig:rima_non-spatial_fold2}
\end{figure}

\begin{figure}[htbp]
    \centering
    \includegraphics[width=1\linewidth]{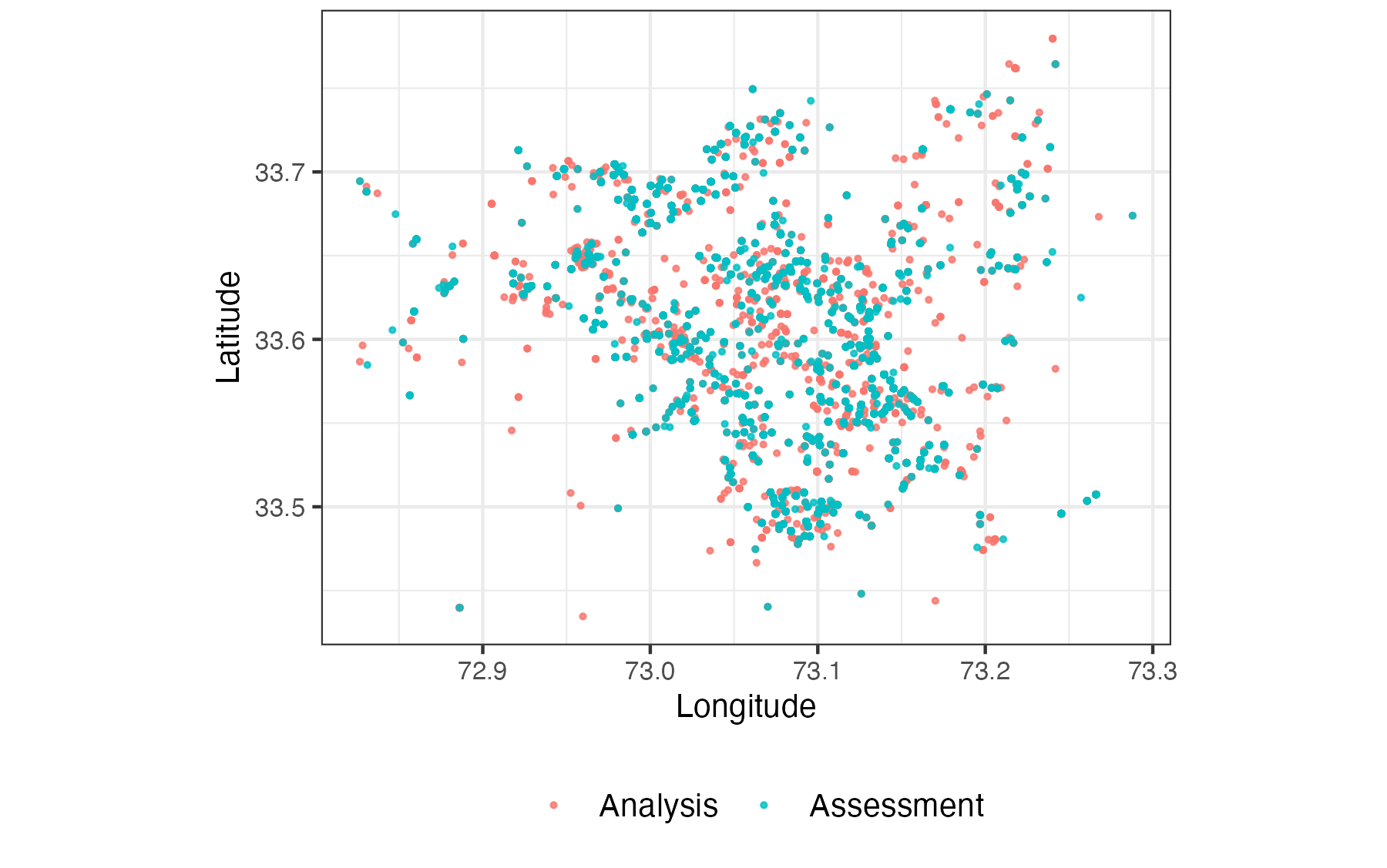}
    \caption{RIMA Non-Spatial Fold 3}
    \label{fig:rima_non-spatial_fold3}
\end{figure}

\begin{figure}[htbp]
    \centering
    \includegraphics[width=1\linewidth]{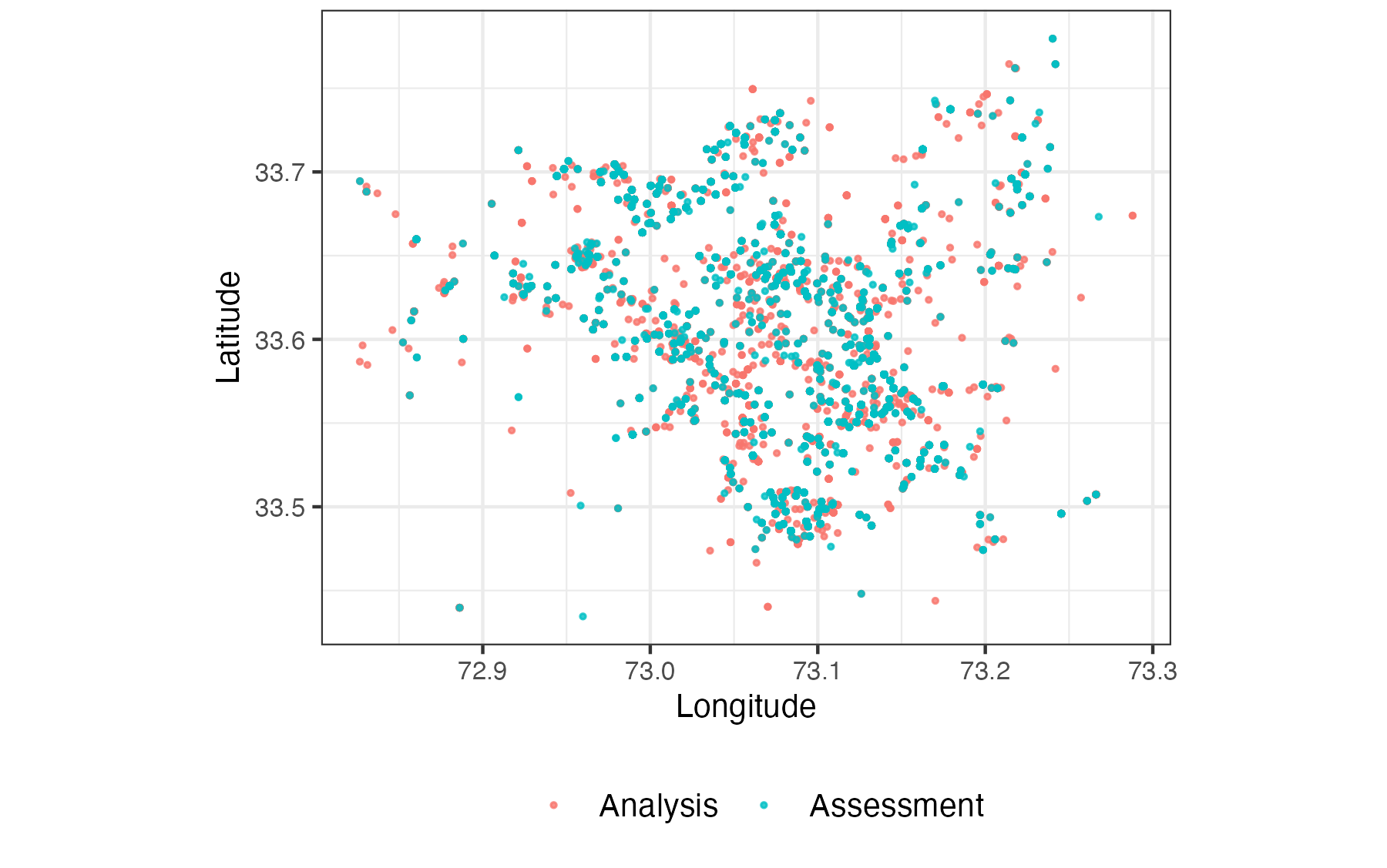}
    \caption{RIMA Non-Spatial Fold 4}
    \label{fig:rima_non-spatial_fold4}
\end{figure}

\begin{figure}[htbp]
    \centering
    \includegraphics[width=1\linewidth]{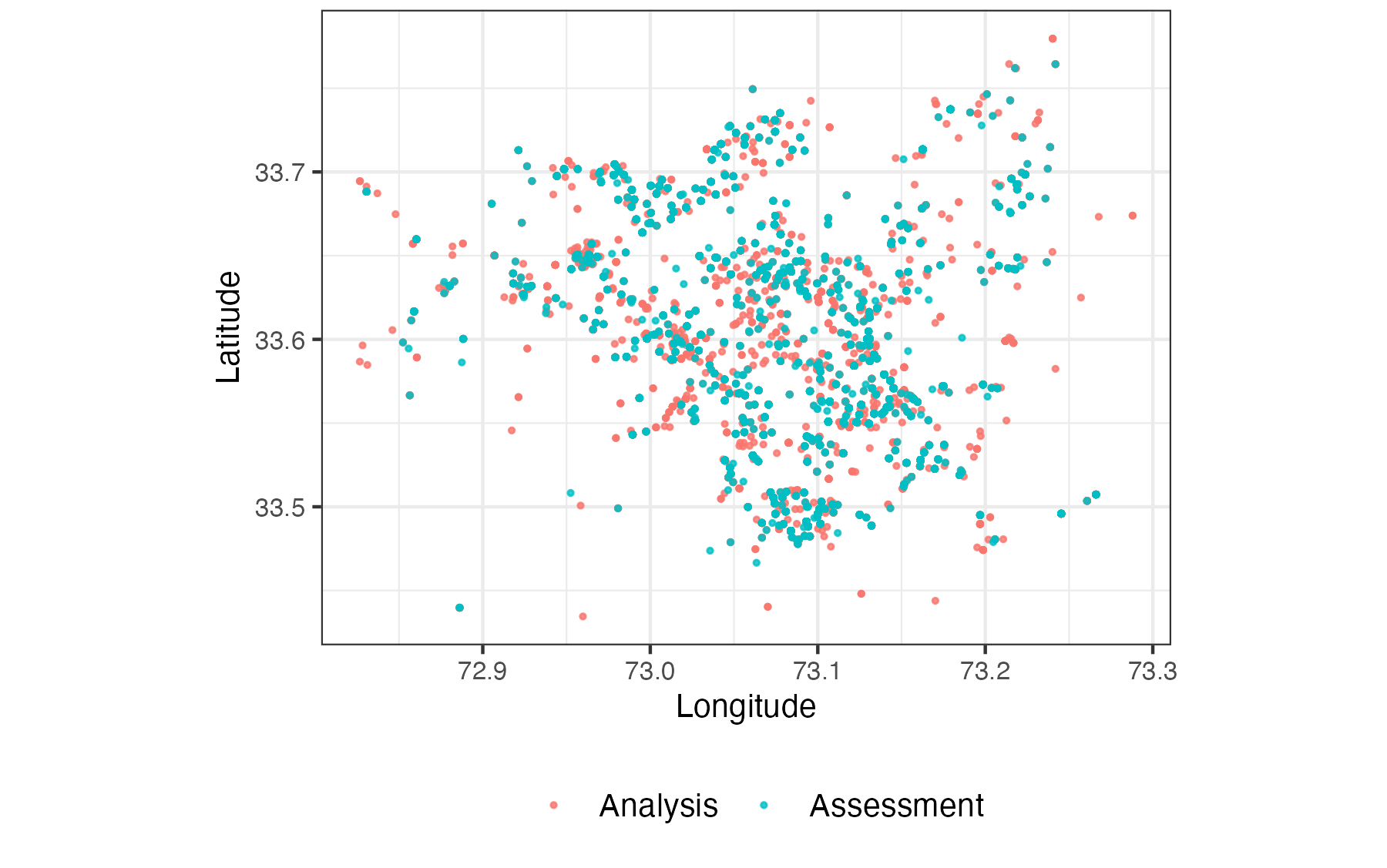}
    \caption{RIMA Non-Spatial Fold 5}
    \label{fig:rima_non-spatial_fold5}
\end{figure}

\begin{figure}[htbp]
    \centering
    \includegraphics[width=1\linewidth]{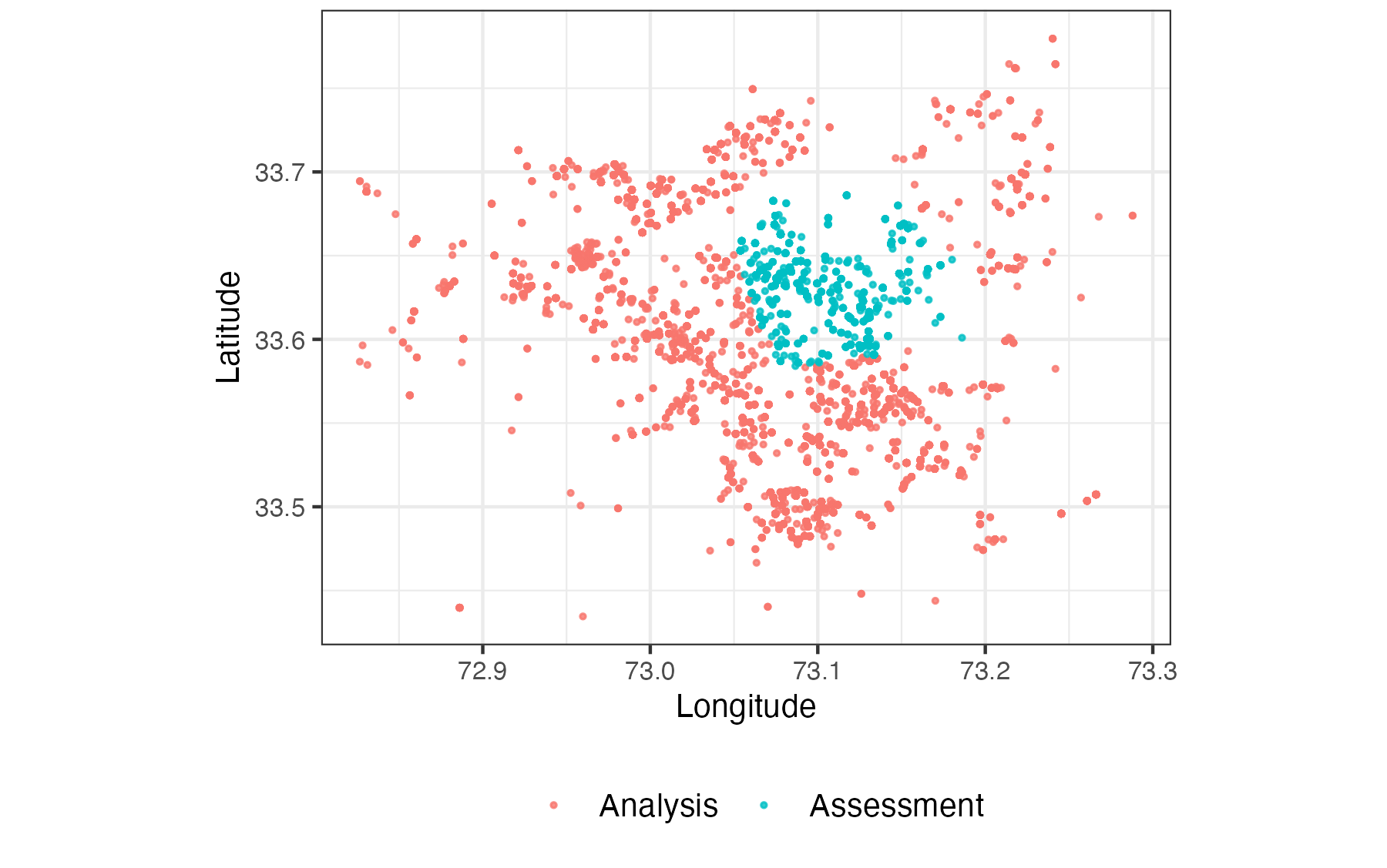}
    \caption{RIMA Spatial Fold 2}
    \label{fig:rima_spatial_fold2}
\end{figure}

\begin{figure}[htbp]
    \centering
    \includegraphics[width=1\linewidth]{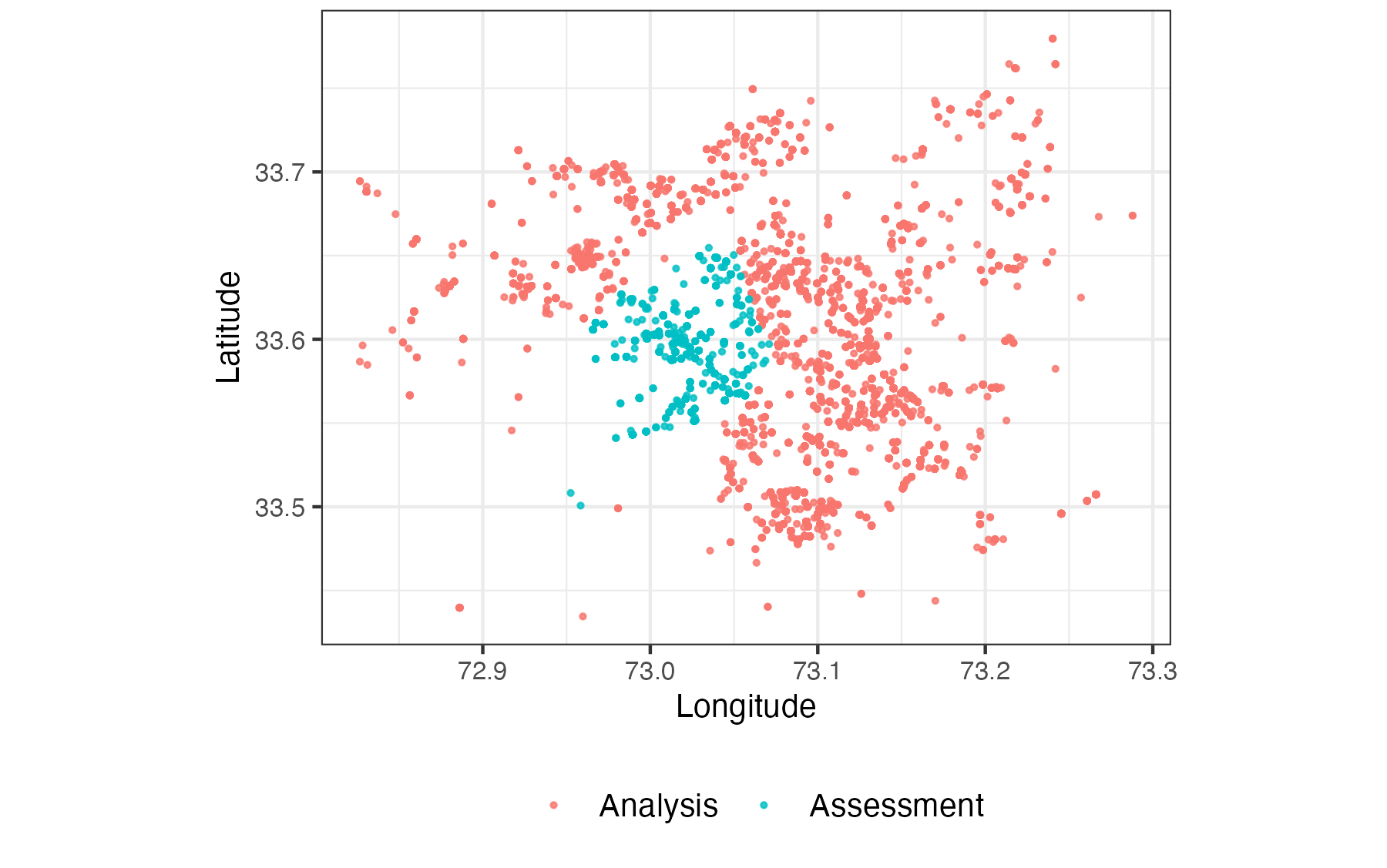}
    \caption{RIMA Spatial Fold 3}
    \label{fig:rima_spatial_fold3}
\end{figure}

\begin{figure}[htbp]
    \centering
    \includegraphics[width=1\linewidth]{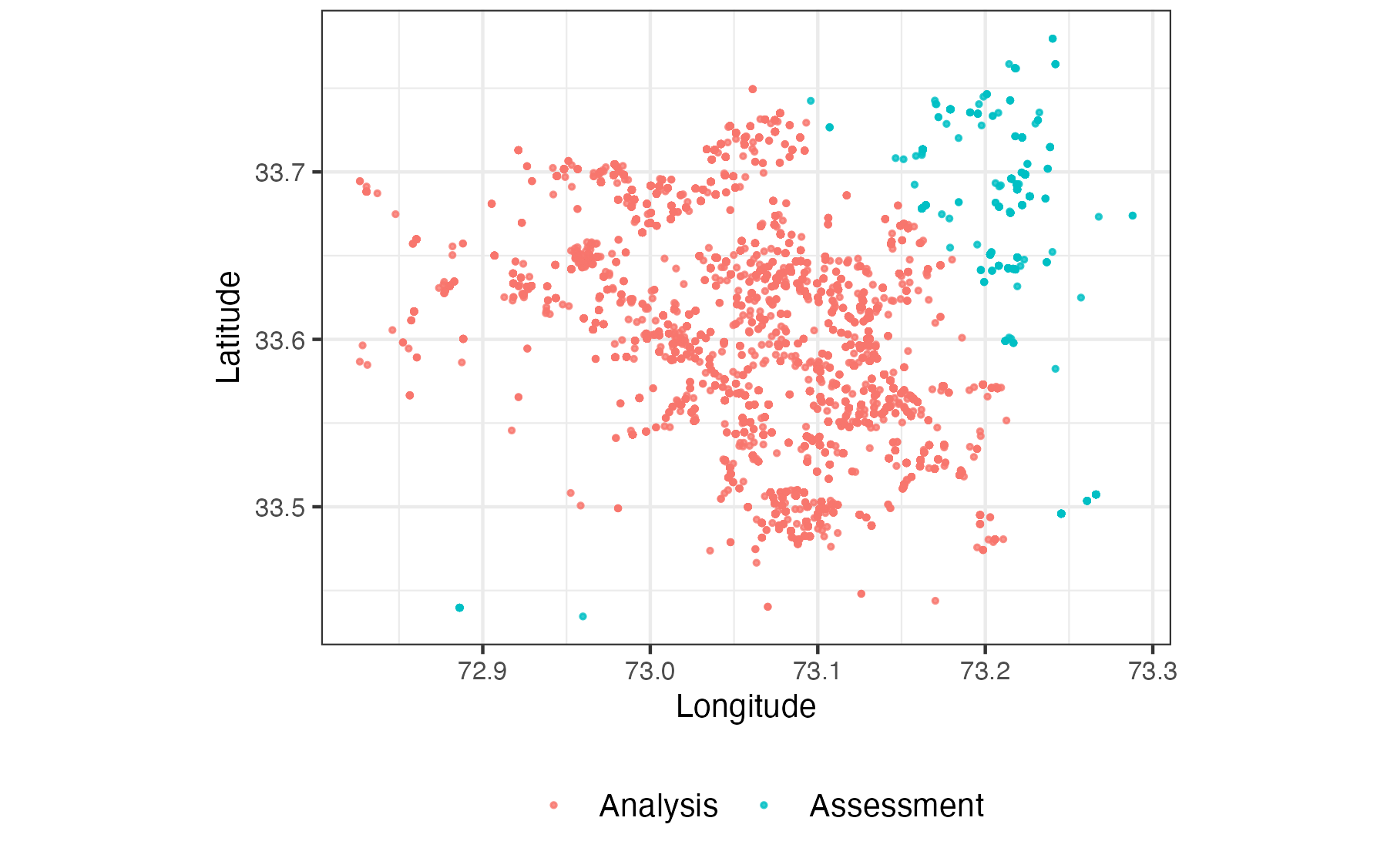}
    \caption{RIMA Spatial Fold 4}
    \label{fig:rima_spatial_fold4}
\end{figure}

\begin{figure}[htbp]
    \centering
    \includegraphics[width=1\linewidth]{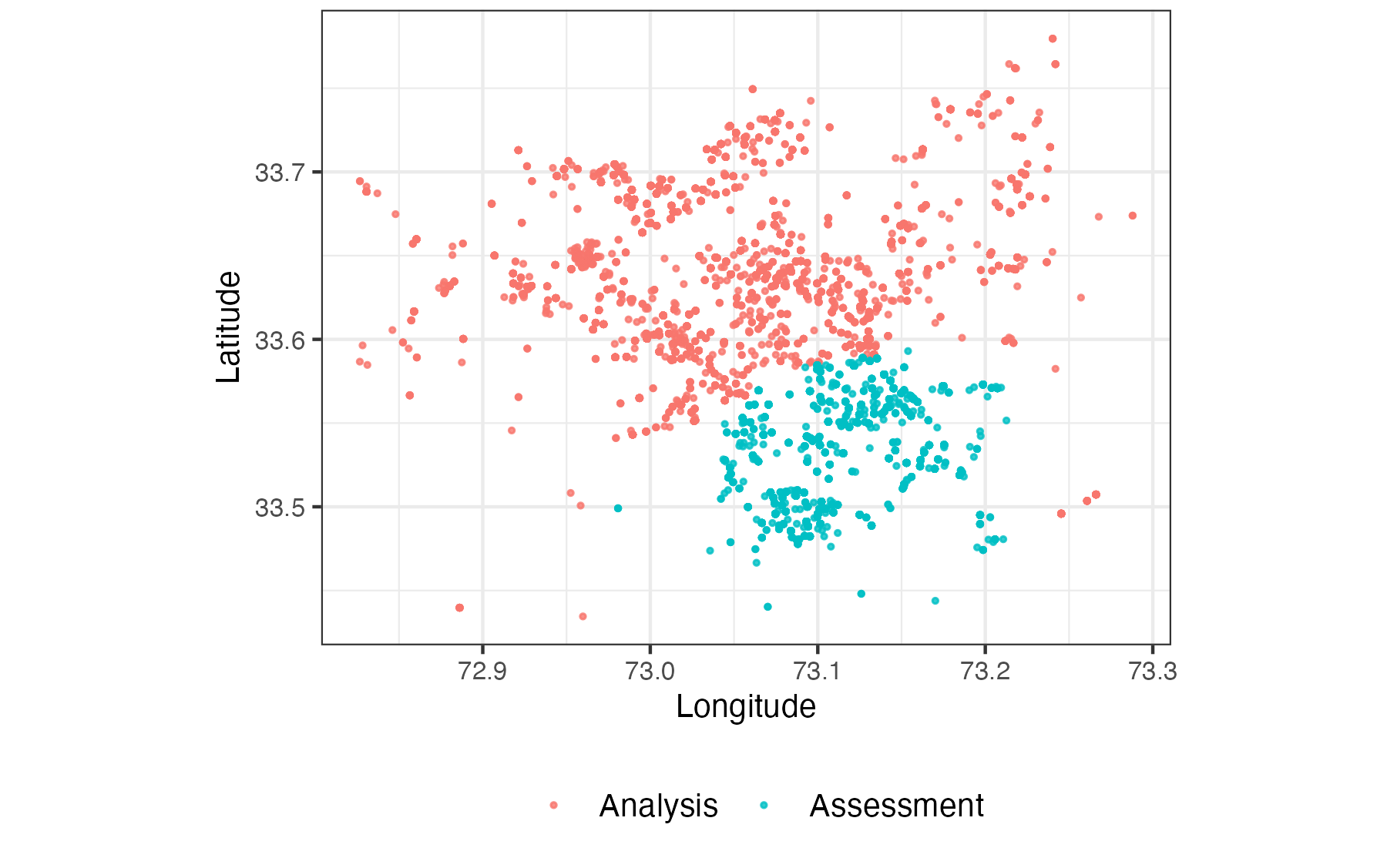}
    \caption{RIMA Spatial Fold 5}
    \label{fig:rima_spatial_fold5}
\end{figure}


\begin{figure}[htbp]
    \centering
    \includegraphics[width=1\linewidth]{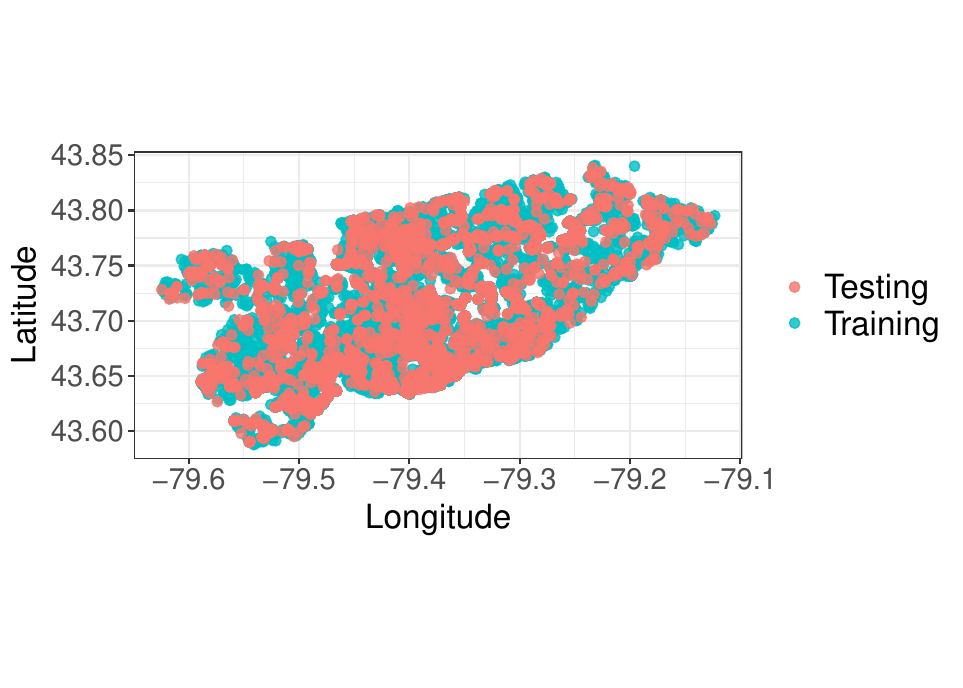}
    \caption{Toronto Non-Spatial Fold 2}
    \label{fig:gta_non-spatial_fold2}
\end{figure}

\begin{figure}[htbp]
    \centering
    \includegraphics[width=1\linewidth]{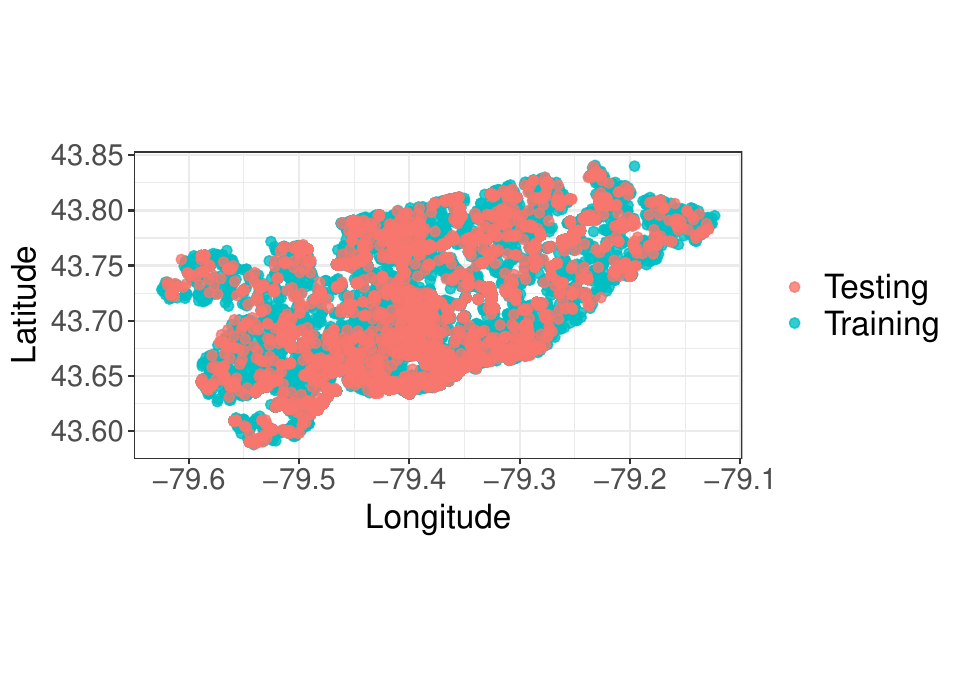}
    \caption{Toronto Non-Spatial Fold 3}
    \label{fig:gta_non-spatial_fold3}
\end{figure}

\begin{figure}[htbp]
    \centering
    \includegraphics[width=1\linewidth]{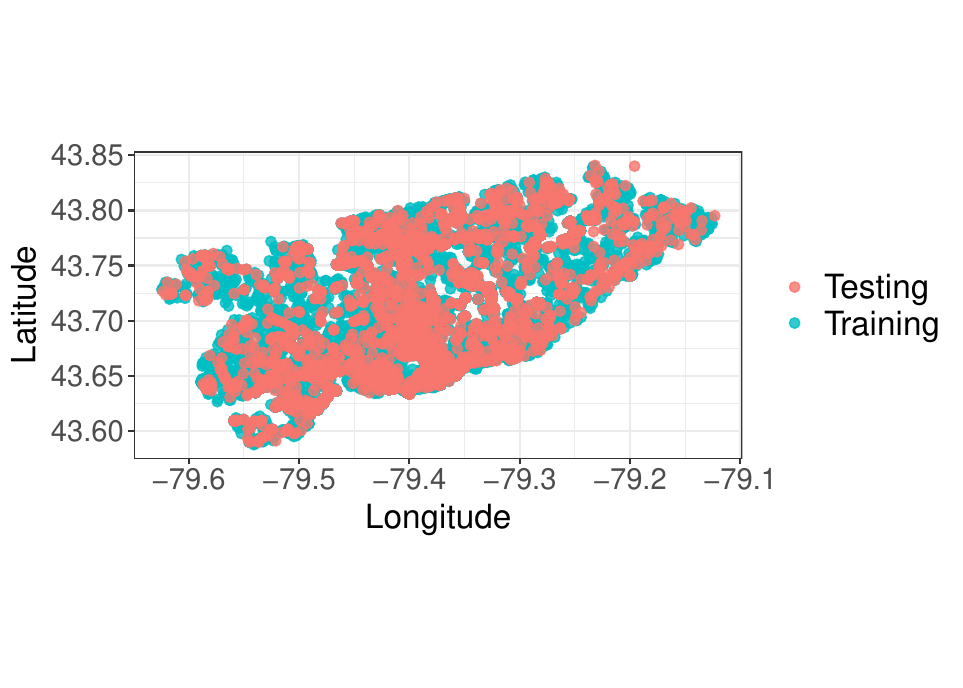}
    \caption{Toronto Non-Spatial Fold 4}
    \label{fig:gta_non-spatial_fold4}
\end{figure}

\begin{figure}[htbp]
    \centering
    \includegraphics[width=1\linewidth]{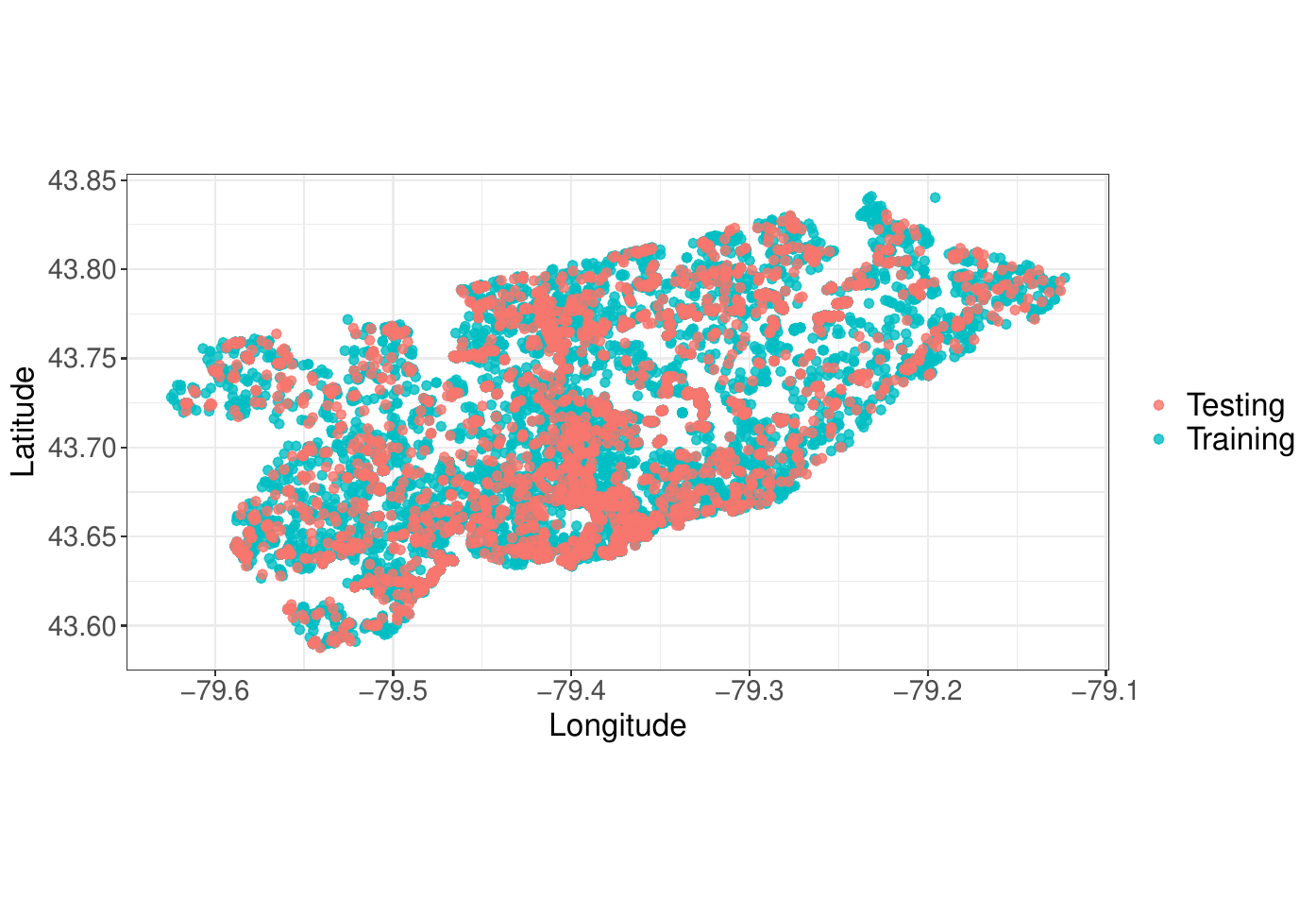}
    \caption{Toronto Non-Spatial Fold 5}
    \label{fig:gta_non-spatial_fold5}
\end{figure}

\begin{figure}[htbp]
    \centering
    \includegraphics[width=1\linewidth]{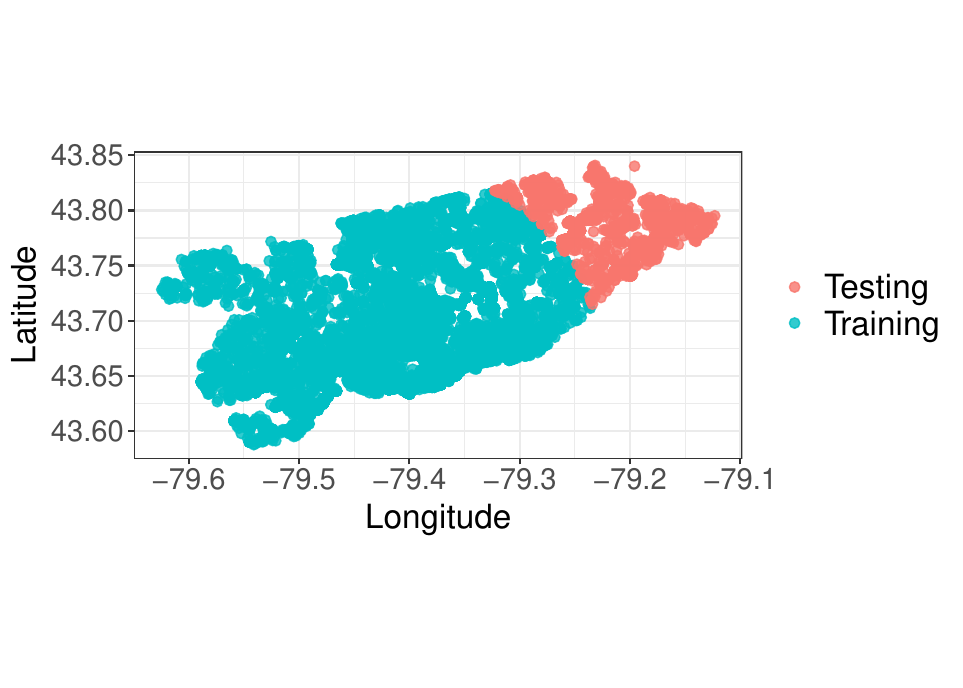}
    \caption{Toronto Spatial Fold 2}
    \label{fig:gta_spatial_fold2}
\end{figure}

\begin{figure}[htbp]
    \centering
    \includegraphics[width=1\linewidth]{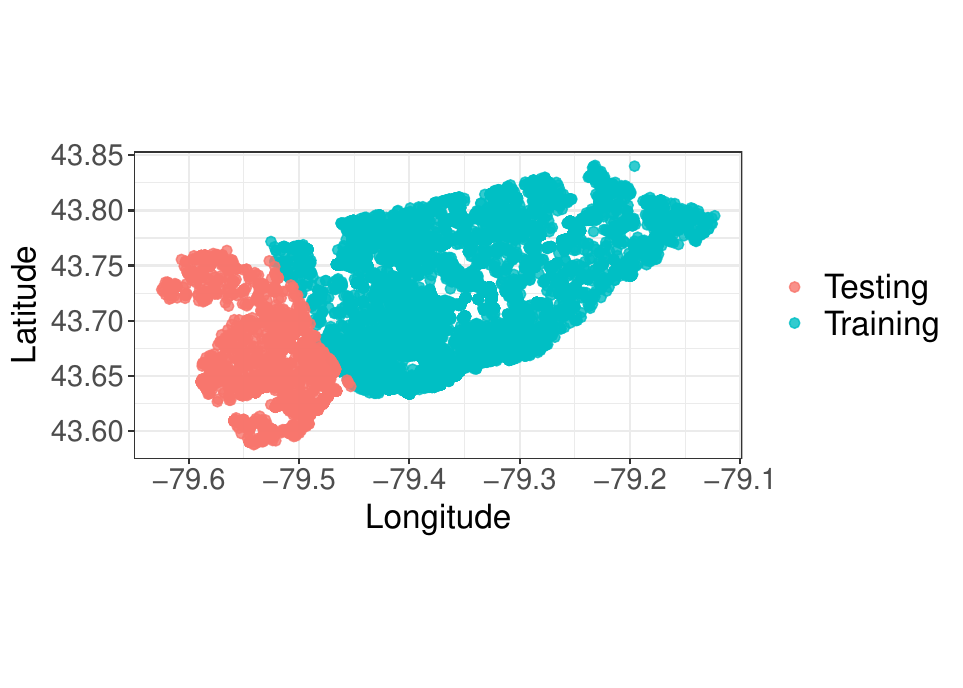}
    \caption{Toronto Spatial Fold 3}
    \label{fig:gta_spatial_fold3}
\end{figure}

\begin{figure}[htbp]
    \centering
    \includegraphics[width=1\linewidth]{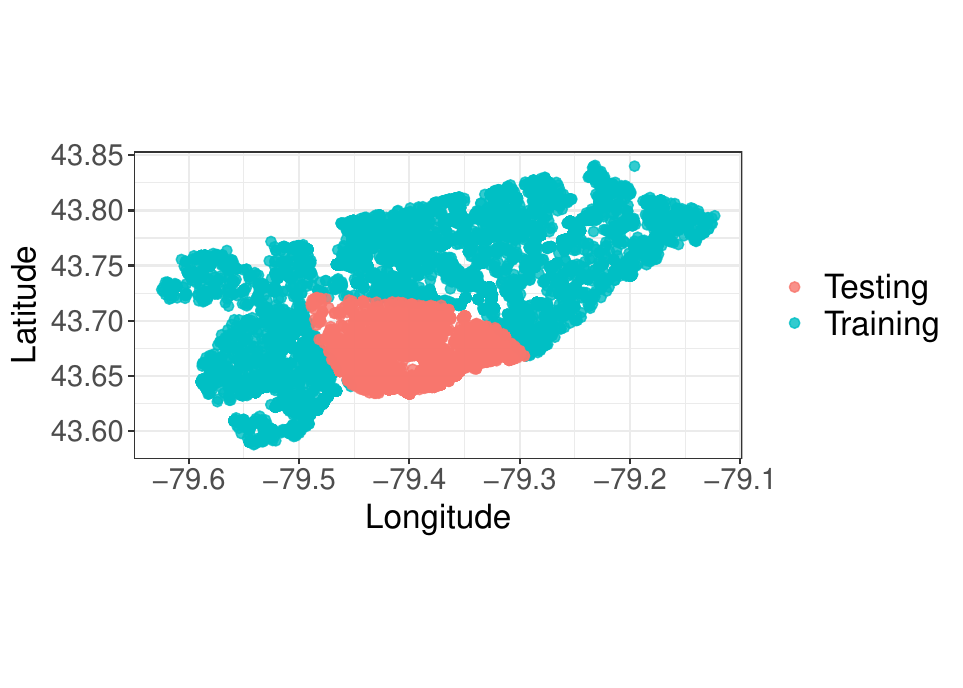}
    \caption{Toronto Spatial Fold 4}
    \label{fig:gta_spatial_fold4}
\end{figure}

\begin{figure}[htbp]
    \centering
    \includegraphics[width=1\linewidth]{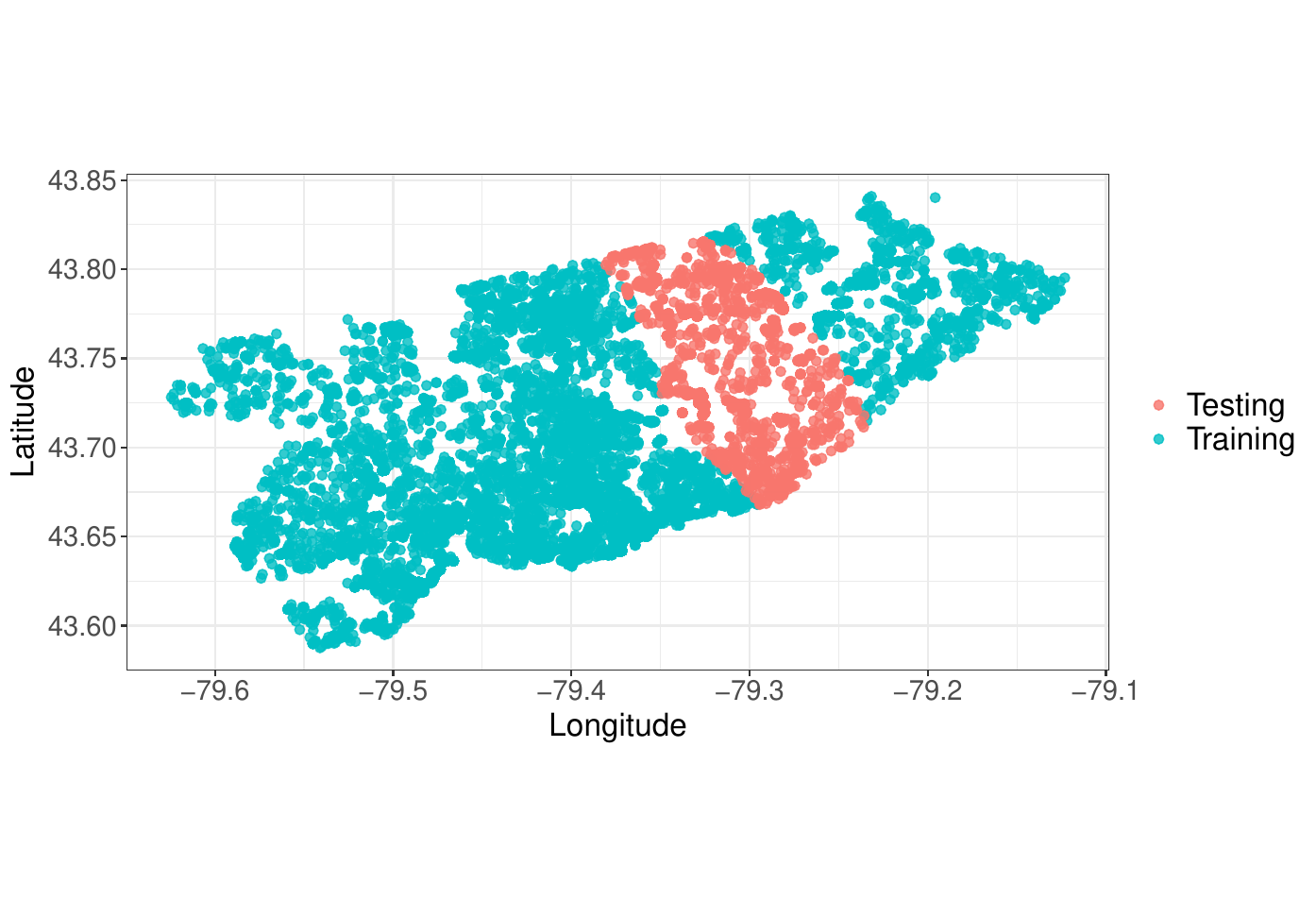}
    \caption{Toronto Spatial Fold 5}
    \label{fig:gta_spatial_fold5}
\end{figure}

\FloatBarrier

\renewcommand\thefigure{\thesection B\arabic{figure}}
\setcounter{figure}{0} 

\section{Appendix B. RIMA - Box Plots of the Spatial and Non-Spatial Folds}

\begin{figure}[htbp]
    \centering
    \includegraphics[width=1\linewidth]{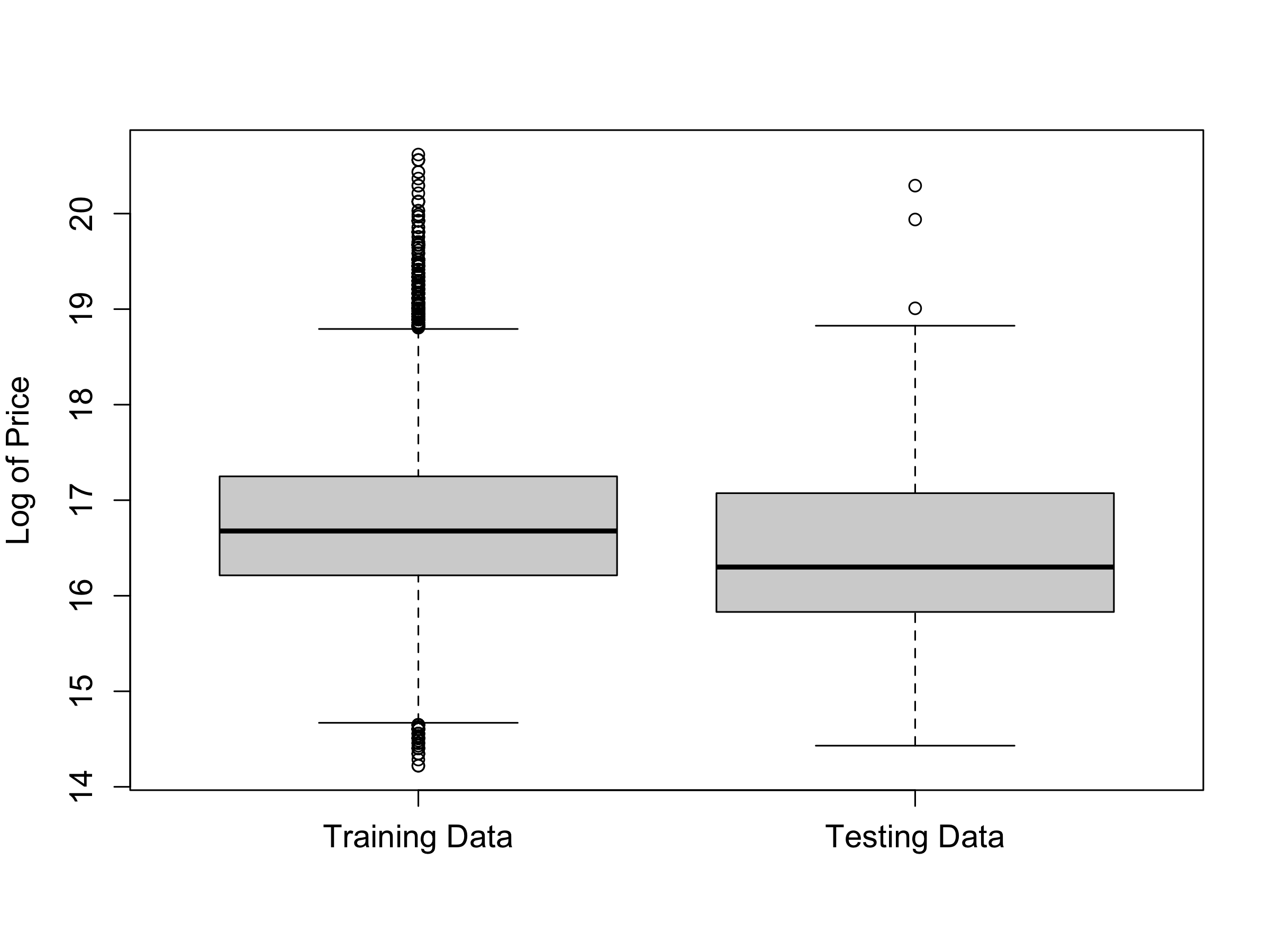}
    \caption{Distribution of Prices in RIMA Spatial Fold 2}
    \label{fig:rima_spatial_boxplot2}
\end{figure}

\begin{figure}[htbp]
    \centering
    \includegraphics[width=1\linewidth]{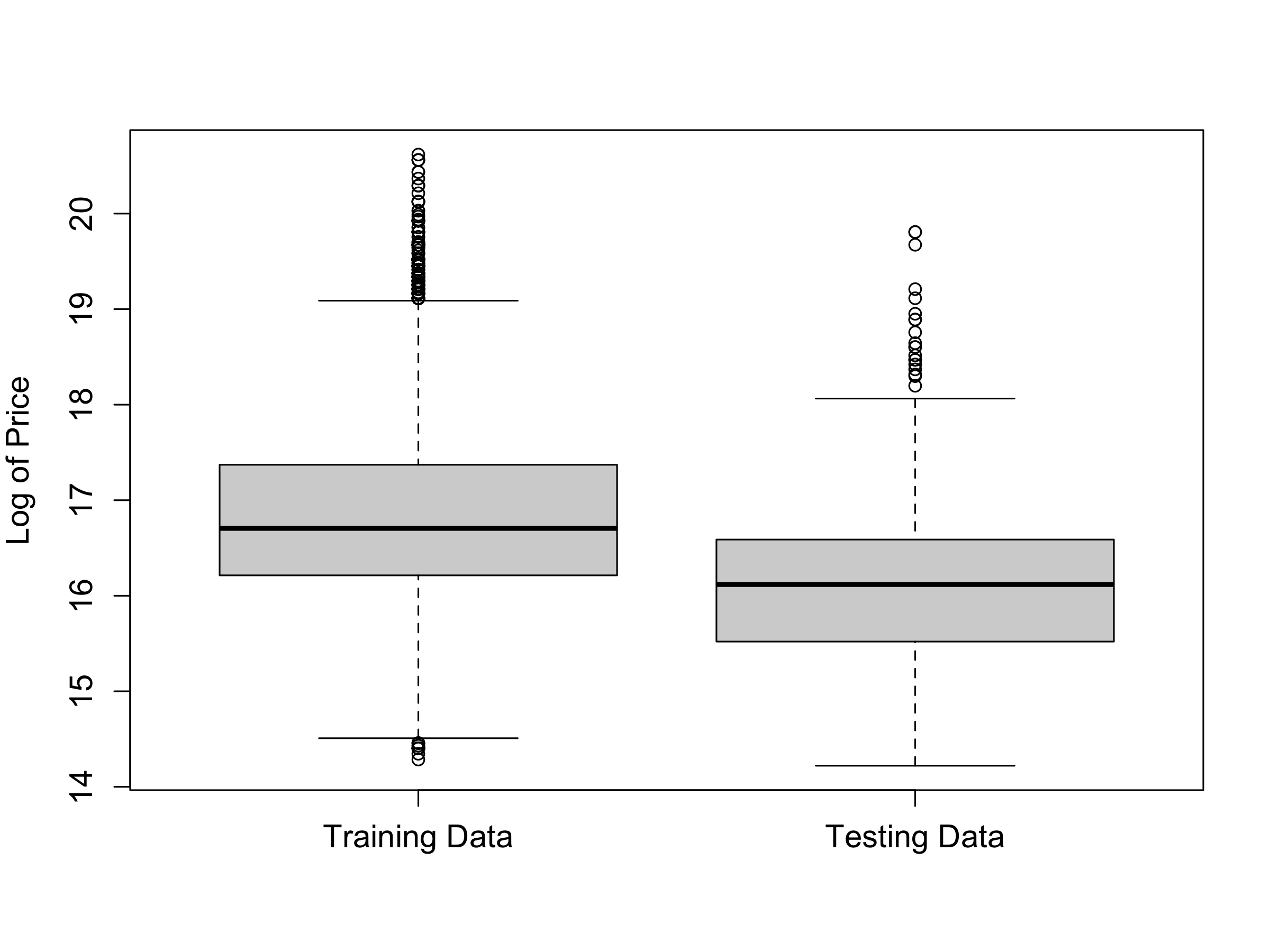}
    \caption{Distribution of Prices in RIMA Spatial Fold 3}
    \label{fig:rima_spatial_boxplot3}
\end{figure}

\begin{figure}[htbp]
    \centering
    \includegraphics[width=1\linewidth]{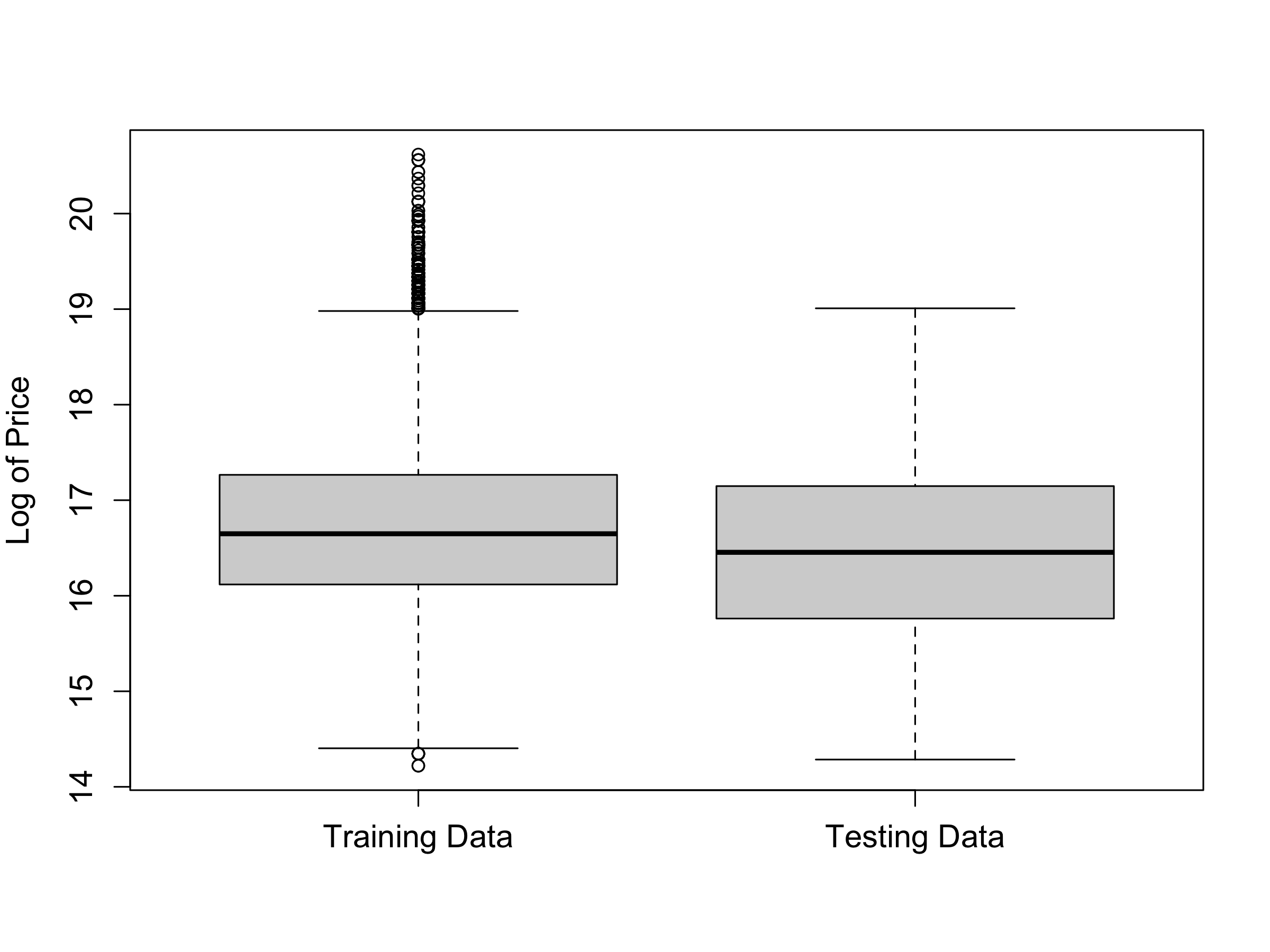}
    \caption{Distribution of Prices in RIMA Spatial Fold 4}
    \label{fig:rima_spatial_boxplot4}
\end{figure}

\begin{figure}[htbp]
    \centering
    \includegraphics[width=1\linewidth]{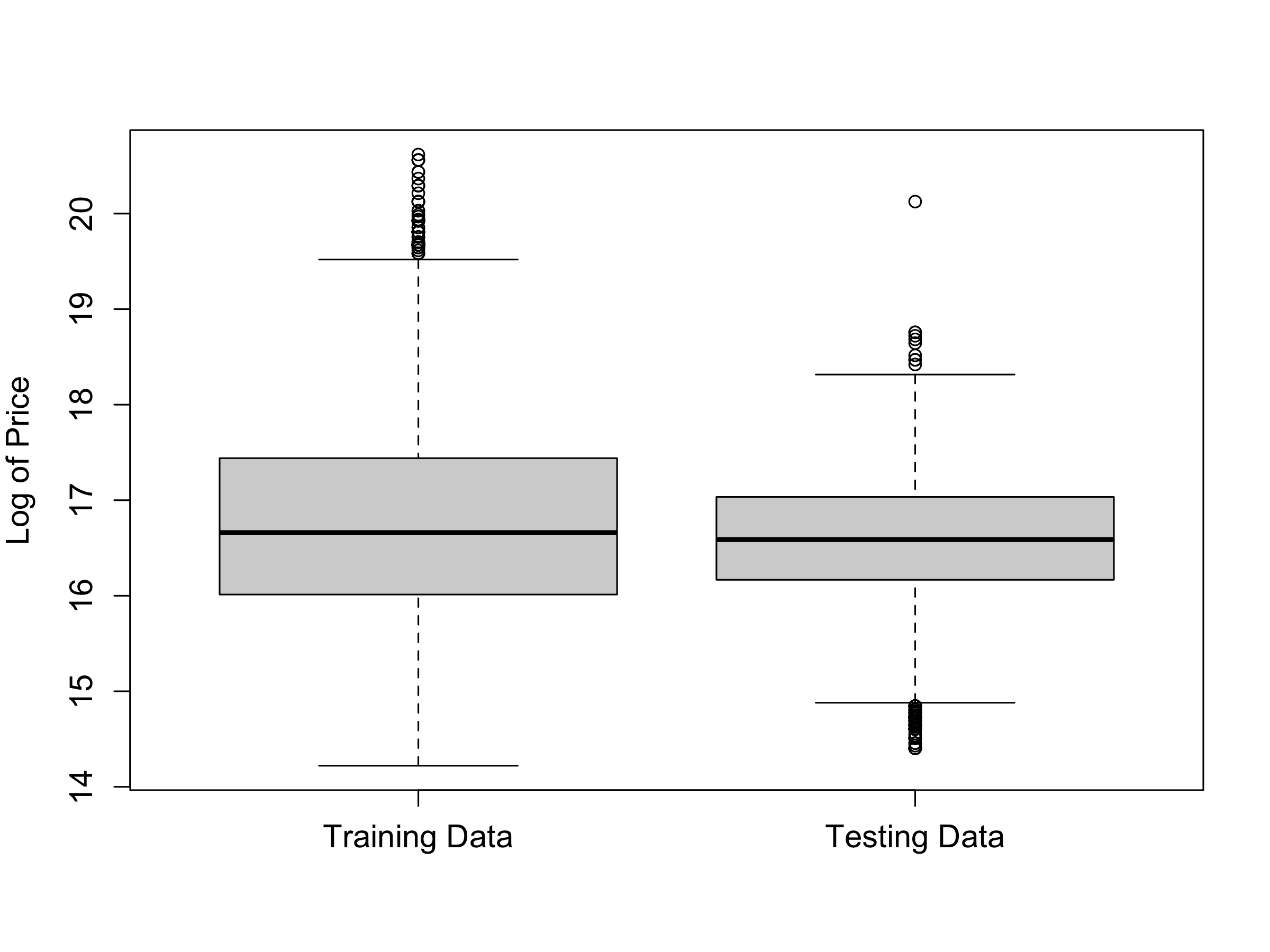}
    \caption{Distribution of Prices in RIMA Spatial Fold 5}
    \label{fig:rima_spatial_boxplot5}
\end{figure}

\begin{figure}[htbp]
    \centering
    \includegraphics[width=1\linewidth]{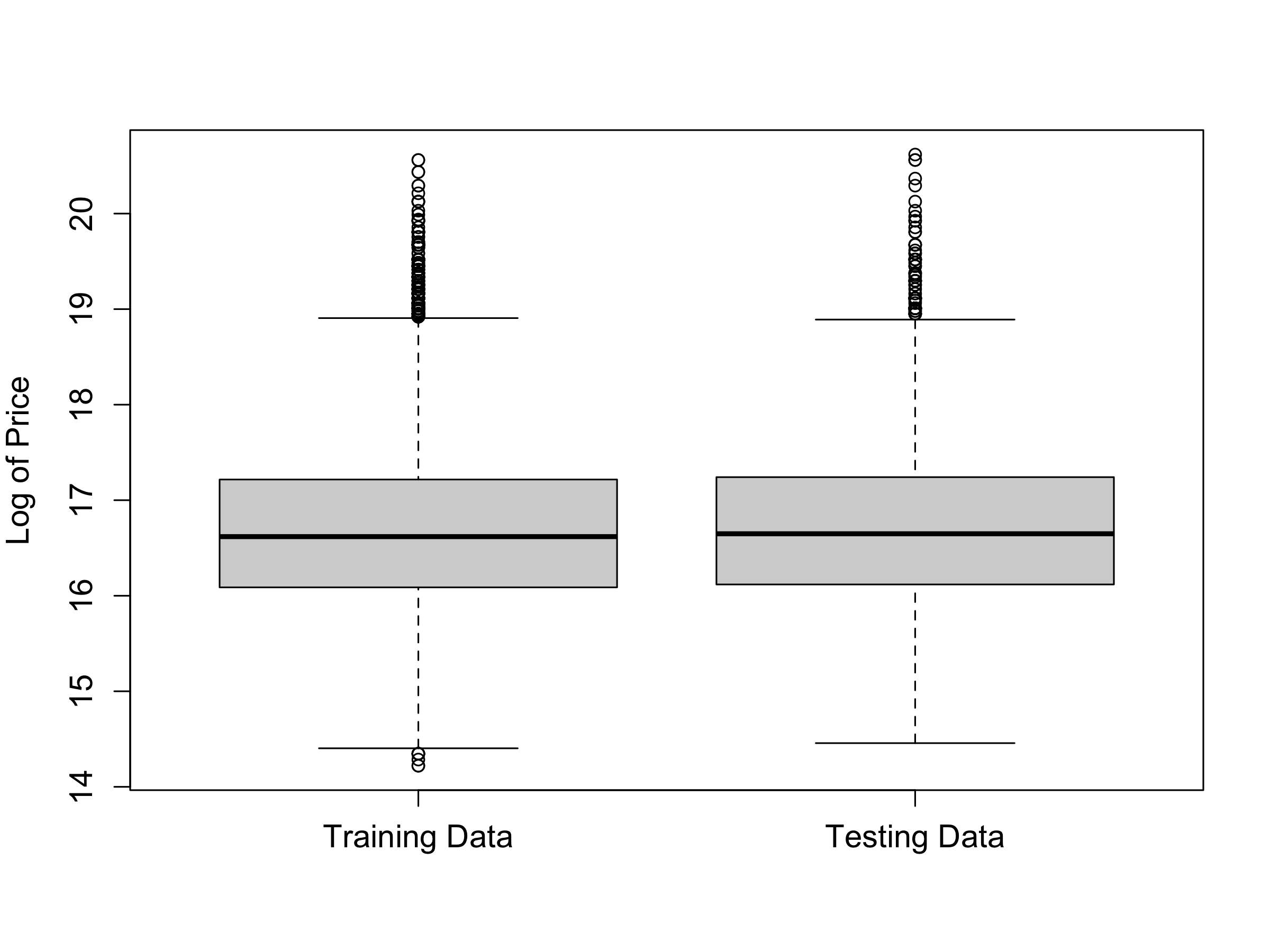}
    \caption{Distribution of Prices in RIMA Non-Spatial Fold 1}
    \label{fig:rima_aspatial_boxplot1}
\end{figure}

\begin{figure}[htbp]
    \centering
    \includegraphics[width=1\linewidth]{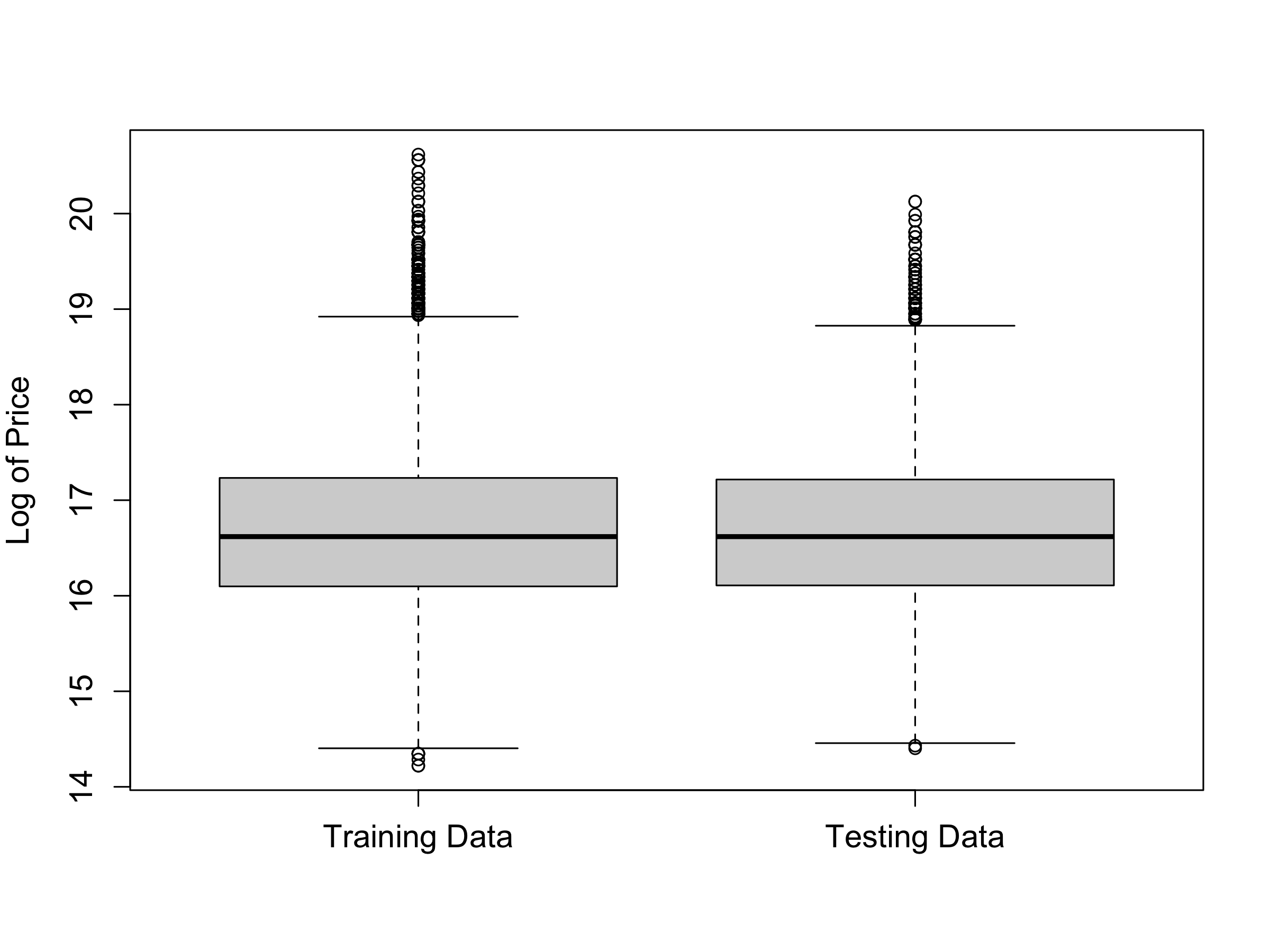}
    \caption{Distribution of Prices in RIMA Non-Spatial Fold 2}
    \label{fig:rima_aspatial_boxplot2}
\end{figure}

\begin{figure}[htbp]
    \centering
    \includegraphics[width=1\linewidth]{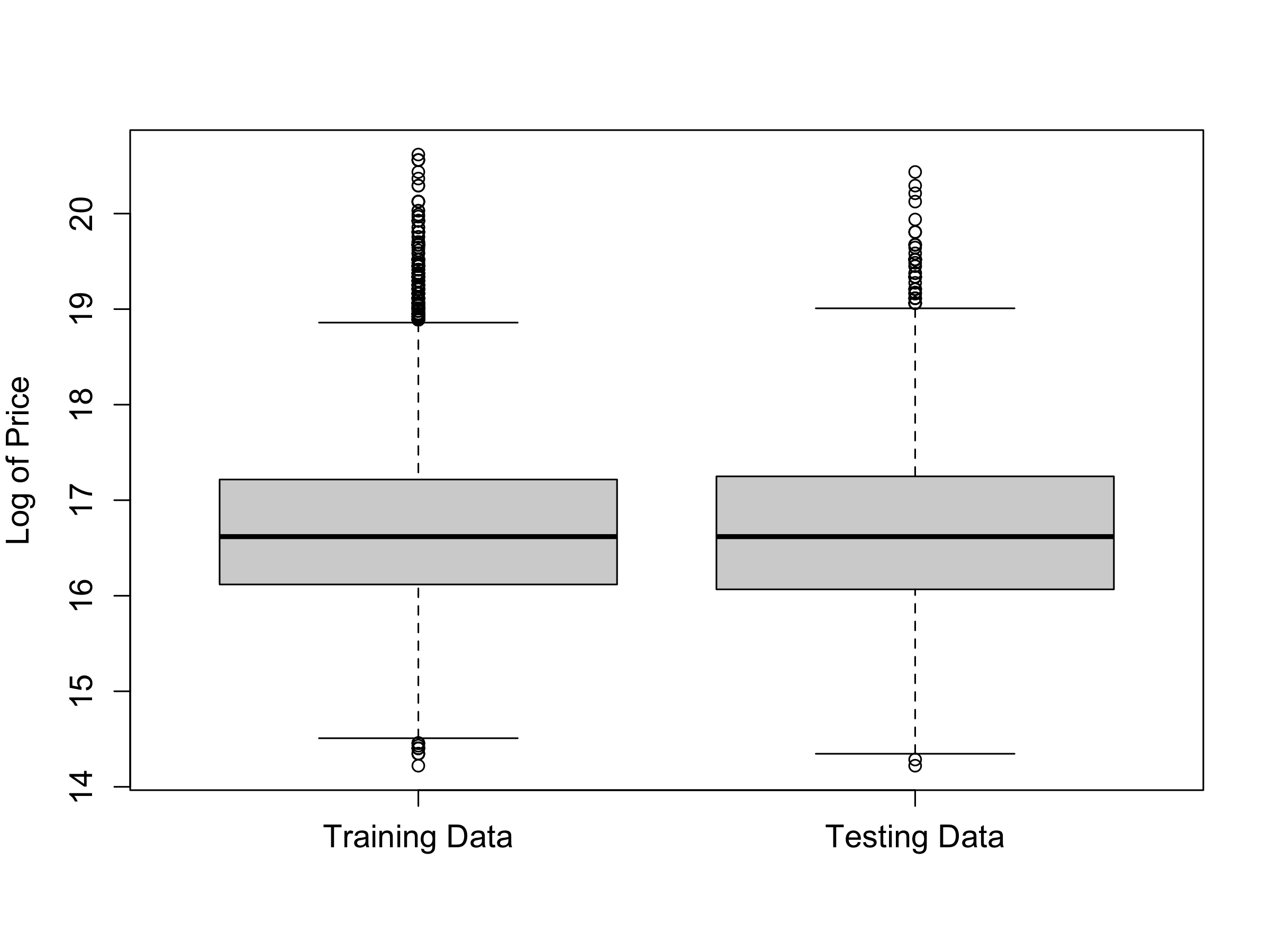}
    \caption{Distribution of Prices in RIMA Non-Spatial Fold 3}
    \label{fig:rima_aspatial_boxplot3}
\end{figure}

\begin{figure}[htbp]
    \centering
    \includegraphics[width=1\linewidth]{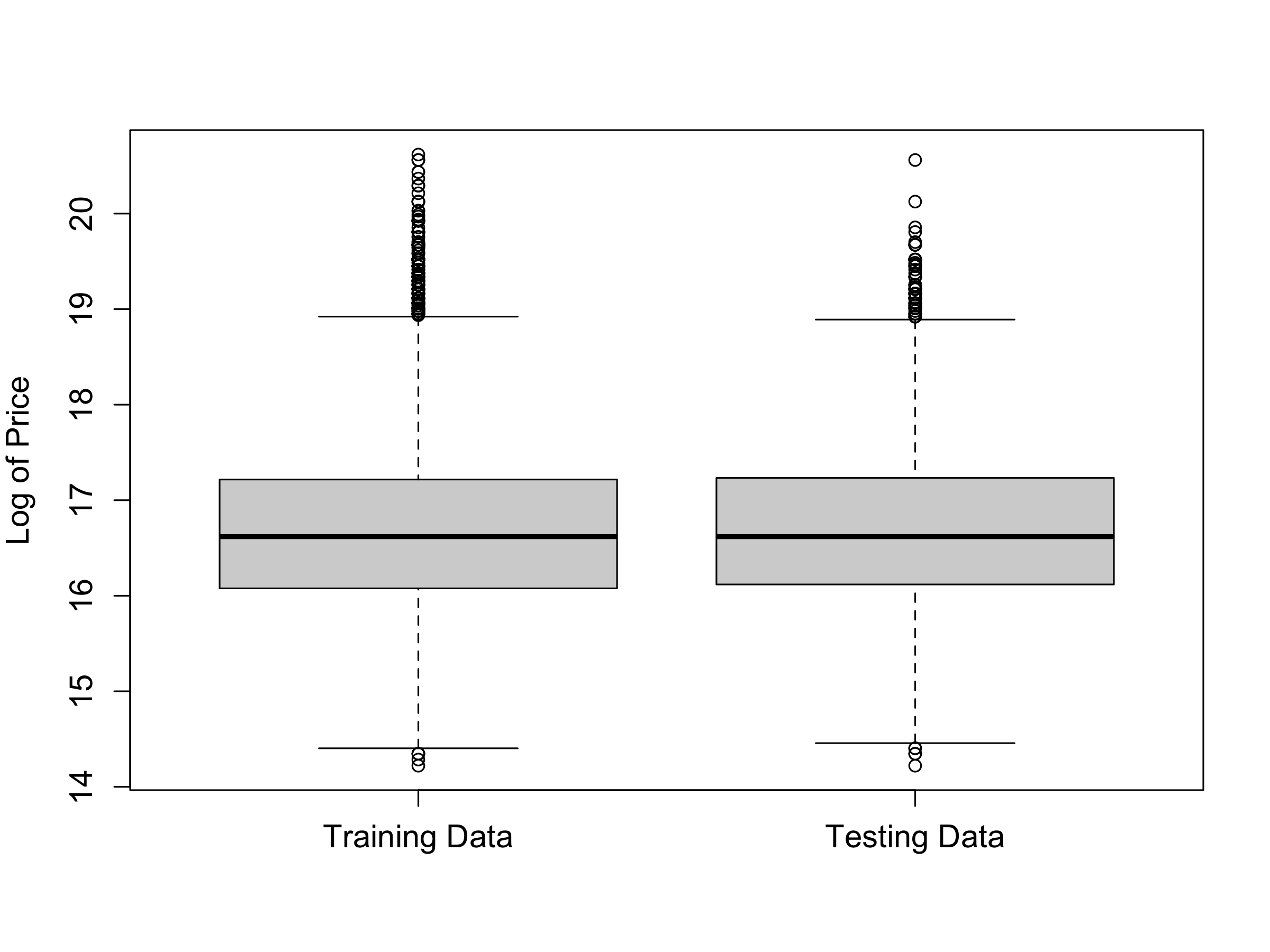}
    \caption{Distribution of Prices in RIMA Non-Spatial Fold 4}
    \label{fig:rima_aspatial_boxplot4}
\end{figure}

\begin{figure}[htbp]
    \centering
    \includegraphics[width=1\linewidth]{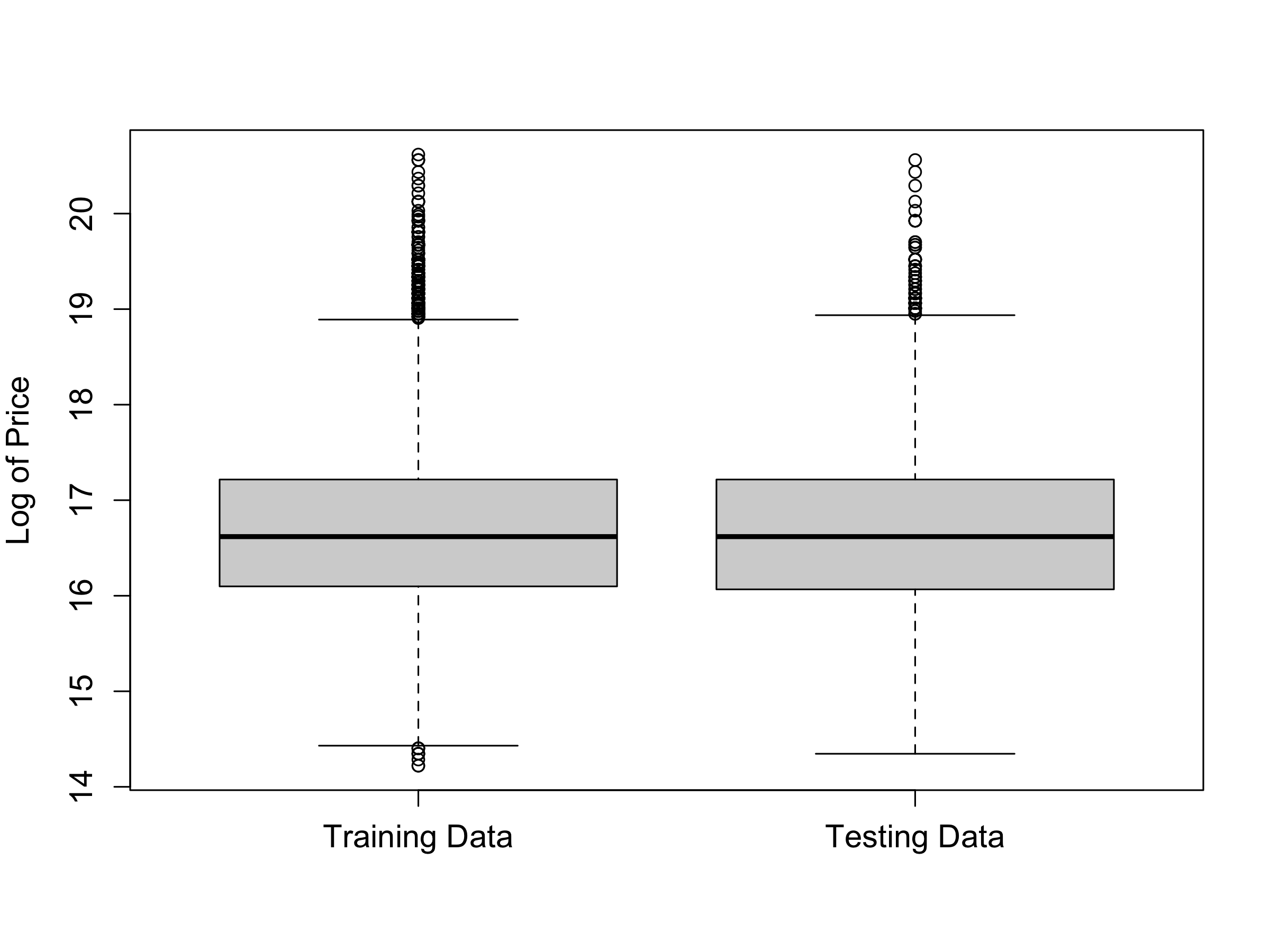}
    \caption{Distribution of Prices in RIMA Non-Spatial Fold 5}
    \label{fig:rima_aspatial_boxplot5}
\end{figure}

\FloatBarrier
\renewcommand\thefigure{\thesection C\arabic{figure}}
\setcounter{figure}{0} 
\section{Appendix C. SHAP Plots for Remaining RIMA and Toronto Folds}

\begin{figure*}[]
   \centering
    \includegraphics[width=0.8\linewidth]{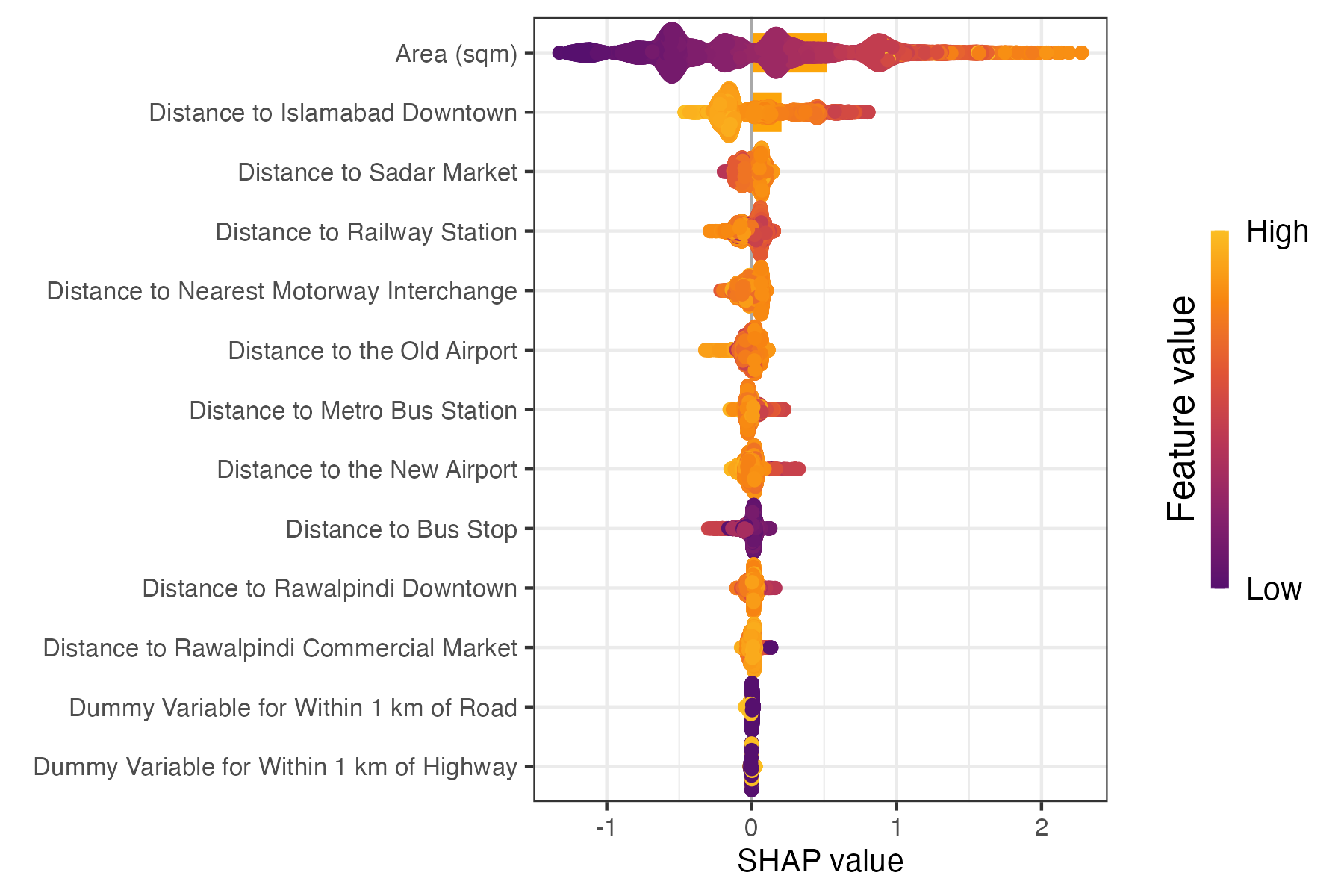}
    \caption{SHAP Values for Non-Spatial Model 2 (RIMA)}
    \label{fig:rima_shap_aspatial2}
\end{figure*}

\begin{figure*}[]
   \centering
    \includegraphics[width=0.8\linewidth]{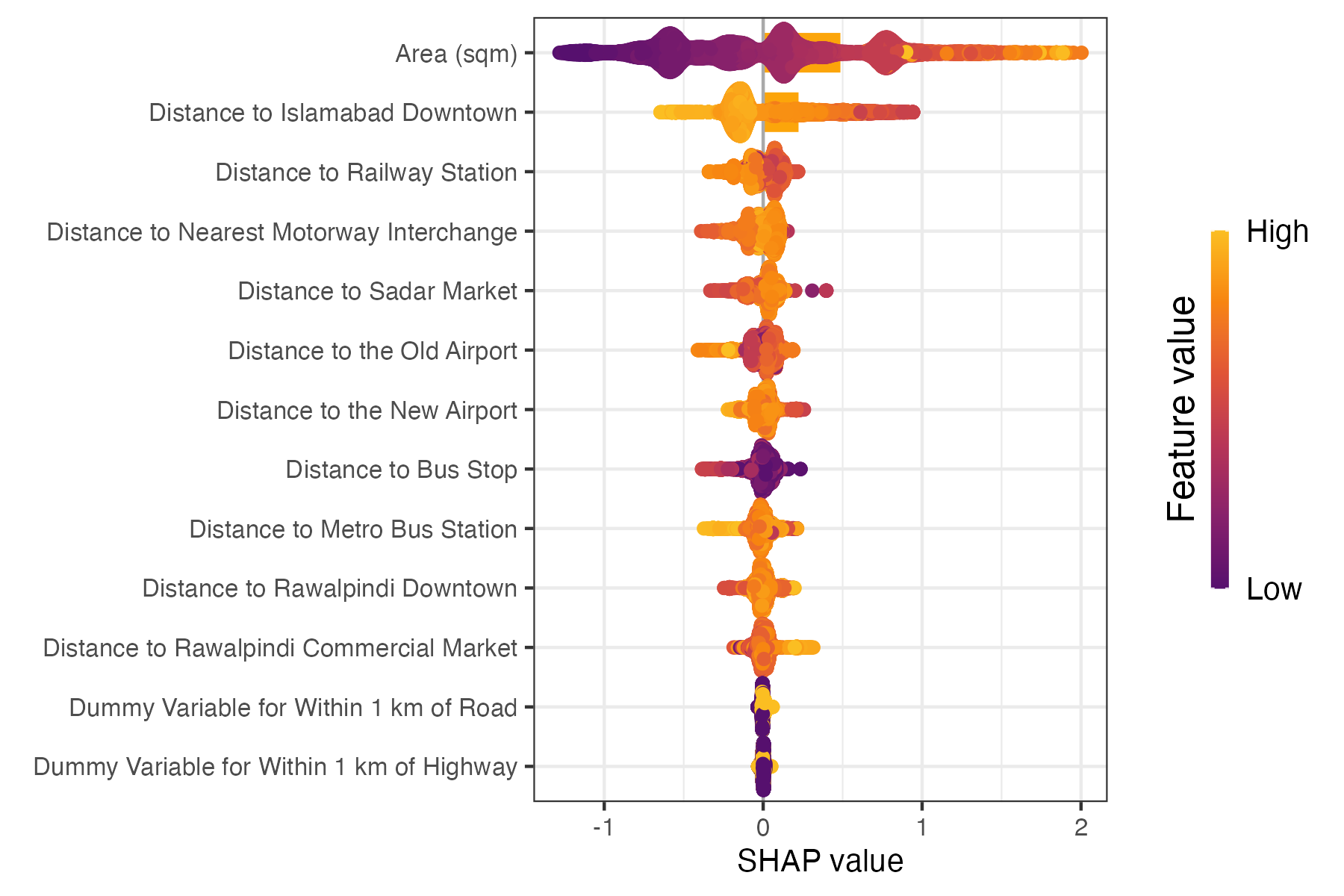}
    \caption{SHAP Values for Spatial Model 2 (RIMA)}
    \label{fig:rima_shap_spatial2}
\end{figure*}

\begin{figure*}[]
   \centering
    \includegraphics[width=0.8\linewidth]{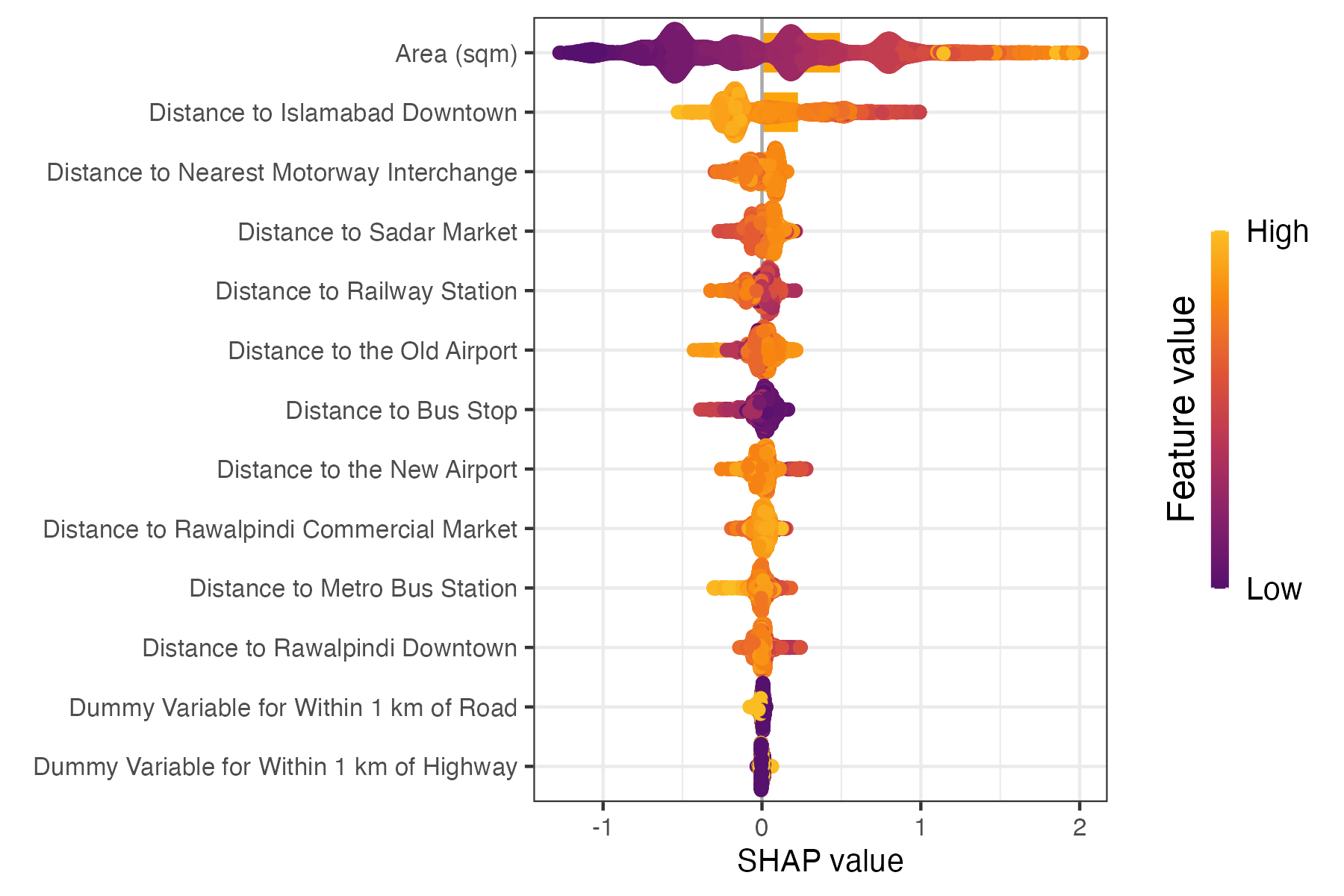}
    \caption{SHAP Values for Non-Spatial Model 3 (RIMA)}
    \label{fig:rima_shap_aspatial3}
\end{figure*}

\begin{figure*}[]
   \centering
    \includegraphics[width=0.8\linewidth]{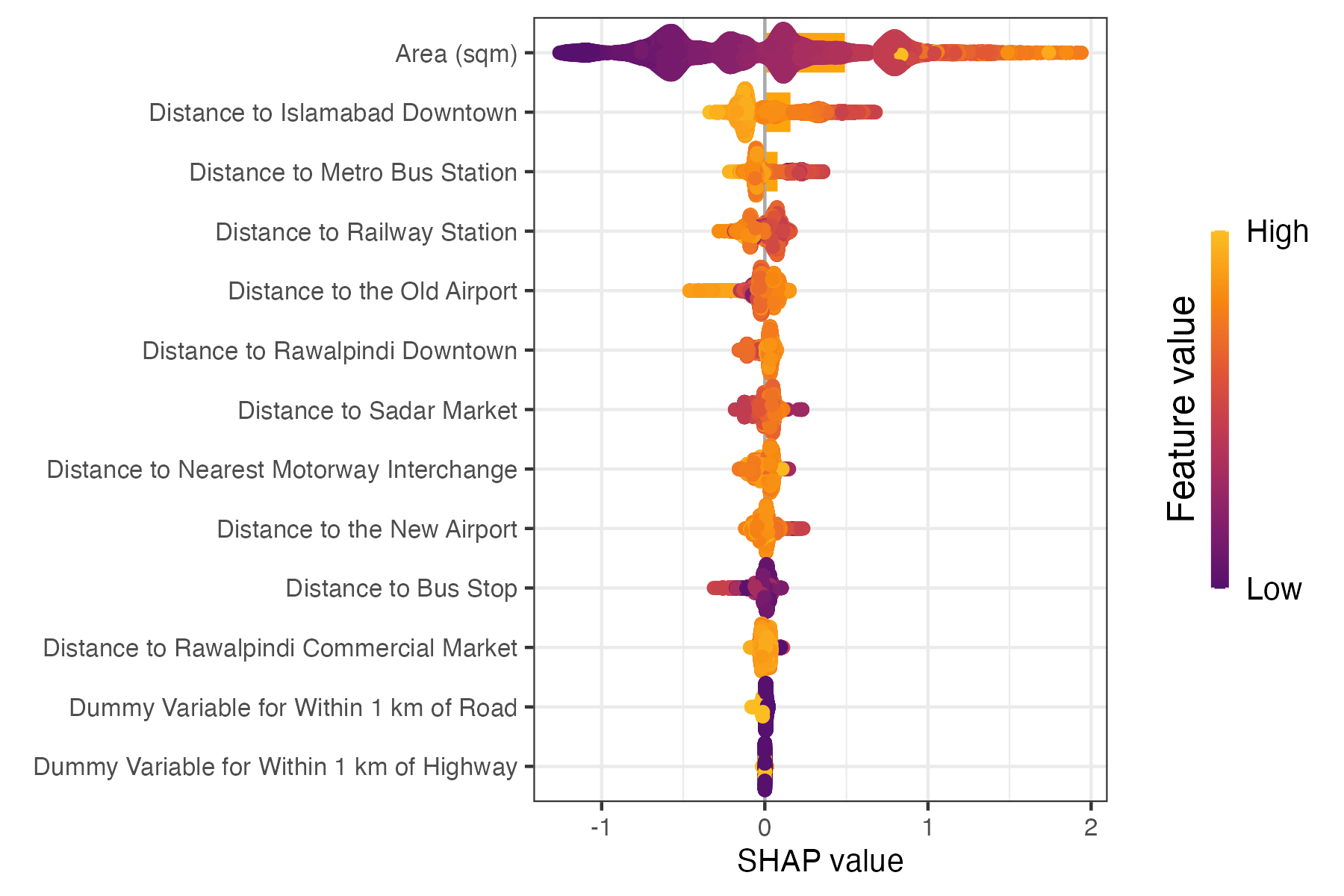}
    \caption{SHAP Values for Spatial Model 3 (RIMA)}
    \label{fig:rima_shap_spatial3}
\end{figure*}

\begin{figure*}[]
   \centering
    \includegraphics[width=0.8\linewidth]{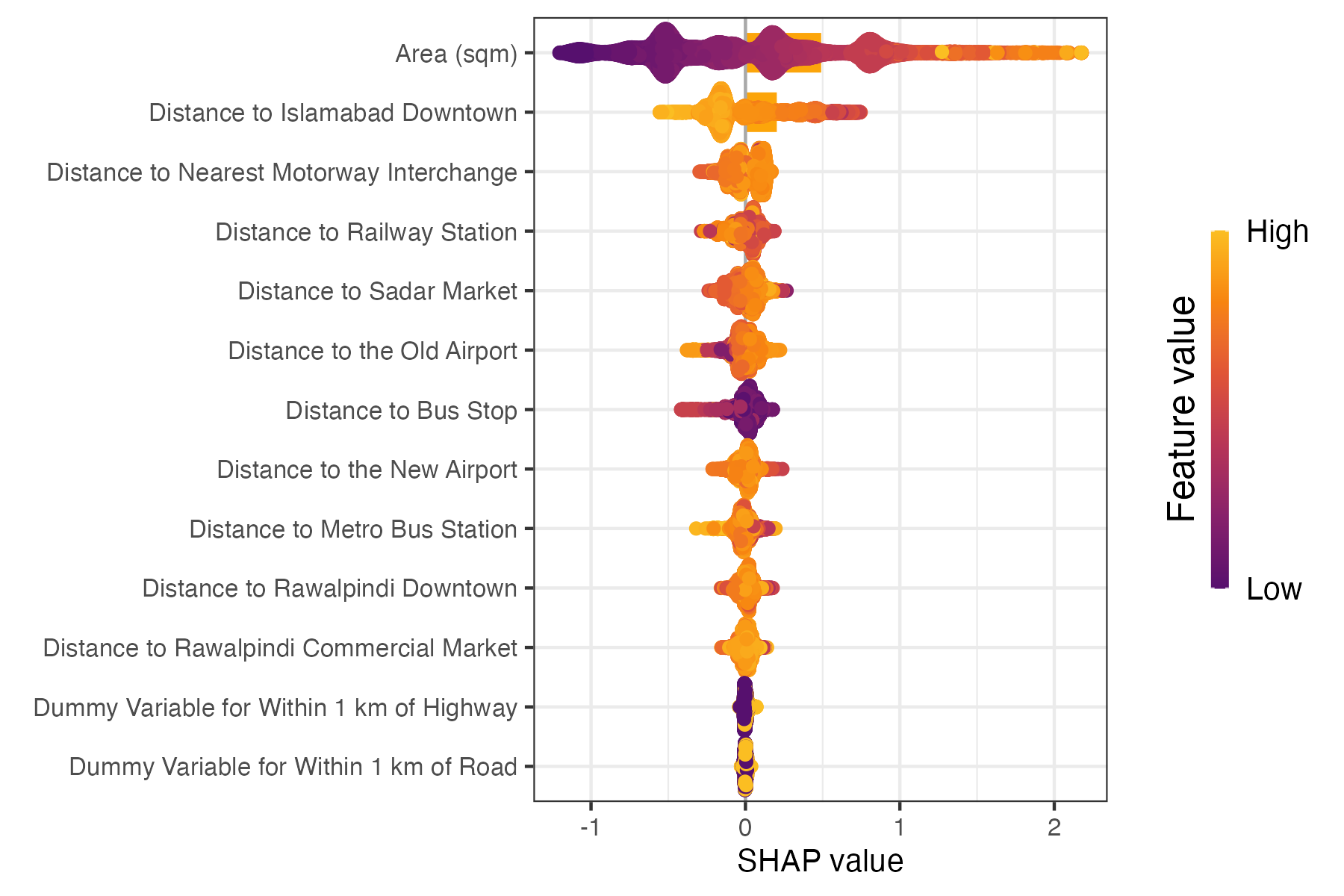}
    \caption{SHAP Values for Non-Spatial Model 4 (RIMA)}
    \label{fig:rima_shap_aspatial4}
\end{figure*}

\begin{figure*}[]
   \centering
    \includegraphics[width=0.8\linewidth]{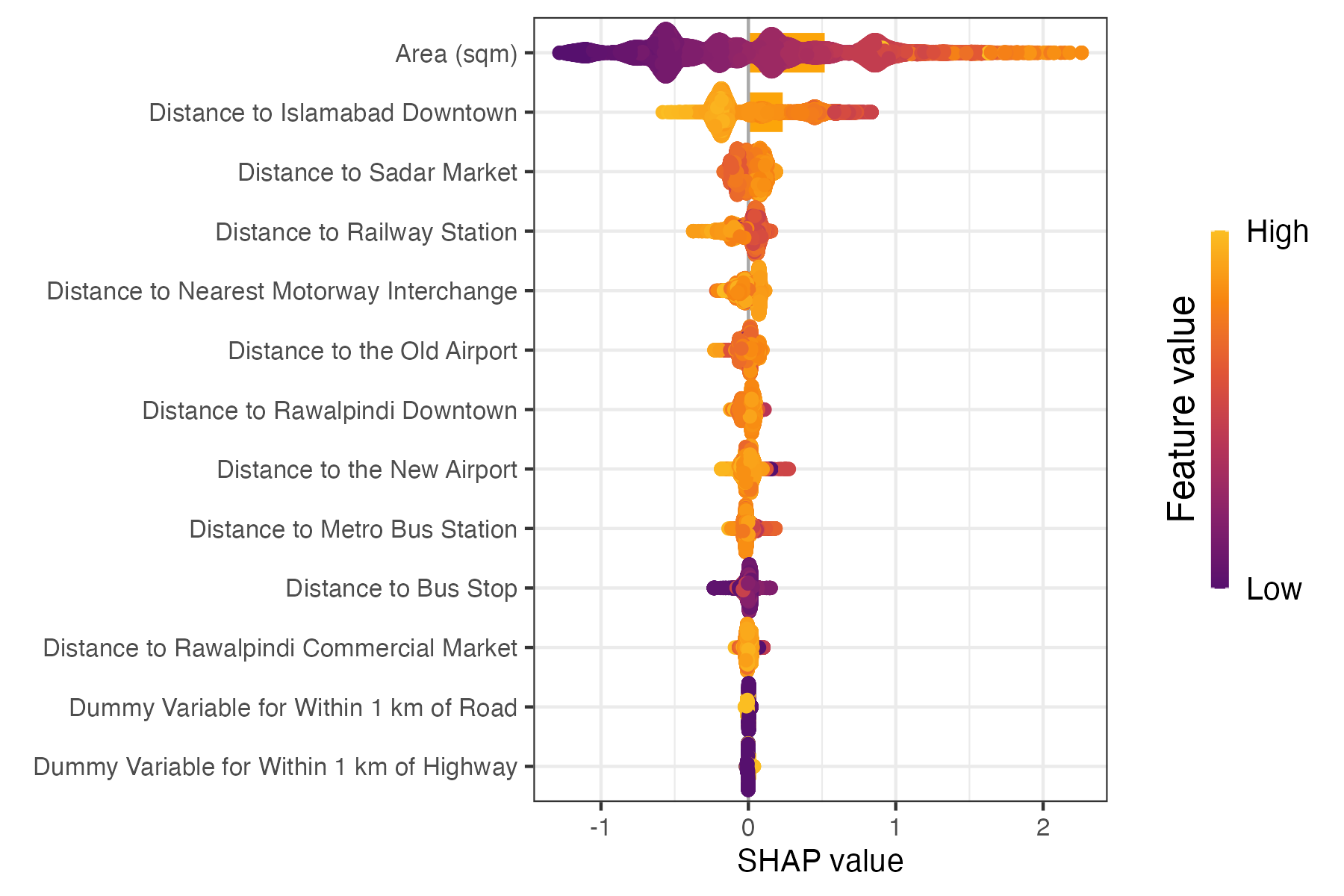}
    \caption{SHAP Values for Spatial Model 4 (RIMA)}
    \label{fig:rima_shap_spatial4}
\end{figure*}

\begin{figure*}[]
   \centering
    \includegraphics[width=0.8\linewidth]{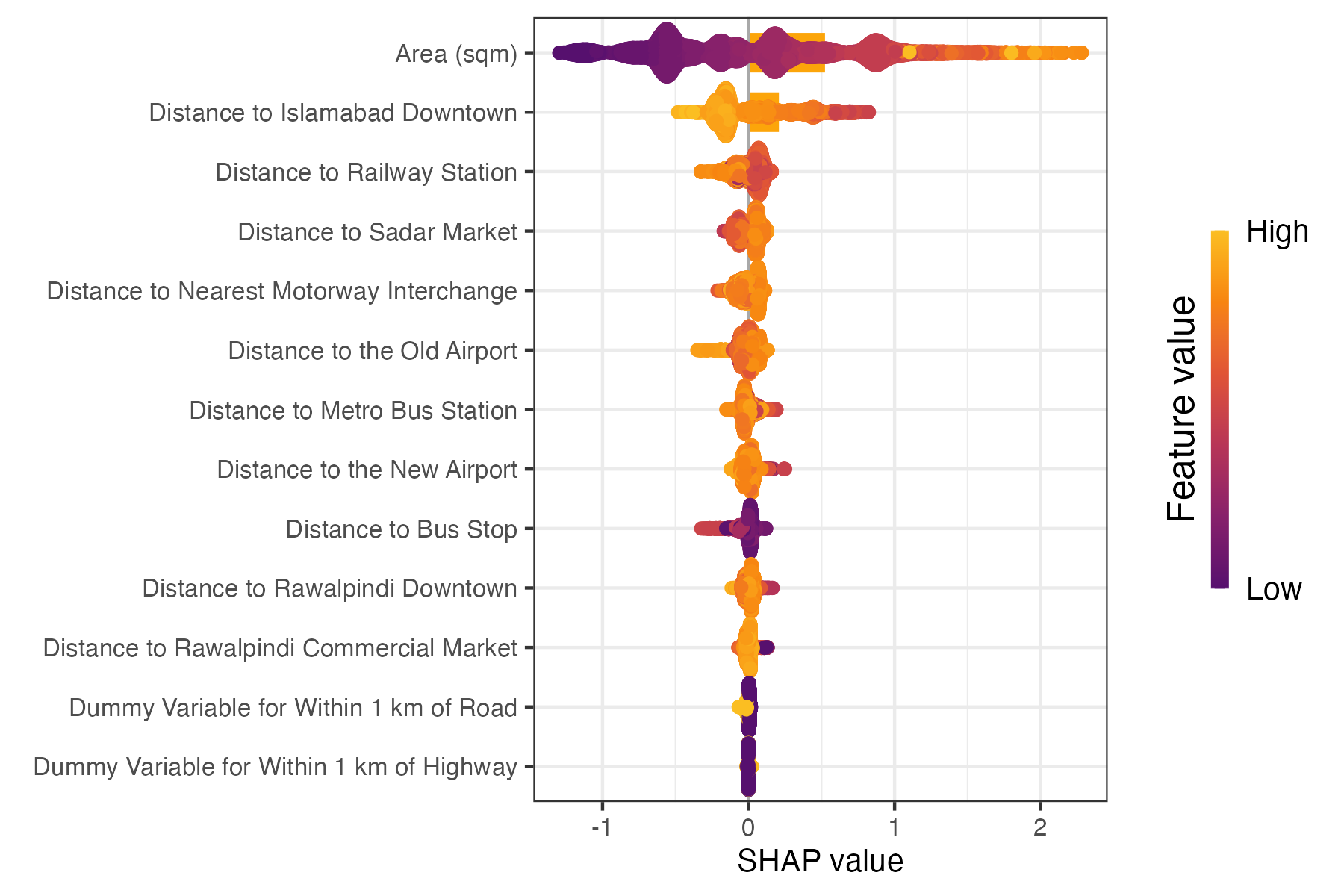}
    \caption{SHAP Values for Non-Spatial Model 5 (RIMA)}
    \label{fig:rima_shap_aspatial5}
\end{figure*}

\begin{figure*}[]
   \centering
    \includegraphics[width=0.8\linewidth]{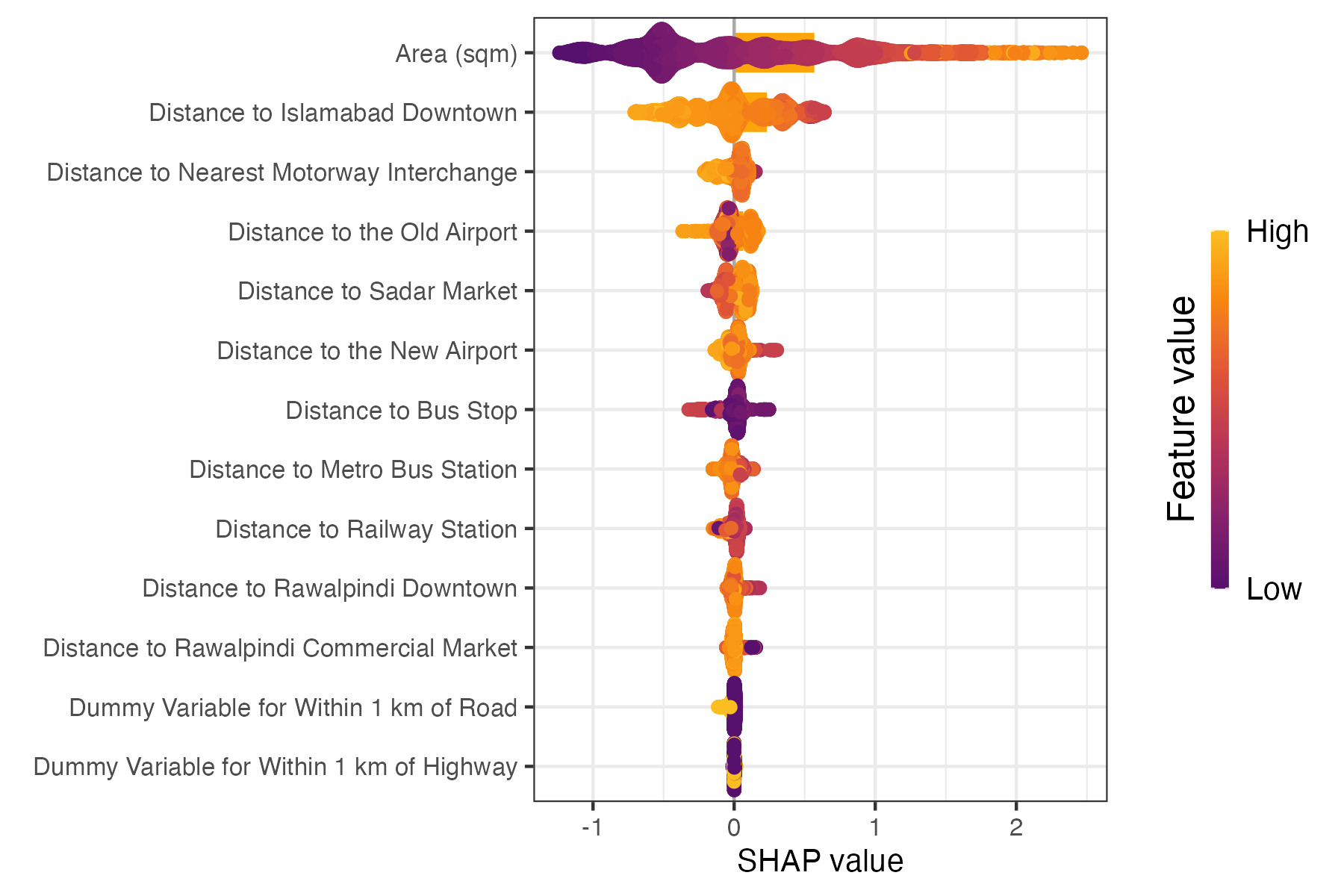}
    \caption{SHAP Values for Spatial Model 5 (RIMA)}
    \label{fig:rima_shap_spatial5}
\end{figure*}
\FloatBarrier


\begin{figure*}[]
   \centering
    \includegraphics[width=0.8\linewidth]{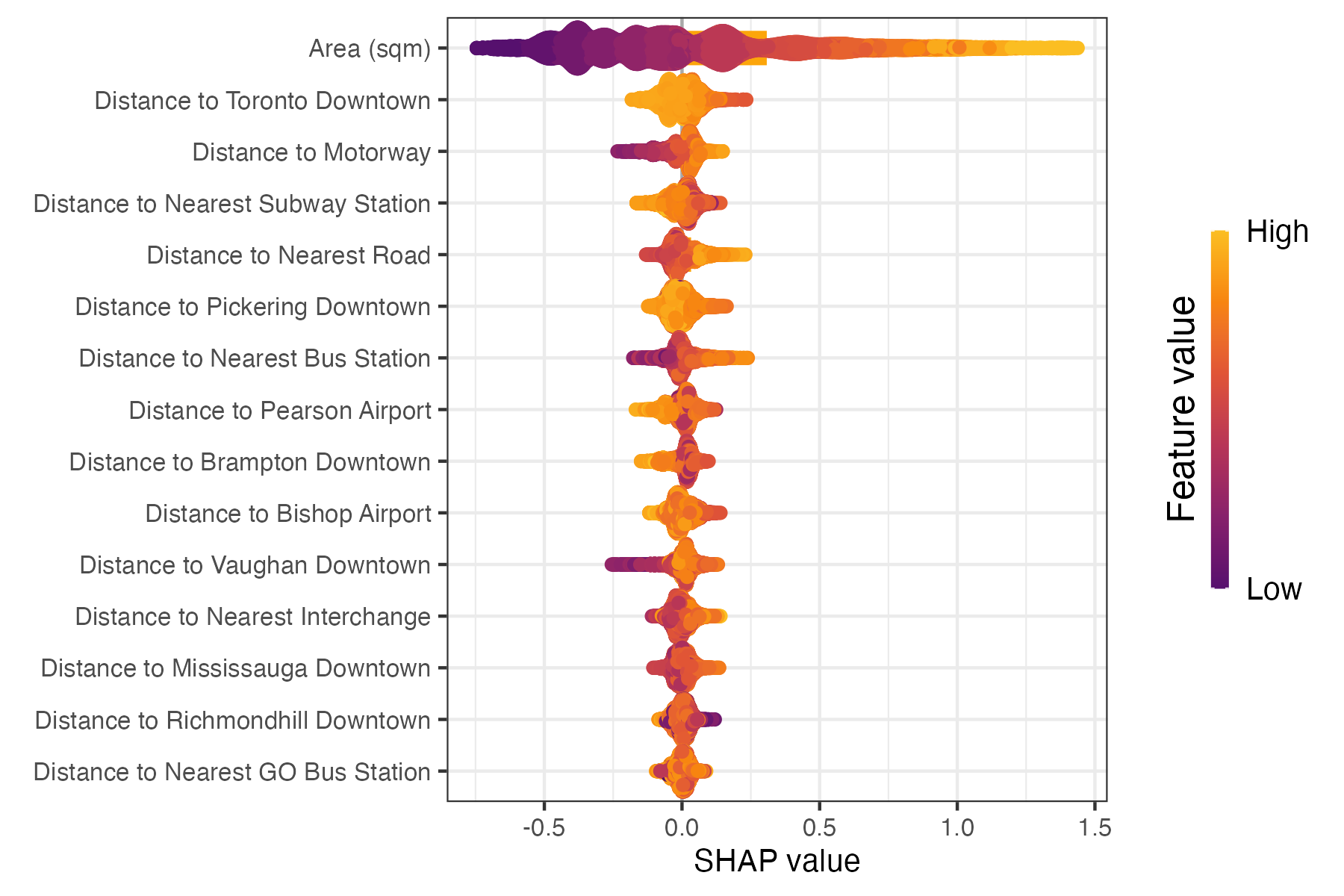}
    \caption{SHAP Values for Non-Spatial Model 2 (Toronto)}
    \label{fig:gta_shap_aspatial2}
\end{figure*}

\begin{figure*}[]
   \centering
    \includegraphics[width=0.8\linewidth]{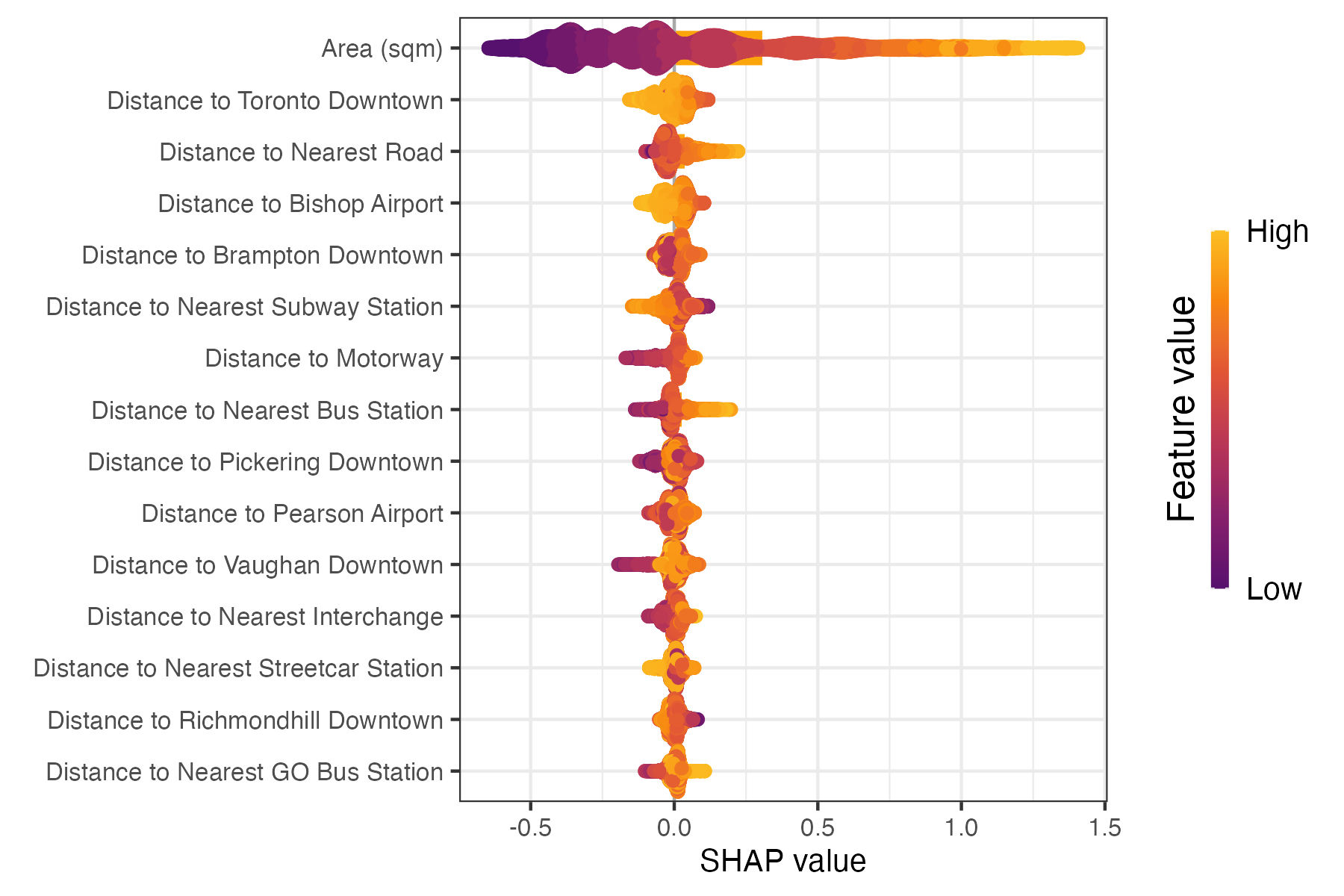}
    \caption{SHAP Values for Spatial Model 2 (Toronto)}
    \label{fig:gta_shap_spatial2}
\end{figure*}

\begin{figure*}[]
   \centering
    \includegraphics[width=0.8\linewidth]{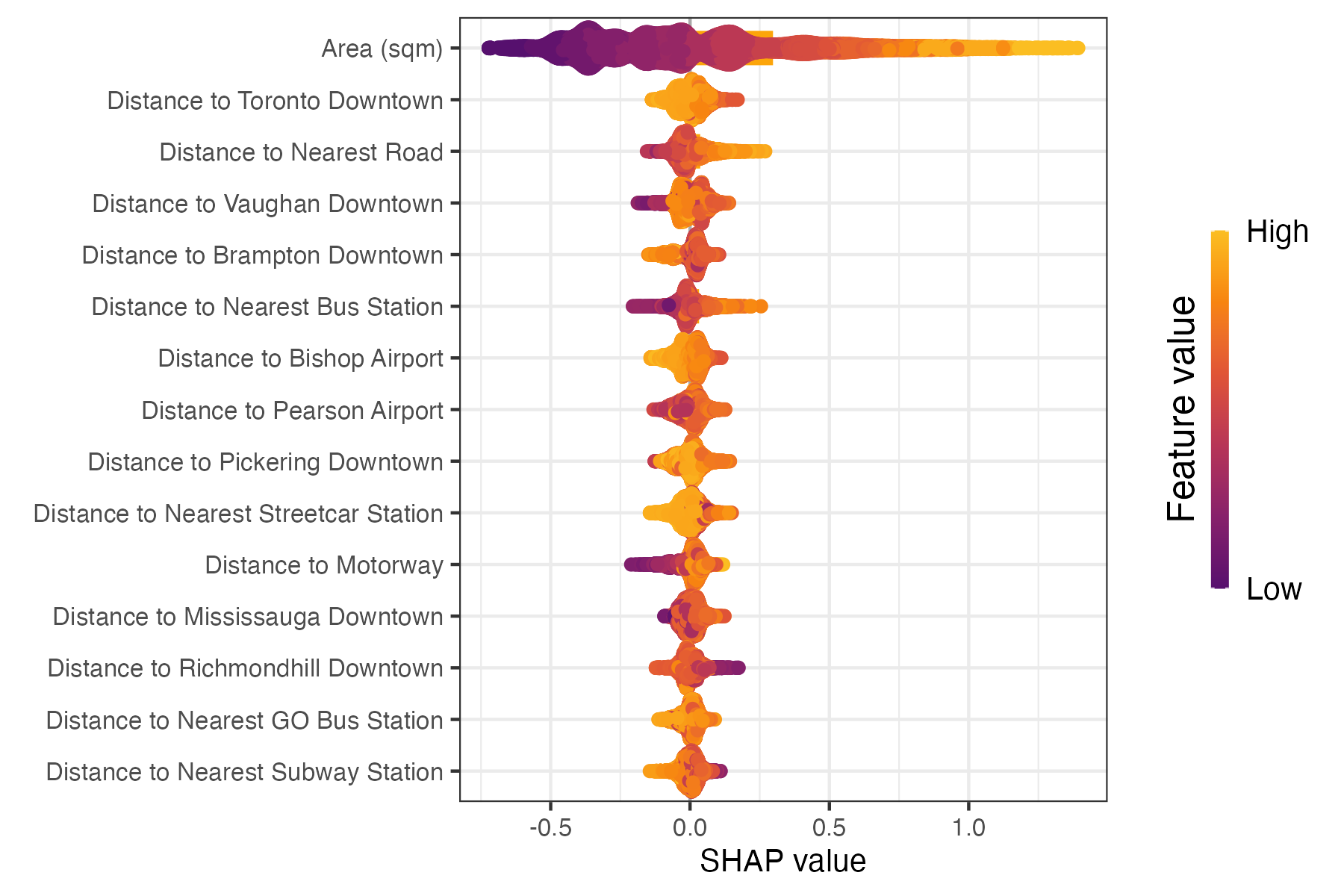}
    \caption{SHAP Values for Non-Spatial Model 3 (Toronto)}
    \label{fig:gta_shap_aspatial3}
\end{figure*}

\begin{figure*}[]
   \centering
    \includegraphics[width=0.8\linewidth]{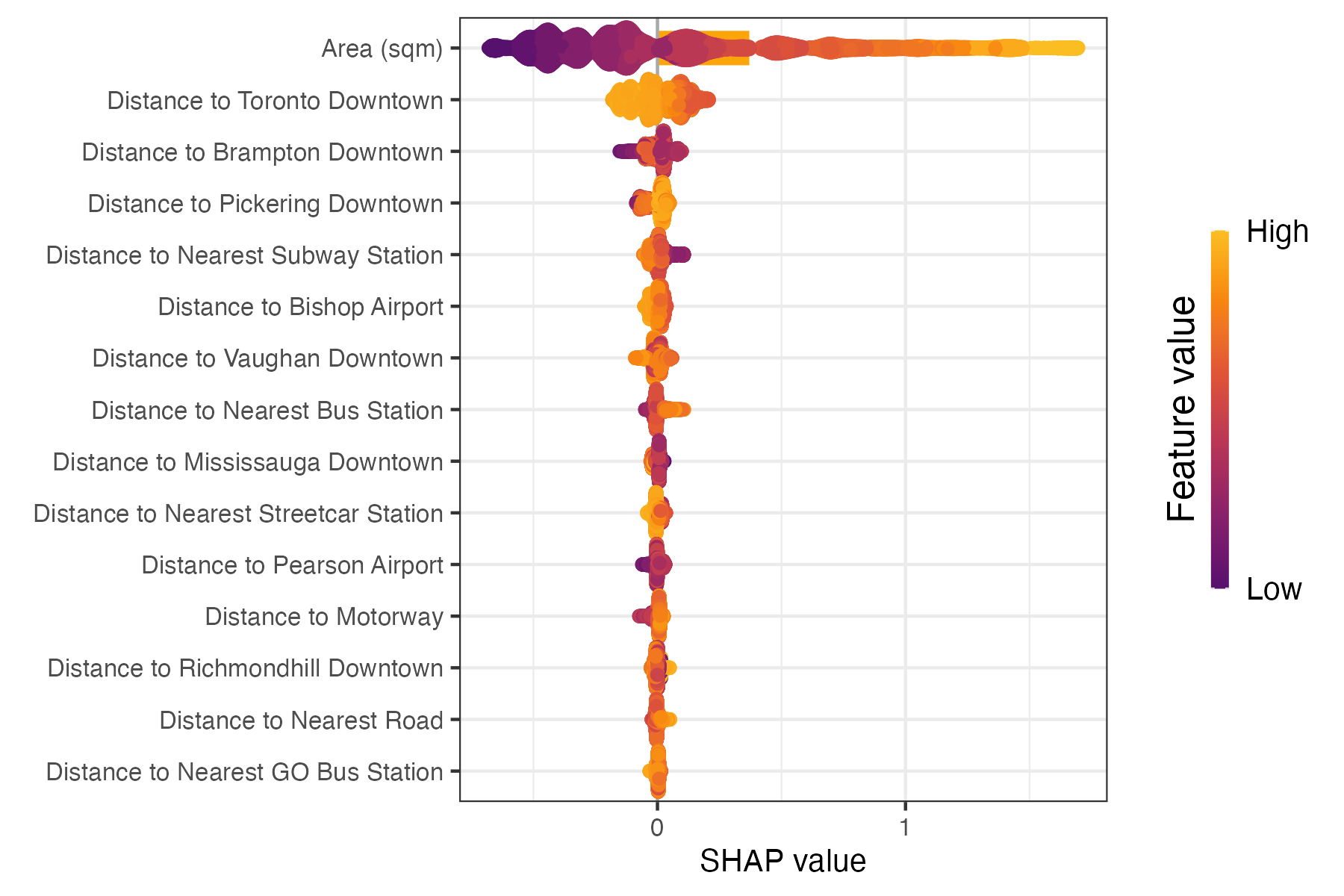}
    \caption{SHAP Values for Spatial Model 3 (Toronto)}
    \label{fig:gta_shap_spatial3}
\end{figure*}

\begin{figure*}[]
   \centering
    \includegraphics[width=0.8\linewidth]{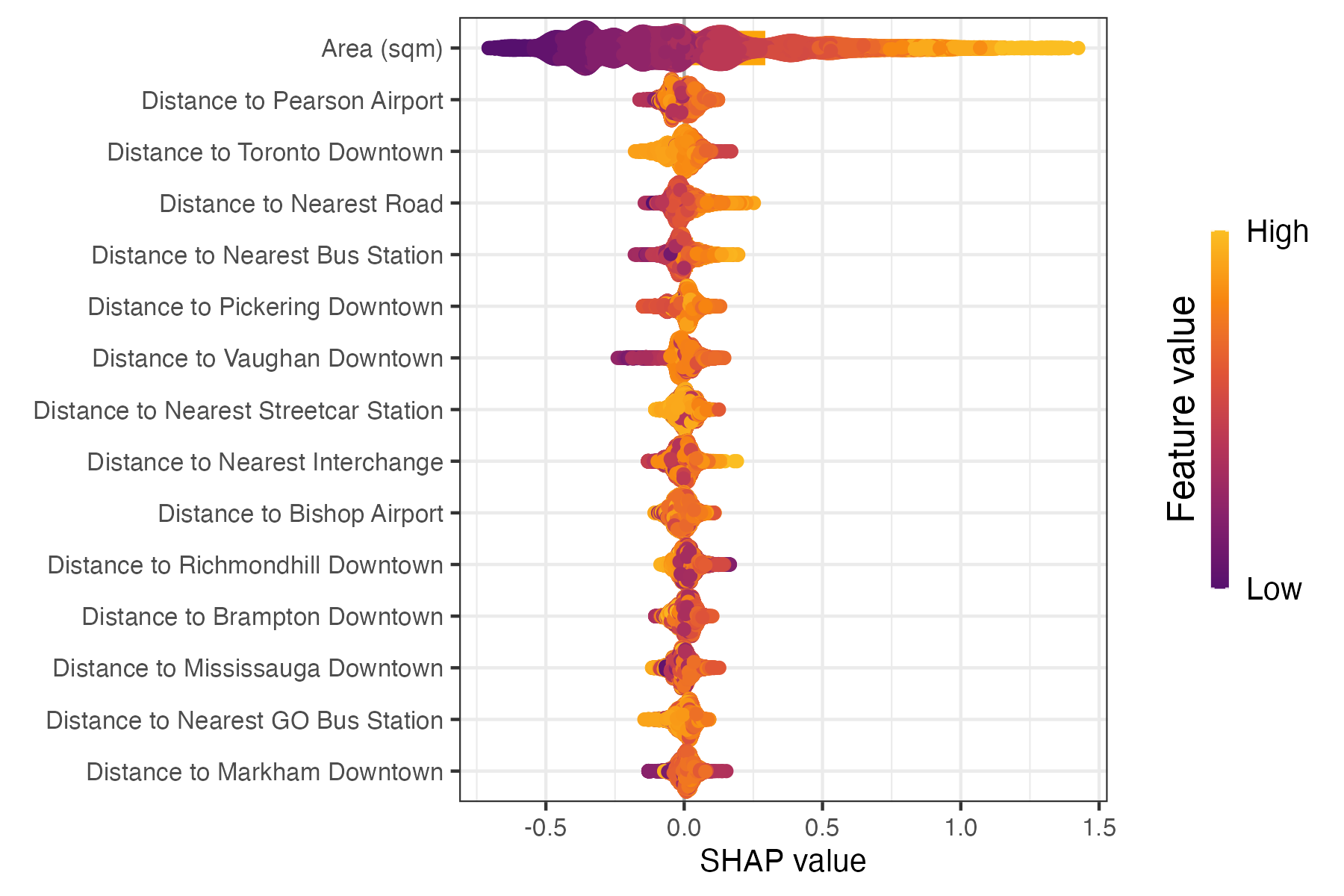}
    \caption{SHAP Values for Non-Spatial Model 4 (Toronto)}
    \label{fig:gta_shap_aspatial4}
\end{figure*}

\begin{figure*}[]
   \centering
    \includegraphics[width=0.8\linewidth]{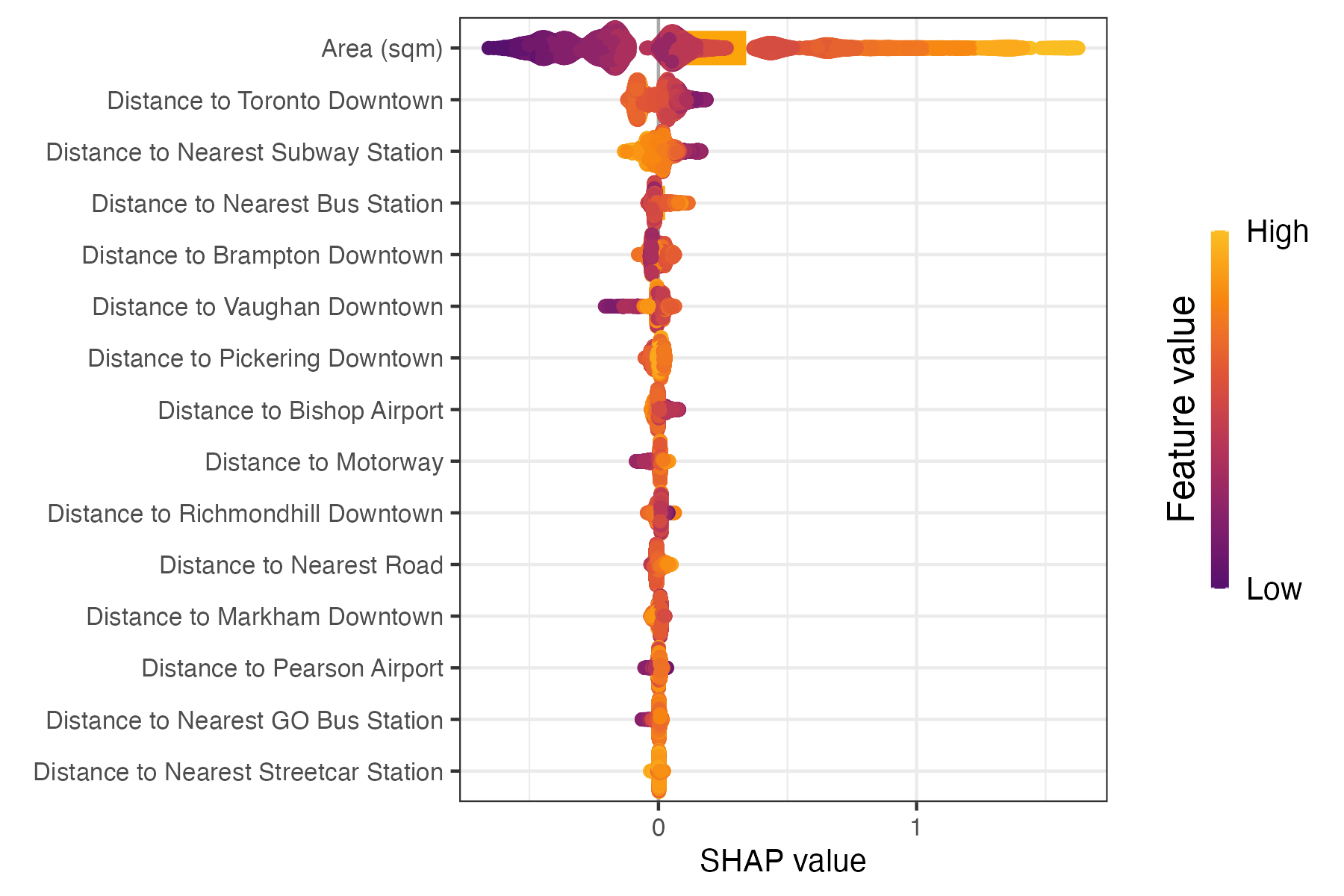}
    \caption{SHAP Values for Spatial Model 4 (Toronto)}
    \label{fig:gta_shap_spatial4}
\end{figure*}

\begin{figure*}[]
   \centering
    \includegraphics[width=0.8\linewidth]{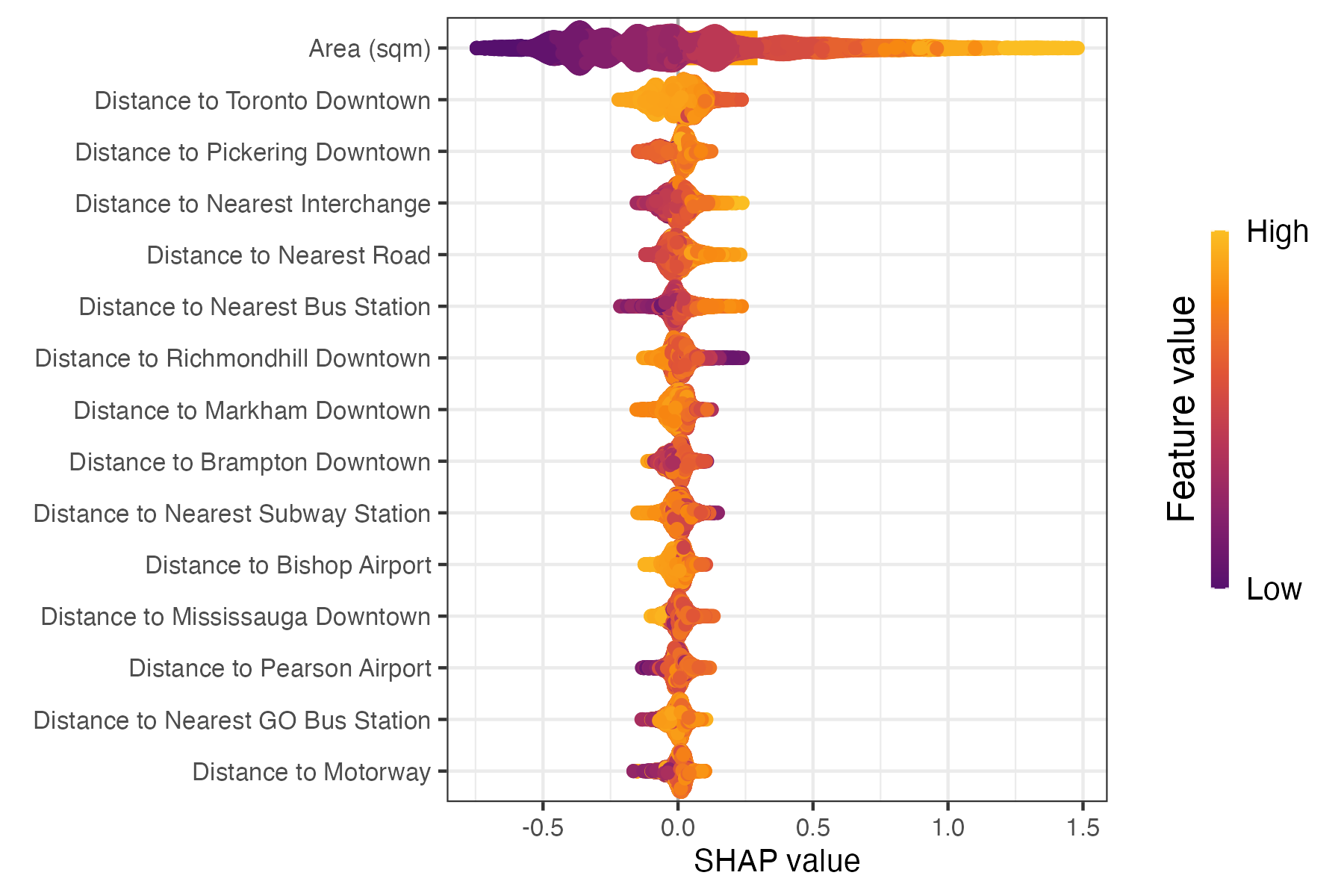}
    \caption{SHAP Values for Non-Spatial Model 5 (Toronto)}
    \label{fig:gta_shap_aspatial5}
\end{figure*}

\begin{figure*}[]
   \centering
    \includegraphics[width=0.8\linewidth]{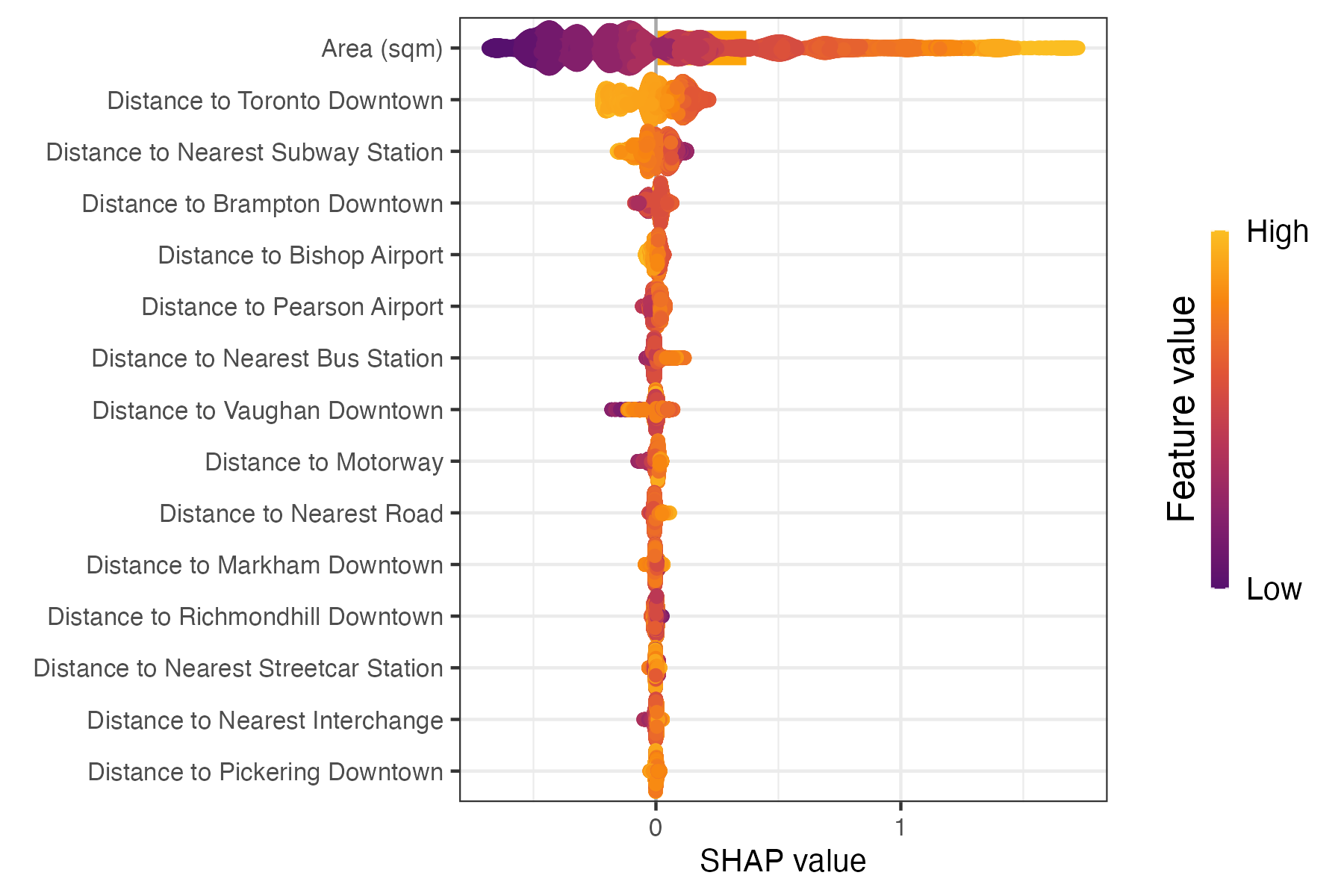}
    \caption{SHAP Values for Spatial Model 5 (Toronto)}
    \label{fig:gta_shap_spatial5}
\end{figure*}

\end{document}